\documentclass[11pt,a4paper,]{article}
\usepackage[left=1in,right=1in,top=1in,bottom=1in]{geometry}
\usepackage{authblk}
\usepackage{palatino} 
\usepackage[numbers, round]{natbib}
\usepackage[table,xcdraw]{xcolor}
\usepackage{booktabs} 

\usepackage{tikz}
\usetikzlibrary{arrows}
\usetikzlibrary{shapes.multipart}
\usepackage{tabularx}
\usepackage{adjustbox}
\usepackage{placeins}
\usepackage{algorithm}
\usepackage{algpseudocode}
\usepackage{amsmath}
\usepackage{blindtext}
\usepackage{bibunits}
\defaultbibliographystyle{naturemag}

\usepackage[all]{nowidow}

\usepackage[utf8]{inputenc} 
\usepackage[T1]{fontenc}    
\usepackage{hyperref}       
\hypersetup{
    colorlinks=false,
    linkcolor=blue,
    filecolor=magenta,      
    urlcolor=cyan,
    pdfpagemode=FullScreen,
    }
\usepackage{url}            
\usepackage{booktabs}       
\usepackage{amsfonts}       
\usepackage{nicefrac}       
\usepackage{microtype}      
\usepackage{xcolor}         
\usepackage{amsmath}
\usepackage{mathtools}
\usepackage{amsthm}
\usepackage{amssymb}
\usepackage{bbm}
\usepackage{wrapfig}
\usepackage{subcaption}
\usepackage[font=small,labelfont=bf]{caption}  
\usepackage{enumitem}

\usepackage[suppress]{color-edits}

\usepackage{xltabular} 

\usepackage{xcolor}
\usepackage{hyperref}
\hypersetup{
    colorlinks,
    linkcolor={red!50!black},
    citecolor={blue!50!black},
    urlcolor={blue!80!black}
}
\let\oldfootnote\footnote
\renewcommand{\footnote}[1]{%
  {\hypersetup{linkcolor=black}\oldfootnote{#1}}%
}
\usepackage{amsfonts}

\global\hyphenpenalty=1000

\newcommand{\SelectNode}{\textrm{SelectNode}}
\newcommand{\ExpandNode}{\textrm{ExpandNode}}
\newcommand{\Backpropagate}{\textrm{Backpropagate}}
\newcommand{\DepthCharge}{\textrm{DepthCharge}}
\newcommand{\PlacePiece}{\textrm{PlacePiece}}
\newcommand{\HasTerminated}{\textrm{HasTerminated}}
\newcommand{\LegalMoves}{\textrm{LegalMoves}}
\newcommand{\RandMove}{\textrm{RandMove}}
\newcommand{\Node}{\textrm{Node}}
\newcommand{\IsWin}{\textrm{IsWin}}
\newcommand{\IsDraw}{\textrm{IsDraw}}

\DeclareMathOperator{\argmax}{arg\,max}

\usepackage{xcolor}

\usepackage{tocloft}  
\usepackage{changes}
\definechangesauthor[name={Revisions}, color=red]{rev}

\title{\textbf{People use fast and flat simulation to reason about new games}}

\makeatletter
\renewenvironment{abstract}{%
  \vspace{-15pt}%
  \centerline{\large\bf Abstract}
  \setlength{\leftmargini}{30pt}%
  \begin{quote}%
    \small%
}{%
  \par%
  \end{quote}%
  \vskip 1ex%
}
\makeatother

\newcounter{extfigure}
\newcounter{exttable}

\newenvironment{extfigure}[1][h!]{%
    \stepcounter{extfigure}%
    \begin{figure}[#1]%
}{%
    \end{figure}%
}

\newif\ifnaturefinal

\newcommand{\figimage}[2][width=1.0\linewidth]{%
  \ifnaturefinal\else\includegraphics[#1]{#2}\fi
}

\begin{document}

\author[$\dagger$\,1,2,3]{Katherine M. Collins\footnote{Corresponding author: {katiemc@mit.edu}}}
\author[$\dagger$\,1]{Cedegao E. Zhang}
\author[$\dagger$\,1,4]{Lionel Wong}
\author[$\dagger$\,1]{\\Mauricio Barba da Costa}
\author[$\dagger$\,5]{Graham Todd}
\author[3,6]{Adrian Weller}
\author[1]{Samuel J. Cheyette}
\author[2]{\\Thomas L. Griffiths}
\author[1]{Joshua B. Tenenbaum}

\affil[1]{Massachusetts Institute of Technology, Cambridge, MA, United States}
\affil[2]{Princeton University, Princeton, NJ, United States}
\affil[3]{University of Cambridge, Cambridge, United Kingdom}
\affil[4]{Stanford University, Stanford, CA, United States}
\affil[5]{New York University, New York City, NY, United States}
\affil[6]{The Alan Turing Institute, London, United Kingdom}
\affil[$\dagger$]{indicates equal contributions.}

\date{}

\maketitle

\makeatletter
\newcounter{bibcontcnt}
\let\@oldthebibliography\thebibliography
\let\@oldendthebibliography\endthebibliography
\renewcommand{\thebibliography}[1]{%
  \@oldthebibliography{#1}%
  \setcounter{NAT@ctr}{\value{bibcontcnt}}%
}
\renewcommand{\endthebibliography}{%
  \setcounter{bibcontcnt}{\value{NAT@ctr}}%
  \@oldendthebibliography%
}
\makeatother

\begin{bibunit}[naturemag]
\begin{abstract}

Games have long been a microcosm for studying planning and reasoning in both natural and artificial intelligence (AI), often focusing on expert-level or even super-human play~\citep{chase1973mind, campbell2002deep, gobet2004moves, silver2016mastering, silver2017mastering_go, van2023expertise}. But real life also pushes human intelligence along a different frontier, requiring people to flexibly navigate decision-making problems that they have never thought about before. Here, we use \textit{novice} gameplay to study how people reason about new problem settings. Through a series of large-scale behavioral studies with over $1000$ participants and $121$ two-player strategic board games (almost all novel to our participants), we show that people are systematic and adaptively rational in how they play a game for the first time, or evaluate a game (e.g., how fair or how fun it is likely to be) before they have played it even once. We explain these capacities via a computational cognitive model that we call the ``Intuitive Gamer'', a model based on mechanisms of \textit{fast and flat (depth-limited) goal-directed probabilistic simulation}. Our work offers new insights into how people rapidly evaluate, act, and make suggestions when encountering novel problems, and could inform the design of more flexible and human-like AI systems that can determine not just how to solve new tasks, but whether a task is worth thinking about at all. 

\end{abstract}

Games occupy a special space in psychology and and computer science~\citep{cleveland1907psychology, shannon1950, newell1958chess, chase1973mind, campbell2002deep, gobet2004moves, mnih2015human, silver2016mastering, tsividis2021human, yannakakis2018artificial, van2023expertise} for good reason: games are systems of rules and reward that hold a mirror to the structures, patterns and challenges that reality confronts us with, across familiar and merely possible experiences. Mancala lets us sow and capture abstract resources. Chess lets us strategize over bloodless battlefields.  Go lets us encircle and claim territories with silent stone troops. A good game lets us thrill to uncertainties and losses that we might shy from in real life.

\begin{figure}
    \centering
    \figimage{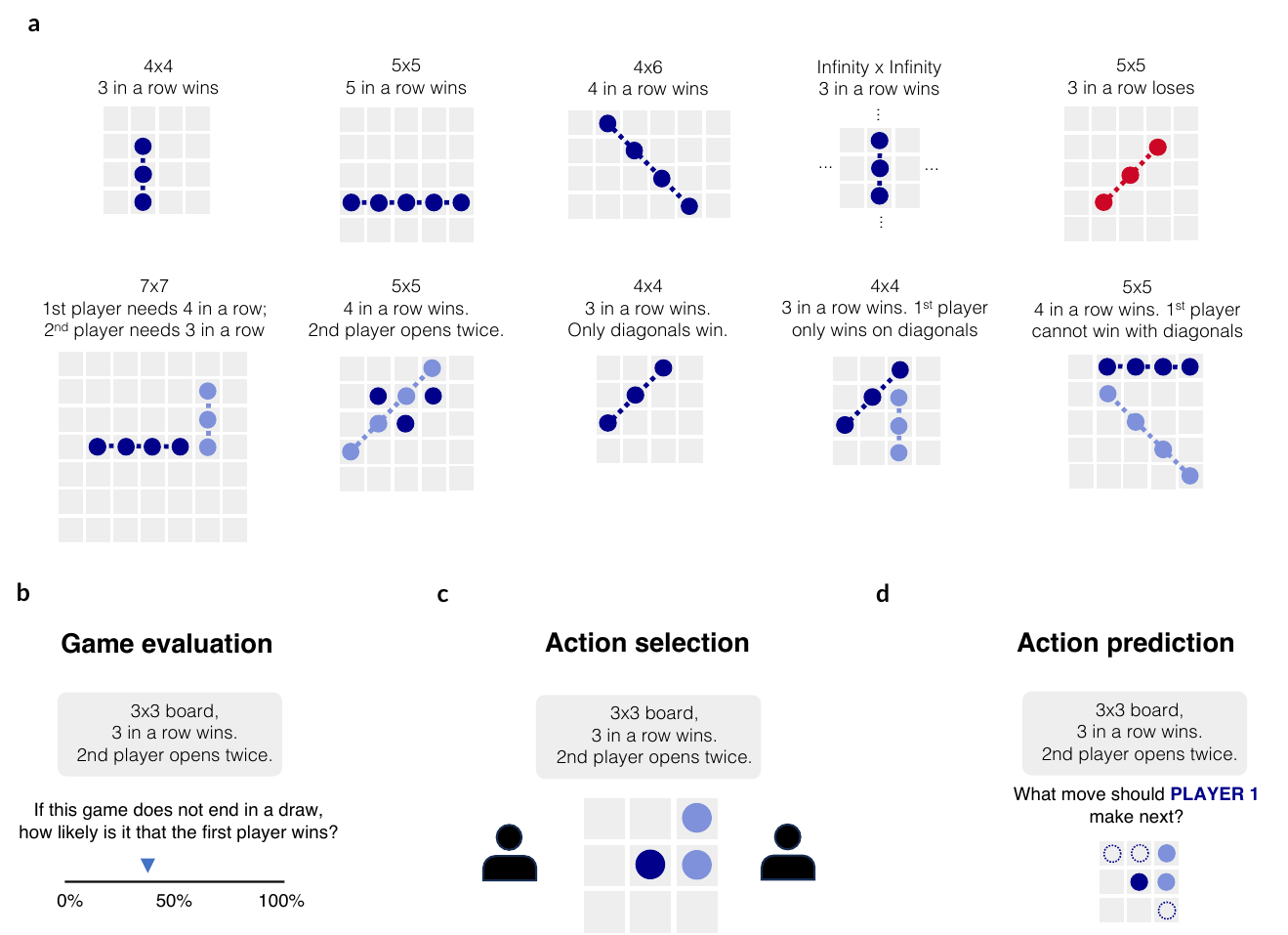} 
    \caption{\textbf{Our novel game dataset and suite of game tasks}. \textbf{a,} Ten example games from our $121$ game dataset. Games vary in board sizes and rules, such as what it takes to win and how many pieces any player can make on their opening move. \textbf{b-d,} We assess people's reasoning about novel game through three behavioral studies designed to test how people \textbf{b,} reason about games before they even play a single game; \textbf{c,} how people decide what actions to make in their first instance of play; and \textbf{d,} how people predict others should play when watching them play.}
    \label{fig:splash-main}
\end{figure}

Substantial work in cognitive science focuses on the nature and development of expertise, using games as a testbed ~\citep{simon1988skill, gobet2004moves, van2023expertise}. Human experts, correspondingly, have long served as goalposts in the quest to build intelligent machines, with a particular focus on emulating and ultimately beating humans at their own games, using more game-playing data and raw computational horsepower than any human expert could acquire in their lifetimes \citep{campbell2002deep, silver2016mastering, silver2017mastering, meta2022human}. 

But people are not experts at most of the problems they encounter throughout their lives, or most systems they are thrown into before they need to decide on an action. Human cognition is flexible enough to consider many potential problems, and many potential systems of rule and reward. Thinking about whether a system of rules and reward makes sense, whether to participate, or what actions to take first---across a wide range of novel situations---is arguably at least as important for everyday human cognition (if far less studied) than the cognitive processes that lead a small number of humans to become expert players in any one game. 

Here we study the abilities of \textit{novice} game reasoners across a range of over $100$ novel games from a subclass of strategic grid-based board games that are not entirely unlike games they have likely seen before, e.g., Tic-Tac-Toe or Connect-4, but vary in their rules and underlying dynamics (see examples in Figure~\ref{fig:splash-main}a and Extended Data Table~\ref{tab:stimuli}). In a series of large-scale behavioral studies each with hundreds of participants, we assess people's thinking about novel games in three reasoning settings (Figure~\ref{fig:splash-main}b-d): game evaluation (determining how rewarding a game is likely to be before playing); action selection (choosing moves the first time playing with another first-time player); and action prediction (judging the likelihood of moves other first-time players might make, without having played directly themselves). Our setting is different from much prior work on games, which usually focuses on one or a small set of games and many rounds of play in that game; here, we consider many games and little (even no) experience. 

\begin{figure}[]
    \centering
        \figimage{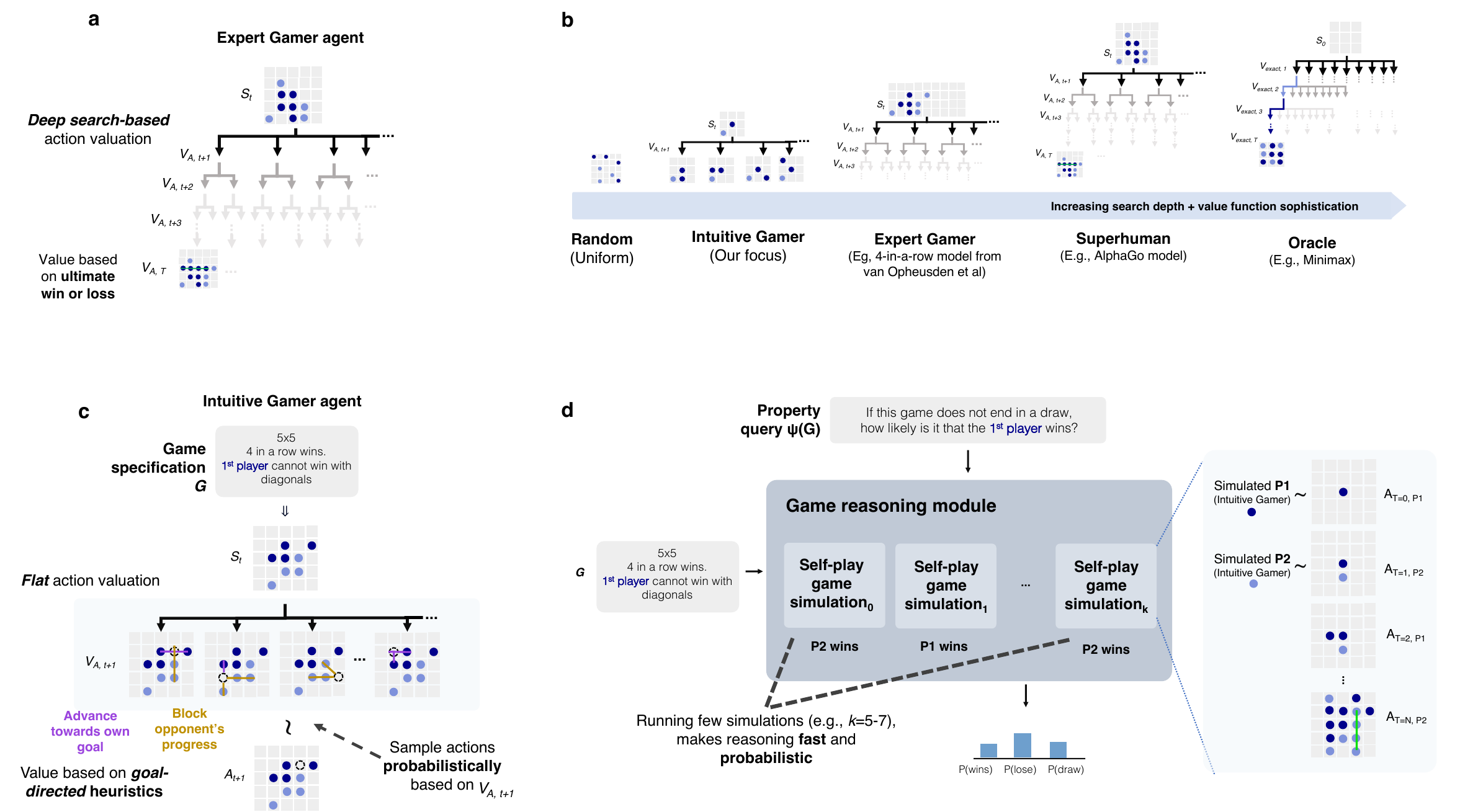} 
    \caption{\textbf{The Intuitive Gamer model, compared to prior models of game reasoning.} \textbf{a,} Prior work modeling expert gameplay often involves deep tree search to determine what move to make given a board state $S_t$~\citep{silver2016mastering, van2023expertise}. It is unlikely that novice human reasoners conduct such computationally expensive search and state evaluation before deciding whether to engage with the problem at all. \textbf{b,} What might novice reasoners be doing instead? Gameplay agents can differ in the amount of compute and expertise brought to bear to reason about any game (differing in search depth and value sophistication). Our proposal is that people reasoning about problems with which they have no experience, sit at the lower-end of this spectrum, but not \textit{the} lowest. \textbf{c,} The Intuitive Gamer conducts depth-limited (``flat'') search with game-general abstract goal-directed value functions that have yet to encode game-specific features. The Intuitive Gamer gameplay agent conducts no more than a single step of lookahead when deciding what action ($A_{t+1}$) to take from a given board state ($S_t$), at a given move turn $t$. The Intuitive Gamer is goal-directed, assessing whether any action would advance the player's own goal (purple) and how much it may block the progress of the opponent's goal (yellow). The final action is selected probabilistically by sampling from a softmax-distribution over the estimated values per action. \textbf{d,} To reason about any new game query $\psi$ for a given game description $\mathcal{G}$, we posit that people conduct only a few ($k$) self-play simulations between the gameplay agent (as depicted in panel \textbf{c}) to answer the query, which could be run to termination or probabilistically stop early. Taken together, the Intuitive Gamer model is fast (low $k$), flat (low-search depth), goal-directed (in the value function), and probabilistic (in action selection)---and involves mental simulations of gameplay. See Methods for a detailed formalism of the model.}
    \label{fig:model-schematic}
\end{figure}

Our core contribution is a computational account of how people reason about these novel problems. Expert models of gameplay typically involve deep tree search with potentially thousands of evaluations of possible states (Figure~\ref{fig:model-schematic}a). It seems unlikely that a novice reasoner is engaging in such intensive search and evaluation before they have extensive---or any---experience. But neither are people likely to be completely unsystematic. We hypothesize that novice thinking occupies a valuable intermediate point between these extremes (Figure~\ref{fig:model-schematic}b): people are non-random in assessing new problems and run computations analogous to those that underlie state-of-the-art AI game systems and cognitive models of expert human gameplay~\citep{van2023expertise, silver2016mastering, silver2017mastering}---but scaled down to a level that is more realistic for everyday thought.

We instantiate and test this hypothesis in an ``Intuitive Gamer'' model that makes judgments and decisions by running mental simulations of gameplay that are (1) fast, (2) flat, (3) goal-directed, and (4) probabilistic. The Intuitive Gamer model accounts very well---and significantly better than alternative models---for how people evaluate these games prior to play. Features computed from the same model can capture people's judgments about the game's objective value, as well as more subjective evaluations like a game's expected funness. Action choices simulated by this same model generally capture how novices actually play in games and how they make predictions about unfolding games they have never played themselves.

\section*{The Intuitive Gamer model}

The capacity to evaluate properties of a new game and decide whether the game is worth engaging with might intuitively feel like it comes prior to developing a policy for actually playing the game. However, our intuition as modelers is the opposite: game properties such as expected values depend on both the rule specifications of a game and the policies adopted by the players. We hypothesize this could be true for novice human game reasoners---before taking a single action in the game---if their policies are extremely simple to construct, fast to execute, and support probabilistic sampling of reasonable action choices. If so, a policy can be queried just a small number of times to simulate gameplay, from which the game reasoner can draw reasonable inferences about the game's properties using these simulated traces. 

We instantiate this hypothesis with our Intuitive Gamer model. The model consists of two modules: a``flat'' game-playing agent, which runs an extremely depth-limited search to selects actions probabilistically based on abstract yet locally evaluable ``goal-directed'' heuristic functions (Figure~\ref{fig:model-schematic}c); and a ``fast'' game reasoner that makes probabilistic inferences about game propositions and their expected values using a small number of simulated games (Figure~\ref{fig:model-schematic}d). 

\paragraph{Player module.} A long tradition of modeling human behavior in games has used some combination of game-state evaluation functions and search over the decision tree of game states \citep{newell_simon1972human, charness1991expertise}, as depicted in Figure~\ref{fig:model-schematic}b, to construct a policy. Our approach broadly falls into this tradition. However, existing accounts typically assume highly tuned or expert-designed value functions and deep simulation~\citep{campbell2002deep, kocsis2006bandit, silver2017mastering, van2023expertise}, which by definition cannot be expected of a novice. Rather, we propose a model that combines a general-purpose abstract value function---which could be easily constructed from understanding the rules of the game---with depth-limited search. Instead of simulating the downstream effects of a move via extensive forward search, the Intuitive Gamer player module uses only a single step of ``look-ahead'' and selects an action probabilistically by evaluating the resulting board states according to a small collection of general heuristics (see Figure~\ref{fig:model-schematic}c). These heuristics are built on the assumption that players understand and pursue their goals (as defined by the game rules) and attempt to prevent their opponents from doing the same, subject to constraints on computational resources~\citep{gigerenzer2004fast, lieder2020resource}. The Intuitive Gamer player module then is a more general, compute-bounded version of prior game-based agents (see Methods).

\paragraph{Reasoning module.} While the Intuitive Gamer player model captures action choice during a game, the Intuitive Gamer reasoning module captures \textit{judgments} about game properties by nesting the player module within a sample-based probabilistic inference procedure (Figure~\ref{fig:model-schematic}d). For any game, the reasoning module infers answers to a query (e.g. \textit{what is the likelihood that the game will end in a win, loss, or draw for a given player?}) by simulating $k$ game playouts and iteratively calling the player module at each turn of self-play. Each simulation ends when the game reaches a win or draw state, or may terminate early with some probability. Computations over the simulated gameplay traces then inform the resulting distribution over the query values (e.g., game outcomes). We assume simulated games are independent, though this could be relaxed in the future (see Discussion). In keeping with our focus on fast, resource-limited reasoning, only a \textit{small number of game simulations} are used (which we estimated empirically, see Results and Methods) in line with prior evidence that people use a relatively small number of simulations to form beliefs and make decisions \citep{griffiths2008categorization, sanborn2010rational, vul2014one, icard2016subjective, zhu2020bayesian, klein1993recognition}.

\paragraph{Alternate models.} 
Our Intuitive Gamer framework naturally extends to model players with more or less competence and motivation. We can think of the Intuitive Gamer as operating under the same mechanisms as more expert models, but scaled down (offering, therefore, a path to scale back up). We can vary parameters in our framework (value function and search depth) to instantiate a version of the kind of cognitive models that have been successful at capturing human experts in a single ``$4$-in-a-row'' game~\citep{van2023expertise}.  We refer to this variant as the ``Expert Gamer,'' which differs from the Intuitive Gamer by conducting a deeper search (approximately depth-$5$) and using a more sophisticated value function (see Methods). On the other end of the spectrum, we consider unmotivated players who select actions uniformly at random (see Figure~\ref{fig:model-schematic}b), which we refer to as the ``Random Gamer''. We also compare reasoning judgments to Monte Carlo Tree Search (MCTS)~\citep{coulom2006efficient, genesereth2014general, silver2016mastering}, an alternative tree-search approach that is even more computationally intensive and popular in AI systems for strong general game playing (see Methods for contrast with our Intuitive Gamer and other cognitive models). Finally, we compare against models that do not involve explicit structured game simulation, operating either over pre-computed linguistic features from the game description alone (see Methods and Supplementary Information 5) or more flexible linguistic computable models, e.g., language models (see Supplementary Information 5).

\paragraph{Resource-rational reasoning.}

The Intuitive Gamer occupies an interesting point on the Pareto frontier of compute efficiency and game reasoning sophistication. On the one hand, the Intuitive Gamer is orders of magnitude faster than more sophisticated game reasoners (MCTS and the Expert Gamer) in wall clock time and number of board states evaluated (see Extended Data Table~\ref{tab:game_resourcerational_reasoning}a and Supplementary Information). The Intuitive Gamer is approximately $700\times$ faster than the Expert Gamer in wall clock time, with approximately $500\times$ fewer board evaluation, and is nearly $40,000\times$ faster than MCTS with almost $10,000\times$ fewer node evaluations, as computed under self-play game simulations for the respective models. On the other hand, while not as close to game-theoretic optimal play as the Expert Gamer or MCTS, game evaluations under the Intuitive Gamer are still well-correlated with game-theoretic optimal analyses and far more aligned with game-theoretic optimal play than random play (see Extended Data Table~\ref{tab:game_resourcerational_reasoning}b and Methods). The Intuitive Gamer's mechanisms of fast, flat goal-directed probabilistic simulation thus constitute a highly efficient and resource-rational approach to game reasoning, which we hypothesize people could plausibly and productively engage when encountering new games.

\paragraph{Games to study novice game reasoning.} 

To explore how people reason about new games within the bounds of a behavioral study, we need a broad collection of games that are \textit{novel} without requiring lengthy demonstrations or explanations. To that end, we implement a range of two-player, grid-based strategy games derived from classic \textit{M-N-K} games (examples include Tic-Tac-Toe and Gomoku where players take turns placing pieces on an $M \times N$ grid trying to make $K$ pieces in a row---a subgenre of games frequently studied in prior work \citep{crowley1993flexible, amir2022adaptive, van2023expertise}. Our set of $121$ distinct games covers a variety of environment specifications (e.g., the size and shape of the board), transition dynamics (e.g., the number of moves a player makes on their turn), and win conditions (e.g., whether completing a line results in a win or a loss). In addition, games vary in their duration and balance under optimal play (see Methods). In each case, however, the core mechanic of the game (i.e., placing a piece into an empty grid cell) remains both consistent and familiar. See examples in Figure~\ref{fig:splash-main}a and Extended Data Table~\ref{tab:stimuli}.

\section*{Reasoning about games before any play}
We first assess how people evaluate a game (Figure~\ref{fig:splash-main}b), from ``just thinking'' about the game alone---before any play. We consider two kinds of game evaluation: how people assess the expected outcomes of a game and a more subjective assessment of whether the game is likely to be engaging (or ``fun'') to play.

\begin{figure}[]
    \centering

 \figimage{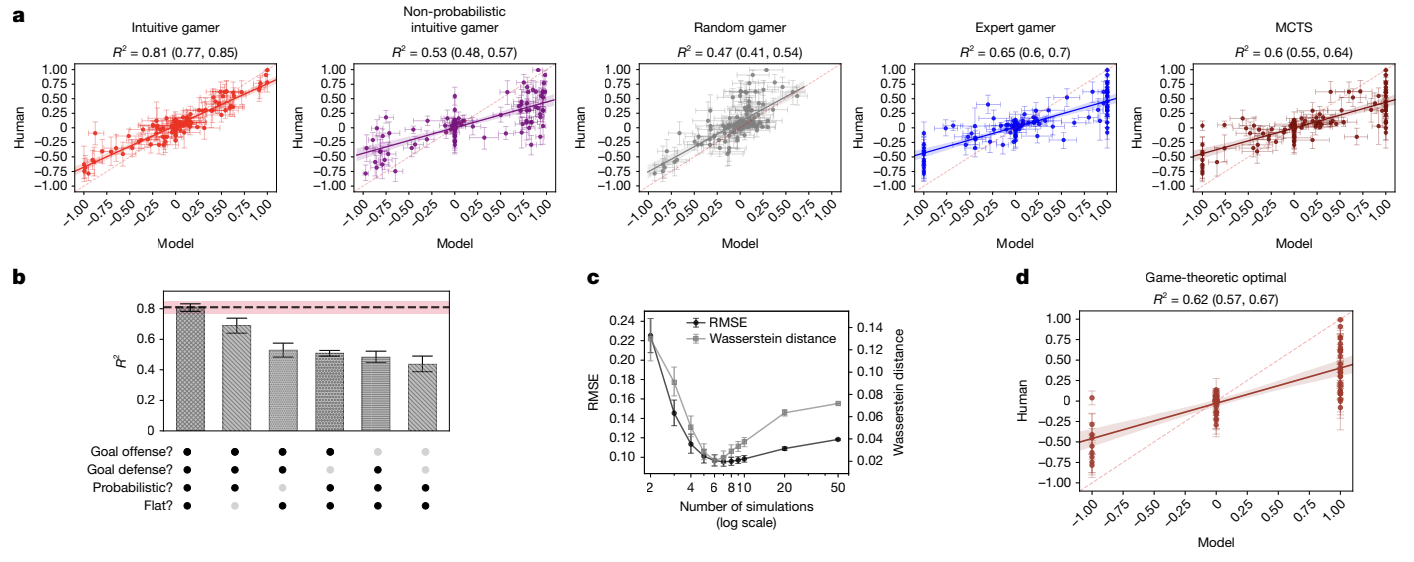} 

        \caption{\textbf{Evaluating games without ever playing.} $238$ participants judged the expected payoff of games drawn from our $121$ game suite. \textbf{a,}  Expected payoff predicted by humans and by alternative models for each game. Each point represents the payoff for one of the $n=121$ game stimuli. Error bars show 95\% confidence intervals around the mean human estimate and the mean model prediction from $k=6$ simulations sampled from 20 simulated participants.  \textbf{b,} Variance explained ($R^2$) by the full Intuitive Gamer and lesioned variants. Removing flatness, probabilistic simulation, or goal-directedness reduces fit to human payoff judgments.  The dashed line indicates the mean and 95\% CI split-half $R^2$ on the human-predicted payoffs, indicating the full Intuitive Gamer model captures essentially all of the explainable variance not due to noise. \textbf{c,}  Fit to the variance of human payoff judgments as a function of the number of simulations $k$. A small number of simulations ($k \approx 5$--$7$; we use $k=6$) best captures the variability in participants' judgments. Fit is measured by RMSE between human and model variance by game and by Wasserstein distance between their variance distributions across games. Full scatterplots are shown in Extended Data Figure~\ref{fig:vary-think-k-dev} and details of our sample complexity analysis are in the Methods. \textbf{d,}  Expected payoff predicted by humans and by game-theoretic optimal analysis for the 78 games (out of 121 games) where an optimal payoff could be computed. Error bars depict 95\% CIs around the bootstrapped mean human prediction per game.}
    \label{fig:splash-judge}
\end{figure}

\paragraph{Is the game likely to be fair?}

We recruited $238$ participants to evaluate the expected outcomes of the game in a ``zero-shot, zero-experience'' experiment. Participants ``just thought'' about the games, estimating both the likelihood of a draw and the likelihood that the first player would win if the game did not end in the draw about games based solely on a natural language description of the rules and depiction of a blank board. These estimates define an expected payoff of the game from the perspective of the first player (see Methods), which in turn encodes an intuitive notion of fairness (i.e., whether a game is biased towards a particular player).

We assessed how each characteristic of the Intuitive Gamer model (fastness, flatness, goal-directedness, and probabilistic simulation) contributes to people's payoff predictions by comparing to alternative models that increase or decrease the sophistication of each component (Figure~\ref{fig:splash-judge}a). The Intuitive Gamer model correlates well with human estimates ($R^2 = 0.81$ [95\% CI: $0.77, 0.85$]), matching the total explainable variance from the human data as estimated by split-half correlations ($R^2 = 0.82$ [95\% CI: $0.77, 0.86$]). The Intuitive Gamer significantly better captures human judgements than alternate models varying in sophistication. It outperforms the random model ($R^2 = 0.47$ [95\% CI: $0.41, 0.54$]), Expert Gamer model ($R^2 = 0.65$ [95\% CI: $0.60, 0.70$]), MCTS baseline ($R^2 = 0.60$ [95\% CI: $0.55, 0.64$]), and a variant of the Intuitive Gamer that is not probabilistic ($R^2 = 0.53$ [95\% CI: $0.48, 0.57$]). 

To more directly probe each component of the Intuitive Gamer model, we conducted a series of experiments where we varied key model components singly and in combination (see Supplementary Information 1 for a complete summary). The base model can be parameterized by a series of binary choices: whether to be goal-directed in value assessments; whether to be probabilistic; whether to be flat. Modifying any of these factors relative to the full model impairs fit to human payoff judgments (see Figure~\ref{fig:splash-judge}b). ``Fastness'' can be further varied by modulating the number of simulations the Intuitive Gamer reasoning modules takes when assessing payoff. A small number of simulations ($k$ more than one, but less than $10$) best captures the variance in human prediction (see Figure~\ref{fig:splash-judge}c and Extended Data Figure~\ref{fig:vary-think-k-dev} and complexity analysis details in Methods). These comparisons suggest that human payoff judgments of novel games are well-modeled by efficient and compute-limited simulation rather than purely random exploration of game states or deeper search and computationally intensive simulation. For simplicity, this estimate assumes that all simulations are independent and run to the end of the game; allowing simulations to probabilistically terminate early (see Methods) yields similar fits in the expected payoff evaluations and number of mental simulations per participant (see Extended Data Figure~\ref{fig:partial}).  

We were able to estimate the optimal game-theoretic payoff for many ($78$ of $121$, see Methods) of the games, allowing us to compare people's subjective judgments to a fully objective game evaluation. Human judgments are reasonable relative to the game-theoretic optimal ($R^2 = 0.62$ [95\% CI: $0.57, 0.67$]). While human judgments generally track the direction of the game-theoretic optimal (e.g., estimating a payoff greater than zero when the first player should definitely win), the Intuitive Gamer provides a superior qualitative and quantitative fit to human judgments. 

\paragraph{Is the game likely to be fun?}

\begin{figure}[]
    \centering

        \figimage{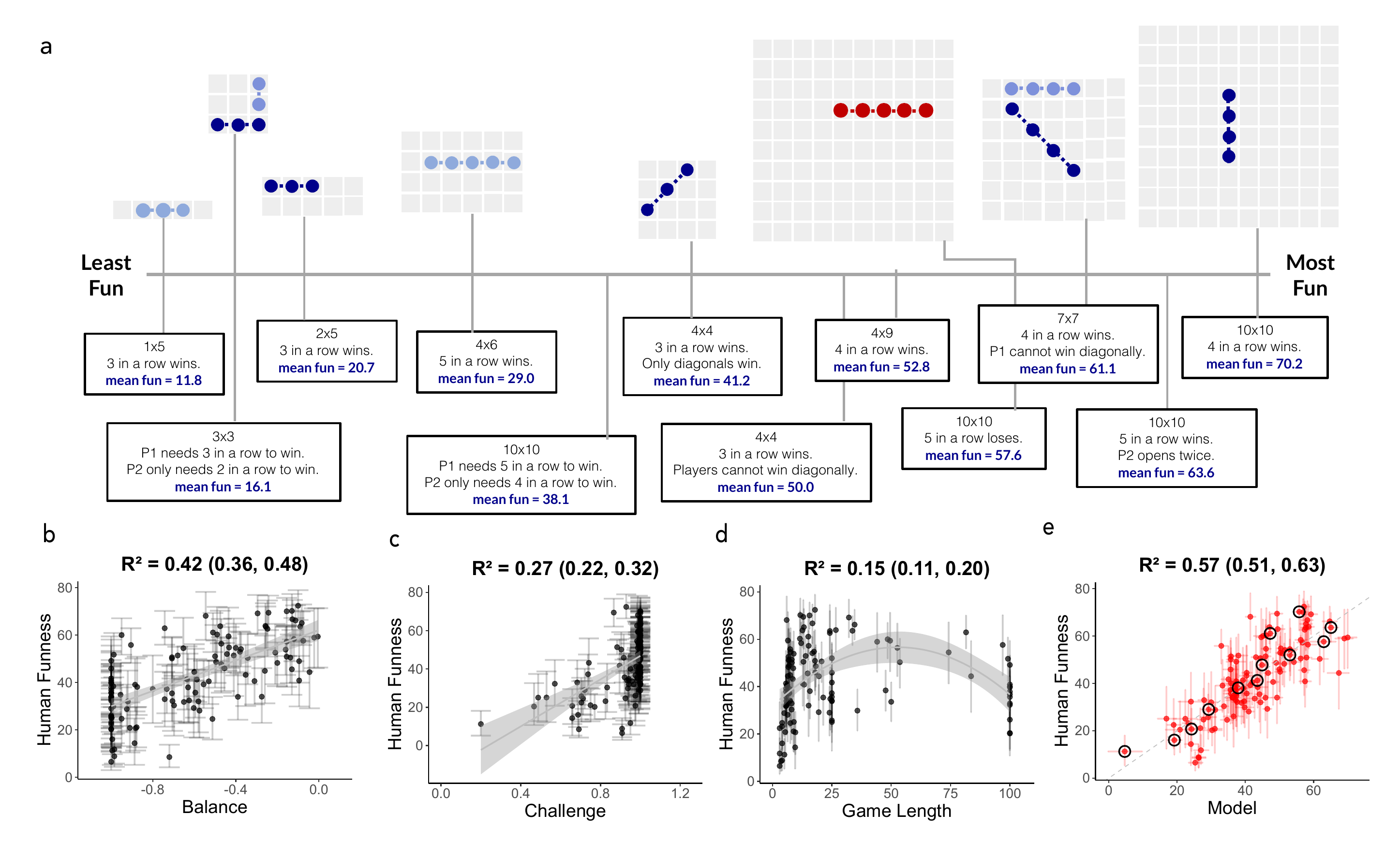} 

    \caption{\textbf{Evaluating whether games are likely to be fun, before ever playing them.} $246$ participants judged whether they find the games fun. \textbf{a,} Representative games rated by people as more or less fun, with position on the spectrum arranged approximately according to the mean of participants' funness ratings. \textbf{(b-d)} These funness ratings are related to several game features that can be read out under Intuitive Gamer model simulations: \textbf{b,} game balance; \textbf{c,} predicted advantage over a random agent given the Intuitive Gamer  model's gameplay (reward for thinking); and \textbf{d,} expected game length, as fit with a quadratic term to account for a non-linear relationship to funness. \textbf{e,} A regression model with all features predicted under the Intuitive Gamer model (balance, reward for thinking, and length) captures much of the explainable variance in the human funness ratings ($R^2 = 0.57$ [95\% CI: $0.51, 0.63$] compared to split-half $R^2 = 0.60$ [95\% CI: $0.51, 0.68$]). Error bars depict 95\% CIs around the human mean funness per game; error bars for the model depict 95\% CI over the predicted mean for each of the fit models (models are fit to each bootstrap subsample). Black circles show games from panel \textbf{a}, highlighting that funness judgments under the Intuitive Gamer generally align with the ordering of funness of games judged by people. A full listing of the games and their average human funness score are in the Supplementary Information 8.} 
    \label{fig:splash-judge-fun}
\end{figure}

We ran a parallel ``zero-shot zero-experience'' study described above on a new group of $246$ participants who were asked to make a subjective estimate of how fun a game is likely to be, before ever playing, to test whether the Intuitive Gamer's reasoning process can also account for how people make more subjective evaluations about new problems. Games varied widely in their judged funness (Figure~\ref{fig:splash-judge-fun}a and Supplementary Information 8), and funness judgments generally varied more across individuals than judgments of expected payoff, which is reasonable given that funness is both more vague and a more subjective evaluation.  Games that were judged most fun tended to have larger boards (often $10 \times 10$) and win conditions connecting a moderate number of pieces in a row (often $4$ or $5$), but there is no simple relation between these features and funness: many $10 \times 10$ games were judged below average (see Extended Data Figure~\ref{fig:vary-10x10}b), and the single most fun game was a $5 \times 5$ board with a misere rule, where $3$ in a row {\em loses}. 

We hypothesized that several features of a game that can be read off easily from our model's fast and flat probabilistic simulations---how \textit{balanced} a game is, how much a game \textit{rewards thinking}, and how \textit{long} a game is---could quantitatively explain much (or even most) of the variance in people's funness judgments. Our measure of balance captures the notion that people prefer games they expect will be decisive (unlikely to end in a draw), where both players have an equal chance to win. Our thinking reward captures the relative advantage of playing strategically (``thinking'') versus moving haphazardly (``without thinking''). The effect of game length is modeled as an inverted U-shape of expected number of moves to a draw or win, reflecting the intuition that longer and more involved games tend to be more fun, unless they extend unnecessarily. (For details of how these game features are computed, see Methods.)

Each of these three features computed under the Intuitive Gamer model captures significant variance in human funness judgments (see Figure~\ref{fig:splash-judge-fun}b-d and Extended Data Figure~\ref{fig:generalization-fun}a-b). A simple regression model that combines these features (including a quadratic game length term) attains $R^2 = 0.57$ [95\% CI: $0.51, 0.63$] (Figure~\ref{fig:splash-judge-fun}e), which approaches the total variance explainable in the human data based on bootstrapped split-half $R^2$ from participants over all games ($R^2 = 0.60$ [95\% CI: $0.51, 0.68$]).  This model also quantitatively explains the qualitative trends we observed in funness judgments, capturing the nonlinear relation between funness, board size and the number $K$ in a row to win (see Extended Data Figure~\ref{fig:vary-10x10}).

Finally, a generalization test (see Methods) showed that these fits are not sensitive to the choice of games, and that game features estimated from Intuitive Gamer simulations better predict funness judgments than those from the Random or Expert Gamer models, using greater or less compute resources (depth of thinking), or models that relied exclusively on surface-level game features (see Extended Data Figure~\ref{fig:generalization-fun}c and Supplementary Information 5.7).

Taken together, these two experiments in which participants ``just think'' in order to rate both objective measures of a game (e.g., its payoff or fairness) and more subjective evaluations (e.g., its enjoyability) lend support to our hypothesis that these kinds of novice judgments are best captured by a reasoning module that is goal-directed but limited in its computational cost (i.e. fast and flat) compared to alternatives that vary in more or less computational demands and game reasoning sophistication.

\section*{Decision-making in first-time gameplay}

Relative to the flat, single-step thinking that our Intuitive Gamer model posits people use to evaluate new games, conventional models of how people actually play games~\citep{van2023expertise} perform much deeper tree search, simulating multiple future turns by both players in order to evaluate potential moves before choosing the best move at each step of a game. But these models have only been evaluated on players who are either experts in a game, or have at least played a game many times. What do human players do when they are playing a game for the very first time? Do they search deeply, as previous models might suggest, or do they use only a fast and flat decision mechanism as our Intuitive Gamer model does---and as our results suggest they do when merely imagining how a game might unfold before playing it?

\begin{figure}[]
    \centering
    \figimage{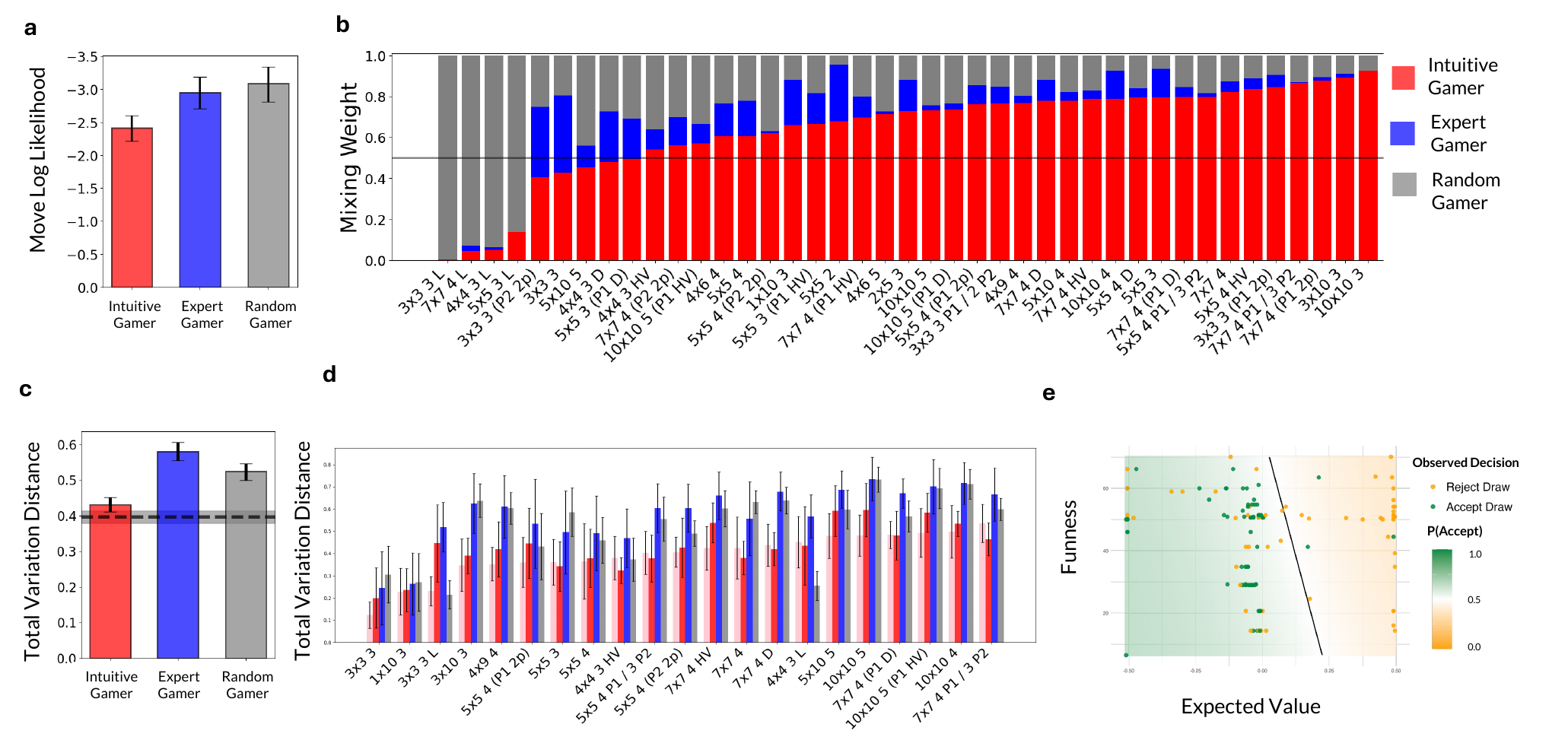} 

    \caption{\textbf{Modeling people's actions and distribution over predicted actions in the first encounter with a new game.} \ \textbf{a,} The average log-likelihood of human moves under three variants of the game player module. Data come from $302$ participants playing a single round each of $5$ novel games (along with Tic-Tac-Toe); across participants 40 of our novel games were tested. The Intuitive Gamer model captures players' moves significantly better than alternatives. Error bars depict 95\% bootstrapped confidence intervals. \ \textbf{b,} The estimated contribution of each model in a probabilistic mixture (admixture model, see Methods) fit to participants' moves for each game. The Intuitive Gamer model is the component that best explains participants' moves in approximately $90\%$ ($37$ of $41$) games; games where people would not be novices (Tic-Tac-Toe) or some particularly atypical games (e.g., misere where the first person to get $K$ in a row loses, listed with an ``L'' like ``$7 \times 7$ $4$ L'') are comparatively less well captured by the Intuitive Gamer compared to alternate models. Game descriptions are lemmatized to show board size and key features (e.g., HV indicates only horizontal and vertical wins are allowed; P2 2p means Player 2 can move twice on their first turn). A full description of lemmatized game codes are in the Methods. \ \textbf{c,} Total Variation Distance (TVD; lower is better) between human and model-predicted distributions over next moves, for the same three models. Judgments come from a new group of $314$ participants who estimated where a participant should move next in frozen videos from the gameplay experiment. The dashed line is the split-half human TVD (effectively the noise ceiling; pink). \ \textbf{d,} Mean TVD by game. The Intuitive Gamer generally best matches human judgment distributions, with similar failures to the play data (e.g., for mis\`ere games). Error bars depict standard deviation over boards for each game. \ \textbf{e,} Decisions of whether to accept or reject a draw, when requested is captured by expected value of the player’s current position, but participants tend to be willing to reject a draw and play out a game with slightly lower expected value if the game is judged to be fun.}

     \label{fig:splash-live-choices-play-watch}
\end{figure}

\paragraph{What moves do people make?} 
To test whether the Intuitive Gamer player module captures how people actually play a game for the first time, we recruited $302$ participants to play each other in a subset of $40$ of the $121$ game variations, plus Tic-Tac-Toe (Figure~\ref{fig:splash-main}c). Each participant played only a single round of any given game. 

Initial evidence that the Intuitive Gamer model captures not only how people think about new games but also how they play them for the first time comes from comparing the distribution of actual game outcomes (first or second player win, or draw) in this gameplay experiment with the outcome distributions to be expected if both players make move decisions according to the model. The model captures most of the variance in actual game payoffs with novice players ($R^2 = 0.72$ [95\% CI: $0.67, 0.76$]) and predicts these payoffs as well as human estimates of expected game payoffs do, from our earlier game evaluation experiment (see Extended Data Figure~\ref{fig:empirical-obs-payoff-v-think}). 

A more direct test is to compare predicted move choices that the Intuitive Gamer model makes at each turn of each game with the actual moves that new players made at the same game steps when they were playing these games for the first time. Because there is a strong element of randomness in people's move choices as well as the model's predictions, we evaluated the likelihood of people's moves under the Intuitive Gamer relative to alternative models that are either more or less sophisticated and computationally intensive, the Expert Gamer and Random models from our previous study, respectively (Figure~\ref{fig:splash-live-choices-play-watch}a). The Intuitive Gamer model best predicts players' moves in aggregate, as measured either by average per-move log likelihoods or per-match log likelihoods (summed over all moves within a match), across $1808$ distinct matches with a total of $9892$ moves. All of these differences are highly significant as measured by paired t-tests (mean differences in total per-match log likelihood for the Intuitive Gamer vs Expert model = $2.83$, $t_{1807} = 29.4$, $p < 0.001$, vs Random model = $3.73$, $t_{1807}$ = $26.4$, $p < 0.001$; mean differences in per-move log likelihood for the Intuitive Gamer vs Expert model = $0.51$, $t_{9891} = 39.9$, $p < 0.001$, vs Random model = $0.68$, $t_{9891}$ = $42.3$, $p < 0.001$).

We also fit these three models of action choice to the distribution of all moves people made in each individual game, and the distribution of all moves an individual player made across the different games they played, using probabilistic (admixture) models (see Supplementary Information). At a per-game level, the Intuitive Gamer model generally but not always fits human play better than the Expert or Random models: it captures more than $50\%$ of the probability distribution of moves in $32$ out of $41$ games, and a plurality of moves in $5$ of the remaining games (Figure~\ref{fig:splash-live-choices-play-watch}b). Individual player's distributions of moves are also best fit by the Intuitive Gamer: it captures more than $50\%$ of the probability distribution of moves for $243$ out of $302$ players, and a plurality of moves in $16$ of the remaining players (Extended Data Figure~\ref{fig:individ-play}). 

Post-hoc exploratory analyses into intermediate stages of gameplay suggest that the Intuitive Gamer model best fits early and mid-stage play, though in later stage play, greater depth variants are more comparable (see Supplementary Information). This could be due to some people being more motivated to think deeper at the end of the game or finding it easier to think deeper with a smaller space of moves to consider (as there are fewer legal moves at the end of the game).   

Additionally, as in the game evaluation experiments, both aspects of goal-directedness in the Intuitive Gamer value function (offensive progress towards your goal, and defensive assessment of how to block your opponents progress) matter for capturing human behavior in novel games (see Extended Data Figure~\ref{fig:ablate-value-comp}). Overall, our results support the hypothesis that when playing a novel game for the first time, people adopt a similar fast, flat and goal-directed approach to choosing their next move as they appear to do in mentally simulating games before they start to play. 

\paragraph{What move should someone else make?} 

In the previous experiment, as in almost all studies of gameplay, participants only made a single choice for each move, and they only played each game once. Yet our model posits that people think about new problems by running \textit{probabilistic} mental simulations letting them form graded expectations about the value of a range of possible actions. To assess how well the Intuitive Gamer player module captures people's probabilistic expectations, we recruited a new group of participants to predict distributions of likely actions when watching videos of other novices playing these games (see Methods; Figure~\ref{fig:splash-main}d).

The Intuitive Gamer model generally better captures the distribution of moves predicted by people compared to alternate models across the $249$ game boards for each match and game stage, as measured by the Total Variance Distance (TVD) between model and people's predicted distribution, and is near the expected noise ceiling (computed via split-half) human judgments for most game boards (see Figure ~\ref{fig:splash-live-choices-play-watch}c). Differences in distance (where lower means closer to people's predicted action distribution) are highly statistically significant as measured by paired t-tests in the TVD per game board for the Intuitive Gamer vs Expert model = $-0.15$, $t_{248} = -15.0$, $p < 0.001$, vs Random model = $-0.09$, $t_{248} = -7.5$, $p < 0.001$. Exceptions for poor fits align with those found in the play experiments (e.g., on misere games; see Figure~\ref{fig:splash-live-choices-play-watch}d and example boards in Figure~\ref{fig:watch-examples}). Fits are robust to choice of distributional measure (see Supplementary Information). People assign similar log probability on the move actually played by the human player to that of the Intuitive Gamer, further highlighting alignment of the Intuitive Gamer to people's probabilistic judgments (see Supplementary Information).

\begin{figure}[]
    \centering
    \figimage{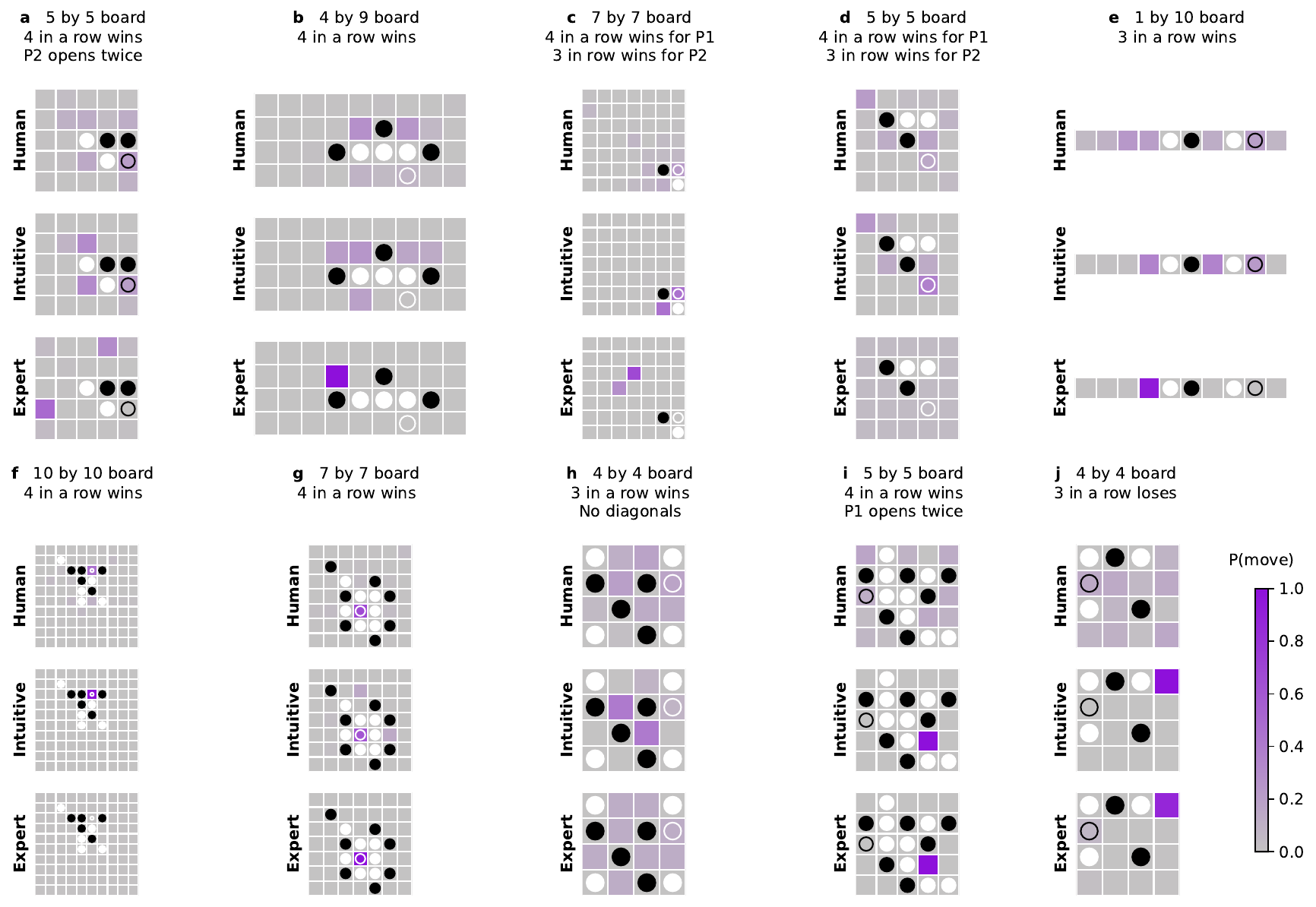} 

     \caption{\textbf{Example human- and model-predicted distributions over the next action in real games.} Each panel (\textbf{a-j}) shows a representative state taken from a different game between two human players, along with the actual move one player took at the next step, and the predictions made by other humans and two models for how that move should have been made. Filled black and white circles show the moves both players took prior to this point in the game. The one open circle on each board indicates where the human participant whose turn it was on this move (white or black) actually played. The top, middle, and bottom boards show the predicted next-move distributions from human participants in the watch-and-predict study, the Intuitive Gamer, and the Expert Gamer, respectively. The purple shading gives the predicted probability of choosing each empty square. Humans often distribute probability over several plausible moves (\textbf{a--e}), though sometimes they strongly favor a single move (\textbf{f,g}); the Intuitive Gamer often captures this pattern. The Expert Gamer model fits human predictions less well, in some cases because it allocates high probability to a small set of high-utility moves which are often counter-intuitive to naive players (\textbf{a-c, e}); and in other cases by allocating diffuse probability across many states when deep search indicates a sure loss under perfect play (\textbf{d, f}). The Intuitive Gamer does not always fit best: the Expert Gamer can sometimes better match human confidence (\textbf{g}), and the Intuitive Gamer can be too sharp in its predictions (\textbf{h}).  In some cases both models are sharper than humans, including crowded boards where people may miss a winning or blocking move (\textbf{i}) and mis\`ere variants (\textbf{j}). The full set of predictions (for $249$ moves across $41$ distinct games) is included in the Supplementary Information.
    }

    \label{fig:watch-examples}
\end{figure}

\paragraph{When is a game worth continuing?}

Game evaluation is important not only when deciding whether a game is worth playing but also for assessing---during play---whether a game is worth continuing to engage with. In our gameplay study, participants had the option to request a draw at any point in play. If they received a draw request from another player, they had the choice of whether to accept the draw and immediately terminate play or reject the draw and keep playing. Over the $1808$ matches participants played, players made $142$ draw requests, of which $83$ were accepted and $59$ were rejected. Upon receiving a draw request, how do people decide whether to keep playing? 

We modeled these choices as probabilistic value-based decisions trading off a player's expected reward of winning with the expected cost of continuing to play, both of which can be computed under the Intuitive Gamer reasoning module conditioning on the current board state (rather than a blank opening board, as in our earlier game evaluation studies; see Methods). In exploratory analyses, we also hypothesized that a player might be willing to keep playing a game even if the expected cost of continuing outweighs the potential objective monetary reward, if playing provides an alternative subjective reward, e.g., if it is likely to be fun. A logistic regression model fit to participants' draw request decisions using these three features (see Methods) captures the intuition that players' game-time evaluations are probabilistically rational under limited resources (time) and sensitive to subjective factors like the engagement of games in addition to their objective value (Figure~\ref{fig:splash-live-choices-play-watch}e). Ablations of the Intuitive Gamer evaluation function lead to significantly worse qualitative and quantitative fits to human draw request choices (see Extended Data Figure~\ref{fig:accept-reject}a and b).

\section*{Discussion}

Systems of rules and rewards govern many aspects of human behavior---from jobs and institutions to games and geopolitics. How can people think reasonably about any such system or problem for the first time, so flexibly and quickly? We formalized and empirically evaluated a computational model of an Intuitive Gamer, which draws on fast and flat goal-directed probabilistic mental simulations to make rich quantitative predictions about how novice players judge the objective value of a previously unseen game, as well as make more subjective evaluations. Our model also captures how people act within these games and does so better than alternative models with more and less computational power. The same model generalizes across these different tasks and the many different games we study.

Our work differs from most prior studies of gameplay in cognitive science and AI which focus on how expert play is achieved, and typically consider only expertise in a single game. In contrast, we study a \textit{class} of $121$ strategic games with varying dynamics and highlight the phase of ``pre-expertise'': the critical first moments before people can rely on the benefits of repeated experience, but when they must nonetheless take reasonable actions and even decide whether to play at all---or continuing playing. Studying pre-expertise across varied domains is important not just for our understanding of cognition generally, but also to construct more accurate models of human behavior in game theory under novel economic mechanisms \citep{myerson1983mechanism, koster2022human, manning2026general, camerer2003behavioral} or alongside unfamiliar agents like AI systems \citep{carroll2019utility, collins2024building}. The Intuitive Gamer approach could inform the design of more efficient AI systems that make reasonable judgments in new multi-agent settings despite using much less compute than conventional expert-level models.

Our work also differs from prior studies of funness in games, which have focused on identifying factors that lead people to experience a game as fun. In contrast, we have focused on modeling the intuitive theories people have about how they would expect to experience a game as fun, prior to playing it for the first time. Features identified in the prior literature, such as strategic depth~\citep{lantz2017depth} and competence~\citep{ryan2006motivational} reflecting the need for challenge, or learning progress~\citep{brandle2025leveling} and thinking progress~\citep{chu2023praise}, do align with those in our model, especially the reward for thinking component. Future work could study whether other aspects of motivation identified in prior gaming literature---autonomy, social connectedness, immersion, and ease of control~\citep{ryan2006motivational}---also contribute to people's intuitions about funness, and whether these could also be modeled using a generalization of our Intuitive Gamer model. 

As a model of human reasoning, our work at present is limited to a subset of perfect information, two-player competitive board games, and these games reflect novelty within a genre that many people are familiar with rather than complete novel experience. It is an open question how the heuristics our model uses could generalize to more complex games in the same class, such as Go or Chess, and to entirely new classes of games. Many extensions are possible, such as generalizing the contiguous-line features that we consider toward connectivity or territory metrics that reflect partial or graded progress towards a game's win conditions. Natural language could be used to suggest short programs expressing such heuristics, as in verifiable code-based rewards in reinforcement learning~~\citep{lehrach2025code}. While we hypothesize that key features of the Intuitive Gamer reasoning module---being fast, with a few probabilistic simulations of play, where play is assumed to involve goal-directed agents---would extend to other game classes, it is an open question how ``flat'' reasoning might be and what more general heuristics might capture intuitive reasoning in the wider range of games that characterize human social life. Understanding peoples' general game reasoning on a wider class of games beyond competitive settings, e.g., cooperative or multi-agent contexts, is necessary to move towards a general model of people's ``intuitive theory of games,'' analogous to the computational models of intuitive theories of physics or intuitive theories of mind that cognitive scientists have constructed~\citep{baker2017rational,ullman2017mind}.

Additionally, there is a need for more fine-grained process-level and individual-level accounts of how people reason about novel games. More work should study how people judge the expected outcome of a game if they stop imagining play before reaching a win or draw, if their simulations are temporally dependent, or when they do not simulate at all---and how models can best capture these complexities. Our model also does not address how people may learn and adapt such heuristics within or across games, limitations that suggest important directions for future work. How mental computation for reasoning about familiar games may shape thinking about a new game~\citep{dasgupta2018remembrance}, especially in high-stakes situations~\citep{klein1993recognition} or when time is limited or thinking is costly~\citep{lieder2020resource,gigerenzer2004fast}, and how in-game experience~\citep{kolb2014experiential, rutledge2014computational} and preferences for different kinds of play and strategy~\citep{nguyen2019games} influence game reasoning will also likely be important for modeling variability in peoples' thinking. 

Finally, people also \textit{make} new games and new problems, either by modifying old ones to make them more challenging or more fun, or creating entirely novel goals and pursuits---which has led to some of humanity’s greatest discoveries and innovations~\citep{chu2023praise}. In ongoing work we are exploring whether models based on fast and flat simulation can also explain the moves people make in these ``meta-games'' of game-creation and modification.

Science itself is such a ``meta-game'': a family of games against nature~\citep{hintikka1981logic}, where scientists set the rules that lead to the most productive research~\citep{laudan1978progress}. Similarly, mathematicians explore conjectures and proof strategies by imagining adding or relaxing the logical rules of how a problem is formulated~\citep{thurston1995proof}. Suppose, more speculatively, that these and other creative activities at the frontiers of human knowledge do in fact share something deeply in common with intuitive reasoning about games, or even children’s play. Then perhaps mechanisms like the fast probabilistic mental simulations studied here could help to explain aspects of how researchers choose to approach a new problem, or which problems to think about in the first place.  Whether scientific thinking can be meaningfully captured by such models is very much an open question, but one we are now better positioned to investigate. At the very least, it feels like a game worth playing.

\putbib[main]  
\bibliographystyle{abbrvnat}
\end{bibunit}

\newpage

\begin{bibunit}[naturemag]

\clearpage
\section*{Methods}

\subsection*{Game construction}

We manually constructed the $121$ two-player competitive strategy game variants played on $M \times N$ grids, where $M$ are the number of rows and $N$ are the number of columns. One goal in creating these games was to ensure there is enough diversity in board size and rule structure as well as systematic variance. To that end, we designed a series of variations on square boards: on $10 \times 10$ boards, we have $K$-in-a-row for $K$ from $2$ to $10$; for 3-in-a-row, we have $M$ by $N=M$ boards from $3$ to $10$; we also include a few $5 \times 5$ boards varying $K$. We additionally included a few other square boards varying in complexity, as well as rectangular boards ranging in size from from $1 \times 5$ to $5 \times 10$, and we have integer $2 \leq n \leq 6$ for $K$-in-a-row. To assess how people reason about games that are not physically realizable, we included three ``infinitely'' sized games with $K = 3, 5, 10$ for $K$-in-a-row. These categories give us $41$ games with typical ``$M$-$N$-$K$ rules'' (e.g., where the players take turns and have the same objective: ``make $K$ in a row, where horizontal, vertical, or diagonal all count''). This set includes the standard Tic-Tac-Toe as well as $4 \times 9$, $4$ in a row wins from ~\cite{van2023expertise}. We then created a number of games with varied game rules. We kept the selection of board sizes and $K$-in-a-row fixed across categories ($10$ games within each category). The selection included $3 \times 3$, $4 \times 4$, $5 \times 5$, and $10 \times 10$ boards and $n \in \{3, 4, 5, 10\}$. We also designed games with more atypical rules, varying the winning conditions (e.g., $K$-in-a-row loses and diagonal connections do not count as wins) and first-mover dynamics (e.g., Player 1 can place two pieces on their first turn). These categories give us an additional $80$ games, totaling the $121$ in our dataset. We manually implemented an automated win condition checker that permits flexible game assessment over all game types.

\paragraph{Game name codes.}

Games are expressed in abbreviated form throughout the manuscript. Games are described by their board size (rows $\times$ columns) and the number $K$ in a row to win. E.g., ``$4 \times 4, 3$'' means the game is played on a $4 \times 4$ grid and the first person to get $3$ pieces in a row wins. Unless otherwise stated, horizontal, vertical, and diagonal all count. If only horizontal and vertical $3$ in a row count, the game would be written as ``$4 \times 4, 3$ HV''. If only diagonal counts, then the game will be written as ``$4 \times 4, 3$ D''. If the constraint only applies to one player (e.g., P1 can only win horizontally and vertically, and P2 can win any way), that is represented as ``$4 \times 4, 3$ (P1 HV)''. If one player can go twice on their turn, e.g., the second player (P2) can play twice, that is written as ``$4 \times 4, 3$ (P2 2p)''. A misere game (where first to K in a row loses) is written with an ``L'', e.g., ``$4 \times 4, 3$ L''. In our dataset, multiple rule modifications cannot co-occur, so each game can be expressed in this abbreviated way.

The full game natural language game descriptions were provided to participants (e.g., as shown in Table~\ref{tab:stimuli}). The game codes are only used for ease of presentation in this manuscript.

\subsection*{Human experiments}

All human experiments were conducted under prior approval from the Institutional Review Board (IRB) at the Massachusetts Institute of Technology through the Computational Cognitive Science Lab. All participants provided informed consent.

\paragraph{Zero-shot outcome evaluation experiment.}
We recruited $238$ participants from Prolific \citep{palan2018prolific} to judge novel games. Each participant was randomly presented with $10$ games sampled from our $121$ diverse game stimuli, as well as regular Tic-Tac-Toe (won by making \textit{$3$ in a row on a $3\times3$ board}) to set baselines for game judgments. We collected approximately $20$ judgments per game stimuli for each game reasoning query. Participants were paid at a base rate of \$12.5/hr with an optional bonus up to \$15/hr; the full experiment approximately took $25$ minutes.

Participants were instructed to evaluate likely game outcomes. Specifically, participants produced judgments on a continuous $0$ to $100$ probability scale to predict the likelihood of a first player win (\textit{if the game does not end in a draw, assuming both players play reasonably, how likely is it that the first player is going to win (not draw)?}) and a draw (\textit{assuming both players play reasonably, how likely is the game to end in a draw?}). Judgments were made using sliders. Both game outcome question sliders appeared on the same page.

Participants produced judgments about each game based on a linguistic game specification. We additionally provided participants with an interactive scratchpad board that they were told they could, but were not required to, use to inform their judgments. The scratchpad was automatically sized to the board of the game specified; infinite boards were presented on a $13 \times 13$ grid with dashed lines indicating the board could continue. The scratchpad permitted automatically placing pieces of different colors (``red'' and ``blue'' to simulate different players); participants could force the the next play to be made by the same player (e.g., color red twice in a row) by pressing the spacebar. Two buttons appeared below the scratchpad, permitting the user to either undo their last move or clear the screen to begin a new ``game''. See Supplementary Information 4 for examples of the experiment interface. Participants were required to consider each game for at least $60$ seconds before being allowed to make their game judgments. Outliers were determined as the $10\%$ of the judgments farthest from the mean judgments of other participants (in terms of the summed distance from the two queries) and filtered out for each game.

After answering all game reasoning queries, we additionally asked participants to create a new grid-based game variant that they would find fun. Participants wrote a linguistic game specification, describing the board size and win conditions. As in the game judgment queries, participants were again provided an optional scratchpad and required to spend $60$ seconds before submitting a response. The scratchpad enabled participants to try out the game they intend to create. After specifying a game, participants were asked to answer the same game reasoning query (either game outcomes or game fun) about their own game. These game-generation responses were not studied here and are actively being explored in a follow-up study. Example screenshots of experiment interfaces are included in Supplementary Information 4. 

\paragraph{Zero-shot funness evaluation experiment.}

We repeated the same methodology above with a new group of $257$ participants, replacing the game outcome questions with a question about game funness. Participants instead assessed the expected funness of the game (\textit{how fun is this game}?) on a confidence scale spanning $0$ (\textit{the least fun of this class of game}) to $100$ (\textit{the most fun of this class of game}). 11 participants were filtered due to having provided non-effortful or AI-generated games in the creation stage, resulting in a total of $246$ participants, and the same outlier filtering was applied per trial as in the outcome evaluation.

\paragraph{Zero-shot human-human play experiment.}

We recruited $302$ participants to play these novel games in a pre-registered experiment. We selected a subset of $40$ games from the full set of $121$ to span a representative range of the gameplay variations (board shapes, board sizes, and win rules) in the original dataset while generally favoring games that would not take very long to play in a live experiment. We randomly constructed $8$ batches of $5$ of these games. Each participant additionally played one round of Tic-Tac-Toe. The order of games was shuffled for each new set of participants.

Participants were automatically paired with another player. We developed our interface using Empirica~\citep{almaatouq2021empirica}, which supports synchronous human-human pairing. Participants played one round from five different games. Players were informed they would get a bonus of \$0.50 for every win. Participants had to spend at least $5$ seconds reading the game description before they began. We appended ``Horizontal, vertical, and diagonal all count'' to all game descriptions where any direction was allowed after we noticed some participants in pilots were confused as to which line directions would result in a win.
Players were randomly assigned to move either first or second and a corresponding piece color (red or blue). Players took turns making moves on the synchronous game interface. Players had no time limit on their turn.

Players were also allowed to request a draw or decide to surrender using buttons at the bottom of the interface. If a player surrendered, the game ended immediately (and that player lost). If a player requested a draw, the other player was allowed to either accept the draw (after which the game ended immediately and no player won) or reject the draw (leading the game to continue being played). Draw requests appeared as a popup banner for the other player.
We include screenshots of the interface in Supplementary Information 4. 

The match ended when either a  player won, a player surrendered, the board filled up completely (draw), or the players agreed to a draw. Both participants were informed about the game outcome. After each match, players made a judgment about either the expected outcomes of that game overall (with a new set of reasonable players) or the game's funness (in a new match against a new player). Each pair of players was randomly assigned to either the outcome or funness rating condition.
Judgments were made on a slider. Players were also presented with a ``frozen'' version of the match on an example board with which they could replay all of the moves they and their opponent had made. Players also indicated how skilled they thought their opponent was at this game (\textit{out of $100$ other random new players, where do you think the opponent you just played would rank in skill for this game?}). After the judgments were made, the players continued to the next, new game. At the end of the study, they filled out a text-based survey providing general information on their strategy and how fun they found the experiment. We filtered out $18$ participants who did not pass our quality control (that is, they provided judgments that were ``standard'' values (near $0$, $50$, or $100$) on $80\%$ or more of judgments) for a total of $284$ subjects.

\paragraph{Watching and predicting play experiment.}

We recruited a new set of $314$ participants, in a pre-registered experiment, to reason about the games zero-shot from only \textit{indirect} experience: watching two other agents play. We selected a subset of $20$ from the games from the previous human-human play study to ensure representation across game rules and dynamics. We also included Tic-Tac-Toe (totaling $21$ games). Participants watched a series of videos of other agents’ gameplay. Each video involved two humans playing each other, sourced from our live human-human gameplay experiment. We sampled $4$ human-human played matches randomly from each of the $21$ games\footnote{Due to a randomized batching error, only $3$ unique matches were sampled for Tic-Tac-Toe; hence, $249$ game boards over the matches from $21$ games and three stages per game.}, after filtering out any matches that ended preemptively from a draw request or surrender. For each match, we sampled three specific boards to be evaluated corresponding to the beginning, middle, and end of the match. For the beginning and end boards we randomly selected either the third or fourth move and the second to last or third to last move, respectively. For the middle board, we selected the median move. We filtered out any match that ended before eight moves. Participants watched one match from five different games, plus Tic-Tac-Toe.

Before each match, participants were informed of the game rules and required to think about the rules for $5$ seconds before the video began. We again appended ``Horizontal, vertical, and diagonal all count'' to all game descriptions where any direction was allowed.  Videos played forward at a fixed rate, as in ~\citep{van2023expertise}. We chose two seconds per move to give viewers enough time to process each move without taking too long overall. Each video was stopped at the three time points described above. At each stopping point, participants indicated their belief over where they thought the acting player should move next. Participants were given five clicks which they could spread across the legal moves on the board to indicate their confidence that the player should move there. We chose five clicks to balance granularity of the elicited belief distribution against the burden on the participants. After each click, the opacity of the cell increased to indicate higher confidence.
Participants were informed of the number of clicks they had left and could reset their clicks by clicking on a button below the interactive board. 

After watching each video and indicating where they thought a player should move at each of the three timepoints, participants were then shown the remainder of the game as a board snapshot (cells indicated where players had moved and the order of play). Participants then answered either the same game outcome or funness judgments about the game overall, as described above. Judgments were made on a slider. Participants also indicated how skilled they thought each of the players was. We filtered out $10$ participants who did not pass our quality control, leaving us with a total of $304$ valid participants.

\subsection*{Problem formulation}

Formally, one can think of a system, problem, or game $G$ in terms of a space of feasible states $\mathcal{S}$; possible actions $\mathcal{A}$; rules $\mathcal{T}$ specifying valid actions and state transitions, and governing the overall dynamics; and goal functions mapping from states $\mathcal{S}$ to possible rewards $\mathcal{R}$. We are interested in how a reasoner infers properties of a new game $\psi(G)$ (e.g. whether a game is likely to end in a draw or have a bias towards a particular player; whether the game is likely to be engaging and fun) as well as a policy $\pi_{G}(a_t \mid s_t)$ for how to play (choosing actions given the current state at time $t$, to achieve their goal, e.g., to win), without experience of actual traces of gameplay and relying instead on \textit{simulated} or imagined traces. Our aim is to model how people \textit{approximate} $\psi(G)$ and $\pi_{G}$ in a way that can (1) take in any $G$ as input, and (2) do so with a limited compute budget and no direct experience with game $G$. 

\subsection*{Intuitive Gamer model specification}

We first describe a formal account of the Intuitive Gamer player module, and then describe the reasoning module. 

\paragraph{Intuitive Gamer player module.} 

Formally, at a given board state ($s_t \in \mathcal{S})$, the Intuitive Gamer player module scores all legal next actions ($a_t \in \mathcal{A}$) according to a measure of \textit{immediate progress} made towards a player's own goal ($U_{\text{self}}$) and a measure of progress \textit{blocked} ($U_{\text{opp}}$) towards the opponent's goal (Figure~\ref{fig:model-schematic}c). Progress is based on the extent to which an action connects more contiguous pieces towards a winning $K$-in-a-row configuration. These two utilities are designed to capture the general intuition that game players aim to make progress towards their goal while preventing their opponents from doing the same.  In addition, we can consider other easily computable heuristic functions that may bear on the value of an action ($U_\text{aux}$, for ``auxiliary''). In our experiments, we consider an auxiliary heuristic based on proximity to the center of the board. It encodes a ``center bias''---a preference for making moves near the center of the board, which allows a piece to participate in more winning terminal configurations. These heuristics are drawn from features used by prior studies in similar games \citep{amir2022adaptive, crowley1993flexible, van2023expertise}, though we generalize them to our broader class of strategic games. The final heuristic value assigned to a given action on a particular board, $\tilde{\mathcal{V}}(s_t, a_t)$, is a sum of the three utility components described above:

\begin{equation}
\tilde{\mathcal{V}}(s_t, a_t) = U_{\text{self}}(s_t, a_t) + U_{\text{opp}}(s_t, a_t) + U_\text{aux}(s_t, a_t).
\end{equation}

Our approach to action evaluation and choice reduces computational cost as it is strongly \textit{local} in both space and time: it considers only the progress immediately stemming from a specific action and does not account for either past or potential future returns. In contrast, it is common for value functions to be action-agnostic (i.e., to depend only on $s_t$ \citep{van2023expertise}) and to explicitly capture some notion of the game's terminal rewards (e.g., via a learned value network \citep{silver2016mastering}). Intuitively, each of these alternatives represents a substantial increase in cognitive load: the former requires scanning over the entire board to evaluate any action and the latter requires mental simulation all the way to the end of the game (or access to a function that is derived from such simulations) before deciding what action to take. 

We next describe our specific implementation of each utility function. As mentioned, we came up with these heuristics via prior literature grounding, as well as expert intuitions. Concepts of progress and blocking are present in the classic study of ~\citep{crowley1993flexible}, and similarly progress patterns and central tendencies are used in the recent ~\citep{van2023expertise}, on which we base our expert model. More generally, features based on connections (or lack thereof), implemented differently based on games, is common in game-playing AI. Most of the authors are expert cognitive modelers and some are expert board game players (and played many Connect-N style games); the combination of progress, blocking, and centering is effectively the first modeling hypothesis that we came up with and judged to be plausible. A more detailed lesion of such heuristics is included in Supplementary Information 3.3. 

The first utility ($U_{\text{self}}$) computes a measure of intermediate ``goal progress'' of the active player, based on a feature $n_1$ defined as the largest line of contiguous pieces created by the action that could be extended to result in a win. This means that actions which extend a line in an illegal direction (e.g., a diagonal line in a game with only horizontal or vertical wins) or that are already blocked by an opponent's piece or the edge of the board do not contribute to goal progress. If the action would result in a win for the active player (i.e., $n_1=K$) then an additional $1$ is added to the feature to magnify the value of winning actions. We note that our choice of feature only rewards actions that form \textit{contiguous} lines of pieces---any piece which is not immediately adjacent to a previously-placed piece from the active player has $n_1 = 0$. While other formulations (such as rewarding actions that form the ends of a line that is empty in the middle, or detecting particular patterns of pieces on the board) are sensible, we choose a simple and easy-to-calculate that function that reflects a straightforward intuition (i.e., making $K$ in a row by first making $K-1$ in a row, $K-2$ in a row, and so on).

The second utility ($U_{\text{opp}}$) computes a measure of ``progress blocked'' for the opponent, based on a feature $n_2$ that is largely symmetric with the goal progress feature above: it is computed as the goal progress the opponent \textit{would} obtain if they made the same move (i.e. with respect to the opponent's allowed winning directions). We subtract $0.5$ from $n_2$ to reflect people's tendency to weigh offense over defense \citep{crowley1993flexible}, so blocking the opponent's $\hat{K}$ in a row is not as good as making a $\hat{K}$ in a row for oneself (but is better than making a $\hat{K} - 1$ in a row). However, if $n_2$ equals the winning $K$, we do not subtract $0.5$ in order to account for the importance of blocking winning threats. As above, there are other reasonable formulations of this feature that we leave to future work.

Finally, the third utility ($U_{\text{aux}}$) is computed as the normalized Euclidean distance between the position of the action and the center of the board, $\xi \in [0, 1]$. This reflects the intuition, applicable across our family of games, that people often place pieces around the center of the board. This tendency may in part be explained by the fact that placing pieces closer to the center of the board allows that piece to participate in the \textit{most possible winning terminal states} for any \textit{$K$-in-a-row} win condition. This measure generalizes to arbitrary rectangular boards. E.g., on a $4 \times 6$ board, intuitively the ``middle point'' is the center, and the middle four cells are all closest to the center ($\langle2, 3\rangle$, $\langle2, 4\rangle$, $\langle3, 3\rangle$, $\langle3, 4\rangle$). 

We assume each utility function is represented as an exponentiation of the respective feature:  

\begin{equation}
    \begin{split}
\tilde{\mathcal{V}}(s,a) = w_\text{connect} \times 2^{n_1} + w_\text{block} \times  2^{n_2} + w_\text{center} \times  2^{(1-\xi)}.  
    \end{split}
\end{equation}
The choice to exponentiate some measure of progress for heuristic functions is common in gameplay modeling \cite[e.g.,][]{cao2019uct, sheoran2022solving, lihongxun2025gobang}. We chose base two based on light initial exploration (prior to collecting any human gameplay data), under the goal of keeping our intuitive model simple. Future work should explore other modeling choices to capture how people might represent and combine multiple utility functions, including the role of learning in how people might come to flexibly synthesize these functions.

Moves are selected by following Boltzmann rationality~\citep{luce1959individual, train2009discrete, franke2023softmax}, sampling actions from a softmax function over their estimated value (based on the above goal-directed heuristics). We assume that players have already developed the capacity to account for multiple goals simultaneously, unlike potentially even more naive child-like game reasoners ~\citep{crowley1993flexible}. Concretely, the probability of choosing action $\hat{a}$ at state $s$ is given by:

\begin{equation}
    P(\hat{a} \mid s) = \frac{\exp\left(\tilde{\mathcal{V}}(s,\hat{a}) / \tau \right)}{\sum\limits_{a} \exp\left( \tilde{\mathcal{V}}(s, a) / \tau \right)}.
\end{equation}

We fix temperature $(\tau)$ at $1$ for our game reasoning and action experiments (with the exception of marginalizing over $\tau$ only for the admixture analyses in the ``play'' experiment). We set weights of each component $(w)$ to $1$ for all experiments. We use the same settings for the Expert Gamer model (as it uses the same value function and softmax-based action selection). We find that equal weights is a reasonable fit for the Intuitive Gamer model to human payoff predictions (see Supplementary Information 3.3). When lesioning a component of the value function, we set its weight to zero.

\paragraph{Intuitive Gamer reasoning module.}

The Intuitive Gamer player module is queried as part of the game reasoning module. The Intuitive Gamer reasoning module nests player module-based simulations to compute a series of gameplay traces ($\{(s^0_0, s^0_1, ..., s^0_T), ...,  (s^k_0, s^k_1, ..., s^k_{T'})\}$). These simulations involve self-play between the same player module type (with the exception of the funness reward for thinking computation, see below), where each agent is trying to make progress towards their own goal (which may be different, e.g., Player $1$ trying to make $4$ in a row only diagonally, while Player $2$ can win in any direction). From these simulations, a probabilistic judgment of the expected outcomes can be made. That is, for each game $G$, $k$ game simulations are sampled, producing a set of $k$ outcomes $o \in \{-1, 0, 1\}$ encoding whether the first player won ($1$), lost ($-1$), or the game ended in a draw ($0$). From these outcomes, we can compute a payoff of a game $G$: 

\begin{equation}
    \psi(G) = (1) P(\text{win} \mid \neg\text{draw}) \cdot P(\neg \text{draw}) + (-1) \cdot P(\text{lose} \mid \neg\text{draw}) \cdot P(\neg \text{draw}) + (0) \cdot P(\text{draw}).
\end{equation}

\begin{equation}
    \psi(G) = P(\neg \text{draw}) \cdot \left[ P(\text{win} \mid \neg \text{draw}) - P(\text{lose} \mid \neg\text{draw})\right].
\end{equation}

Our game reasoning module is constructed to represent the reasoning of any one participant. Therefore, we draw $k$ simulated matches for $N=20$ simulated participants (as each game has, generally, $20$ participant responses; see ``Analysis methods'' below for details on selecting $k$).

\paragraph{Partial game simulations.}
Our primary game reasoning module assumes mental simulations are conducted until the end of the game is reached: either a player attains the win condition, or all open board positions are filled and the game is called a draw. While many of our games do not take many moves to reach an end, people may not mentally simulate all the way to the end of the game in each of their $k$ simulations. We conduct a preliminary exploration into the impact of partial game simulations in fitting peoples' game fairness evaluations by modeling the possibility that people halt any one of their $k$ simulations early. We sample a ``stopping time'' uniformly from 1 to the size of the board and end the game after that many turns. To determine the outcome of games that are stopped early, we apply a simple rule and treat each of them as a draw (allowing us to explore encoding a weak prior towards games ending in draws; see Supplementary Information 5.4). We repeat the same exploration of variance (see ``Analysis methods'' below) and find that the optimal number of samples is similarly more than one and less than ten, but may be closer to $k=4$ (see Extended Data Figure~\ref{fig:partial}). 
We leave further analysis of the distributions over stopping times and how people assign outcomes for partial games to future work.

\paragraph{Funness model.}

To estimate a game's funness, we consider three features derived from playouts under the Intuitive Gamer player module: balance, reward for thinking, and game length. We define a game's balance as the difference between the probability distribution of observed outcomes and an ideal outcome distribution relative to a game where exactly half of the games end in wins and losses and none end in draws (measured by the Earth Mover's Distance~\citep{rubner1998metric}). As described in the main text, this feature captures the notion that players prefer decisive games (i.e., those that do not end in draws) that are also balanced across players~\citep{althofer2003computer, browne2008automatic, todd2024gavel}.  We measure game length as the expected number of moves until a game ends in a draw or win, computed from the same Intuitive Gamer simulations. The effect of game length on funness is modeled using a quadratic function, based on the intuition that the most fun games will be neither too short nor too long.  We define a game's reward for thinking as the proportion of simulated games won by the Intuitive Gamer model playing against the Random Gamer player module, i.e., a player that chooses actions uniformly at random from any legal move.  This feature captures the intuition that players prefer games they expect will challenge their thinking ~\citep{chu2023praise, lantz2017depth}, and reward them for at least some strategic thinking rather than just responding arbitrarily or without any strategy.

We use $120$ game simulations for each feature (to align with the $k = 6$ simulations for each of $20$ participants), randomizing whether the Intuitive Gamer plays first against the random agent in the simulations used to assess reward for thinking. Game simulations are run to the end of the game (where either a player wins or the game ends in a draw); future work can explore variants of the funness model where features are computed under partial game simulations. We use the same features and number of simulations when comparing to alternate models (see below in ``Fitting regression models to funness judgments'').

\subsection*{Alternate models}

We next detail several alternate models. 

\paragraph{Expert Gamer.}

We compare our Intuitive Gamer model against an ``Expert Gamer'' model that differs along two key dimensions: (1) sophistication of the value function, and (2) depth of search. This model is closely based on the model of human expert play for 4-in-a-row on a $4 \times 9$ board in \cite{van2023expertise}, which empirically estimates tree search depth from human players after hours of continuous gameplay experience. As in \citep{van2023expertise}, the depth is controlled by a probabilistic ``stopping parameter'' governing how many the search tree is expanded; we run all simulations by expanding the search tree a fixed number ($k_{iterations}=636$) of iterations, where $k_{iterations}=636$ is the empirical mean value of this parameter estimated from the \cite{van2023expertise} data. This setting corresponds to approximately depth-5 search. On each iteration, the Expert Gamer selects a node (corresponding to a board state). We then expand this node, considering all legal actions, wherein we compute the value of each action as outlined below. The Expert Gamer model then conducts best-first search (BFS) over the game tree, using its state value function defined above. Specifically, it probabilistically expands nodes in the search tree by repeatedly sampling actions from a softmax policy governed by temperature $\tau$ over its current action estimates based on these computations, expanding unexplored states in the search tree, and backpropagating utilities estimated at future states. As in \citep{van2023expertise}, any node that ends in a definite win or loss is assigned $\pm \sigma$ based on whether the move would result in a win for the current play ($+$) or loss ($-$). We set $\sigma = 1000$. 

We next detail the Expert Gamer value function. Following \citep{van2023expertise}, we compute the value of any move by looking at features over the \textit{entire} board state, rather than locally circumspect around the possible move position in question (like the Intuitive Gamer model). This is more computationally-intensive, particularly for larger boards. Specifically, for any open legal position $p$, we imagine a state $s'$ that has that position played by the current player. We then sweep over all played positions with that board state. For each played position $p'$, we compute the same novice, partial-progress value function. We then define the value of that state $s'$ as the difference in cumulative value from the played positions by the current player minus the opponent. We set the value function feature weights ($w_\text{connect}$, $w_\text{block}$, $w_\text{center}$) using the fit values from the Intuitive Gamer model. It is worth noting that the Expert Gamer is itself an important contribution---it is a more general version of a relatively deep goal-directed model. We leave validation of how well the Expert Gamer captures human expert reasoning and play for future work.

\paragraph{Random Gamer.}

The Random Gamer player selects actions uniformly over the space of legal moves. Games are simulated to the end (e.g., til either player wins or the game ends in a draw).

\paragraph{Monte Carlo Tree Search.}

We additionally implement a standard upper-confidence-bound Monte Carlo Tree Search (MCTS) \citep{coulom2006efficient, genesereth2014general} agent to act as an approximate gameplay ``oracle'' (pseudocode can be found in Supplementary Information 2). Unlike the Intuitive Game and Expert Gamer models, MCTS does not use game-specific heuristic features and instead estimates intermediate utilities by expanding a search tree guided by repeated random rollouts to terminal states. MCTS algorithms are commonly used to approximate optimal gameplay in arbitrary games \citep{genesereth2014general}, as they are empirically both efficient and accurate. Specifically, the MCTS implementation we consider here uses a large number of tree expansions ($10,000$) to model the behavior of a near-perfect player in our novel games. Due to computational costs, we estimate the expected payoffs using $50$ simulations per game; these samples are then bootstrapped in their fit to people, as with the other models.

\paragraph{Comparison of MCTS and Intuitive Gamer.}

We briefly clarify the relation between MCTS—the planning algorithm behind many AI systems that achieve expert-level gameplay such as AlphaGo and AlphaZero—and the contribution the Intuitive Gamer model offers as a model of human reasoning. One could potentially view the Intuitive Gamer’s player module as a particularly restricted or shallow variant of MCTS (as MCTS is also compatible with constraining computational resources and adopting heuristic functions). However, for several reasons we think it is important to distinguish between these classes of models---recognizing that this is in part a matter of scientific emphasis and interpretation rather than algorithmic innovation.

First, framing MCTS so generally as to include the Intuitive Gamer’s play would also include almost any kind of stochastic tree search, including other models in our comparison set, e.g., the Random Gamer. More deeply, it would leave out  core features of MCTS that have made it so powerful in AI systems as well as the most distinctive features of the Intuitive Gamer as a cognitive model. MCTS made such an impact in AI gameplay and reinforcement learning~\citep{coulom2006efficient, genesereth2014general, silver2016mastering, silver2017mastering} precisely because it could achieve very strong play without the need for heuristics to guide search. Instead, it uses extensive inference-time computation and sophisticated tree traversal arithmetic (backtracking, node visit counting) to effectively explore a very large, unstructured game tree. This is in contrast to the Intuitive Gamer, which does not require any sophisticated algorithms and uses very minimal computation, although it does rely on a small number of abstract heuristics when assessing the value of next moves. Such a design leads the Intuitive Gamer to be a highly efficient if less than optimal player, and one which we believe much better captures how people reason (as evidenced in our behavioral action selection and action prediction studies). As described in the ``Resource-rational reasoning’' section of the Main Text and captured in Extended Data Table~\ref{tab:game_resourcerational_reasoning}b), the Intuitive Gamer is orders of magnitude more efficient than either MCTS (run for $10,000$ iterations, as described above) or the Expert Gamer, in terms of wallclock time and number of board states evaluated.

\paragraph{Variations on the full Intuitive Gamer model.} 

We compare the full Intuitive Gamer model against several variants, that modulate whether it is flat, goal-directed, or probabilistic. The non-flat version of the Intuitive Gamer model performs a deeper search over possible moves when selecting actions. The non-goal-directed version of the Intuitive Gamer model ablates one or more of the value function components (i.e. player progress or blocking progress---see Supplement for ablations of the center-bias component). The non-probabilistic version of the Intuitive Gamer model replaces the softmax element of action selection with a deterministic choice (or equivalently, softmax with temperature zero). We further vary fastness by conducting a sample complexity analysis wherein we modulate the number of simulations drawn ($k$) from the gamer model, over which the game reasoning engine computes the expected payoff (see Extended Data Figure~\ref{fig:vary-think-k-dev}).

\paragraph{Game-theoretic optimal payoffs.}

Many of the games in our set have known game-theoretic ground truth outcome values assuming optimal play from both players. We describe how we filter which games of the $121$ are estimatable via a game-theoretic optimal payoff. We first include those with known game-theoretic optimal payoffs (either definite Player 1 win, definite draw, or definite Player 2 win) according to \cite{uiterwijk2019solving}. Then, to approximate a larger set of games' optimal payoffs, we additionally include games for which MCTS converged to $\{-1, 0, 1\}$ in its payoff predictions (noting that MCTS estimates are not guaranteed to be perfect). This process results in a set of $78$ out of the full $121$ games for which we have estimated game-theoretic payoffs.

\subsection*{Analysis methods}

We next provide additional details on experimental analyses.

\paragraph{Comparing payoff predictions.}

Participants answered two questions in the game outcome reasoning experiment: how likely the first player wins given the game does not end in a draw ($P(\text{P1 wins} \mid \text{not draw})$), and how likely the game is to end in a draw ($P(\text{draw})$). Together, these responses yield all information we need to compute an expected payoff. We compare the expected payoff from participants against those of models and report the $R^2$. Specifically, we bootstrap subsample with replacement from the human participant samples, per game, and compare models against the mean payoff per sample. We also bootstrap subsample $k$ for $20$ simulated participants per model from the pool of model simulations. Unless otherwise stated, we run $1000$ bootstrap samples.

\paragraph{Sample complexity analyses.} 

We select $k$ (the number of simulations sampled for each simulated participant) by inspecting the variance of the payoff predictions over the simulated set of $20$ participants against the empirical variance observed in the human data. We measure the Root Mean Squared Error (RMSE) between the model- and human-predicted variance for each game, as well as the Wasserstein Distance between the distribution over model and human variances (Figure~\ref{fig:splash-judge}c). We compute the latter to better capture the poor match of high $k$ to people's variances (i.e., with increased $k$, variance collapses to zero). We find that that generally approximately $k=5-7$ full game simulations reasonably well-captures human variance (see Figure~\ref{fig:splash-judge}c and Extended Data Figure~\ref{fig:vary-think-k-dev}) under both measures. We report all main results with $k=6$.

\paragraph{Fitting regression models to funness judgments.}

We fit a linear regression model to these features using the \texttt{lmer} package in R. Features are normalized to zero-mean and unit standard deviation. We fit the models to $1000$ bootstrapped subsamples of the human data for all games. Models are tasked with predicting the mean of participants' funness judgments for each game. We compare models via an ANOVA test and AIC in Supplementary Information 5.7. 

Additionally, we ran a generalization test wherein we fit regression models on $50\%$ ($59$) of the $118$ games studied (infinite boards were removed) and tested each model on the held-out $50\%$ of games. We report both results in Extended Data Figure~\ref{fig:generalization-fun} and include additional details in Supplementary Information 5.7. 

\paragraph{Assessing contribution of non-simulation-based features to funness judgments.}

In exploratory analyses, we assess the potential role of non-simulation-based features that can be read off of the game description alone. We compute a series of binary game traits that capture ways in which a game may differ from the base Tic-Tac-Toe. For instance, a game may not be a $3 \times 3$ board; the game may end with $K \neq 3$ pieces in a row to win; the second player may have a different win condition than the first player. We consider the following binary features: if the game has constrained win conditions (e.g., no diagonals allowed); if the game has asymmetric win conditions (between Player 1 and 2); if the game has asymmetric play dynamics (e.g., Player 2 can place two pieces on their first turn); if the board is not square; $K \neq 3$ pieces in a row to win; if the board is larger than $3 \times 3$; if the game ends when the first player to achieve $K$ in a row loses (i.e. misere variants). We combine the binary features into a single aggregated measure of ``approximate novelty,'' which is the sum of the number of binary features present in any one game. We also explore the incorporation of board size (encoded as the number of rows $\times$ number of columns) as part of non-simulation-based game features that reasoners may consider when assessing the funness of games. We assess how well participants' funness judgments can be explained by these features alone by fitting bootstrapped sets of linear regression models to these features in the same $50/50$ train/test splits over the games. Additionally, we explore the relative benefit of adding all binary features or any of the aggregate features to the simulation-based model over the full set of games (see Supplementary Information 5.7 for more details and additional analyses).

\paragraph{Assessing game-time action selections.} 

To assess how well models capture how people play, we condition the model in question on an intermediate board state of the actual human-versus-human game. All moves in the game are considered (encompassing early-, mid-, and late-stage play). We computed the model's predicted distribution over next moves. We assess the fit of this predicted distribution in two ways. First, we compute the log likelihood of the actual players' moves against those predicted by the model. To handle cases where any model may place at or near zero probability on any given position, thereby skewing the log likelihood, we incorporate a ``slop'' parameter $\alpha$ which captures the probability that a player makes a move from the primary model ($1-\alpha$ that they make a random move on that turn~\citep{nassar2016taming}). We sweep over $\alpha$ between $0.5$ and $0.95$ in increments of $0.05$and compute the average log likelihood for each model over the settings of $\alpha$. We show per-game stage and per-game results in Supplementary Information 6. We conduct a one-sided paired t-test over each individual move as well as the aggregated move likelihood per game, comparing the Intuitive Gamer to the Expert Gamer and Intuitive Gamer to random, respectively. 

\paragraph{Fitting admixture models to actions.}

We then consider two different admixture models: one over all moves made for a game, and one over all moves made for each player. Admixture models captures the notion that a player may play according to either the Intuitive Gamer model, Expert Gamer, or Random on any turn. We jointly fit the weights of the Intuitive Gamer and Expert Gamer model and take the weight of random to be the remaining mass that brings the weights to $1$, over bootstrapped samples over the entire population of players. As the weights are fit jointly across the models, we need to be particularly careful about the magnitude the results for the respective models. To that end, we jointly sweep over the temperature used in the action selection (with temperature ranging from $0.5$ to $3.0$ to produce a distribution over weights). We also fit the admixture over each individual, estimating the relative contribution of each model to the way they play. As each player only plays six games (and one round per game), we fit these weights to all of their moves across all games. We use \texttt{scikit-learn} for fitting, with the SLSQP optimizer. 

\paragraph{Computing aggregate human distribution over predicted next action.}

For our ``watch-and-predict'' experiments, we have access to finer-grained predictions from any \textit{one} participant. We again extract distributions over likely next moves, conditioned on the observed intermediate board state from the novice and alternate models; however, we now compare the distributions directly to the suggested distribution of moves made by participants. We construct an aggregate move distribution over the participants, treating each of the $5$ clicks per participant as contributing some mass to the ``aggregate human'' distribution. 

\paragraph{Comparing model- to human-predicted distributions over next actions.}

We compare distributions using the Total Variation Distance (TVD); we fix temperature at $1$ and sweep over $\alpha$ for both models and people. We conduct a similar one-sided paired t-test over the TVD across the three game stage boards seen per match and compare the Intuitive Gamer and Expert Gamer, and Intuitive Gamer and random. We compute split-half TVDs per stage, per match, per game by randomly splitting the participants' distributions who saw that board, averaging those boards into two aggregate distributions and computing TVD between them. This forms our split-half human estimate which we use to normalize the other models' TVD. We demonstrate the robustness of our results to our choice of distributional measure by repeating the analyses with the Jensen-Shannon Divergence in Supplementary Information 6.2. We additionally run an admixture model over both participant- and game-level predicted watch distributions. These are optimized against the Jensen-Shannon Divergence.

\paragraph{Comparing model- to human-predicted distributions over next actions to people's played actions.}

We secondarily assess how well the distribution over suggested moves made by the watcher captures how people actually played by treating the suggested distribution of clicks by the human watcher akin to the move distributions from the models. We again assess the likelihood of the move played by the active player to the moves suggested by the predictors, the Intuitive Gamer model, and alternate models by sweeping over and averaging out a slippage parameter ($\alpha$). We conduct these analyses in aggregate (over all board stages, matches, and games), as well as at a per game level (over board stages and matches).

\paragraph{Modeling draw requests.}

We take initial steps to analyze evaluations made during the game by looking at players' decisions of whether to accept or reject a draw, when offered. Draw requests could take place at any point during the game. For each board where a draw request was made, we run $40$ simulations to the end of the game under the Intuitive Gamer agent model. We compute the expected payoff over bootstrapped subsampled sets of $k=6$ outcomes (simulating a single player; previously computed under one of our $\psi(G)$ game reasoning queries), as well as the expected remaining length of the game $\ell$ from state $s_t$. We compute expected payoff here with respect to the player who is deciding whether or not to accept the draw request. Players receive a bonus payout of $\$0.5$ only if they win (nothing if they lose or draw); we can then compute an expected value that is the payoff $\times 0.5$. For each set of bonus-adjusted expected payoffs and expected match length remaining, we fit a logistic regression model to predict whether a player would accept or reject the draw request, based on expected payoff and length, as well as the average predicted game funness predicted by people in the ``just think'' game reasoning experiment. We fit the logistic regression model using the \texttt{glm} R package. We also explore the same computations under different agent models varying in depth (e.g., the Expert Gamer model) and goal-directedness of the value function (e.g., ablating the goal progress component). 

We show the decision boundaries in Extended Data Figure~\ref{fig:accept-reject}a and report the bootstrapped parameter fits in Extended Figure~\ref{fig:accept-reject}b. For plotting, we use a collapsed measure of expected value which we call the ``expected value of continuing'' ($\text{C}$) from state $s_t$: 

\begin{equation}
  \text{C}(s_t) = (\psi(G) \times 0.5) - (\beta_{\text{length}}/\beta_{\text{utility}}) \cdot \ell.
\end{equation}

This measure combines the expected value (bonus-adjusted payoff) with the opportunity cost of continuing (based on expected length remaining).

\putbib[main]
\end{bibunit}

\section*{Data Availability}

Anonymized human data game evaluation, action selection, and action prediction data is hosted at our \href{https://github.com/collinskatie/intuitive-game-reasoning.git}{GitHub repository}. 

\section*{Code Availability}
Code available at our \href{https://github.com/collinskatie/intuitive-game-reasoning.git}{GitHub repository}.

\section*{Acknowledgments}

We thank Max Kleiman-Weiner, Laura Schulz, Rebecca Saxe, Mira Bernstein, Scott Stern, Ilia Sucholutsky, Lance Ying, Junyi Chu, Tracey Mills, Jacob Andreas, Tyler Brooke-Wilson, Judy Fan, Yuka Machino, Sydney Levine, Ionatan Kuperwajs, Umang Bhatt, Wei Ji Ma, Kevin Smith, Nathaniel Daw, Gabriel Poesia, and Ben Prystawski for comments on the manuscript or helpful conversations that informed this work. Thank you to the family and friends of the authors for many fun game nights. We also thank our reviewers for their very helpful suggestions in revising our manuscript.

KMC discloses support from from the NSF SBE SPRF (2507321) and the Cambridge Trust and King's College Cambridge. JBT and CEZ disclose support from AFOSR (FA9550-22-1-0387) and the ONR Science of AI program (N00014-23-1-2355). JBT additionally discloses support from a Schmidt AI2050 Fellowship and the Siegel Family Quest for Intelligence at MIT. LCW discloses support from a Stanford HAI Fellowship. GT discloses support from the National Science Foundation Graduate Research Fellowship under Grant DGE-2234660. TLG discloses support from the NOMIS Foundation. AW discloses support  from  a  Turing  AI  Fellowship  under grant  EP/V025279/1, The Alan Turing Institute, and the Leverhulme Trust via CFI. This work is supported (in part) by ELSA - European Lighthouse on Secure and Safe AI funded by the European Union under grant agreement No. 101070617. CZ, MB, and SC disclose no relevant funding. Views and opinions expressed are however those of the author(s) only and do not necessarily reflect those of the European Union or European Commission.

\section*{Contribution Statement}

All of the authors contributed to project framing and editing the manuscript. KMC, CEZ, LW, JBT conceived of and designed the “just think” experiment. KMC, CEZ, LW, MB, GT, TLG, and JBT conceived of and designed the play and watch studies. KMC, CEZ, LW, JBT developed the Intuitive Gamer model framework. CEZ developed the Intuitive Gamer heuristics. KMC, CEZ, LW, MB, and GT implemented the models. CEZ designed the 121 games. KMC implemented the human experiment interfaces and ran human data collection. KMC, CEZ, MB, and GT implemented and ran the analyses. KMC, CEZ, LW, MB, GT, and JBT drafted the paper. JBT supervised the project and acquired funding.

\section*{Correspondence}

All correspondence and requests for materials should be directed to Katherine M. Collins (\texttt{katiemc@mit.edu}). 
\section*{Competing Interests}

The authors declare no competing interests.


\newpage 

\begin{extfigure}[h!]
    \centering
    \figimage{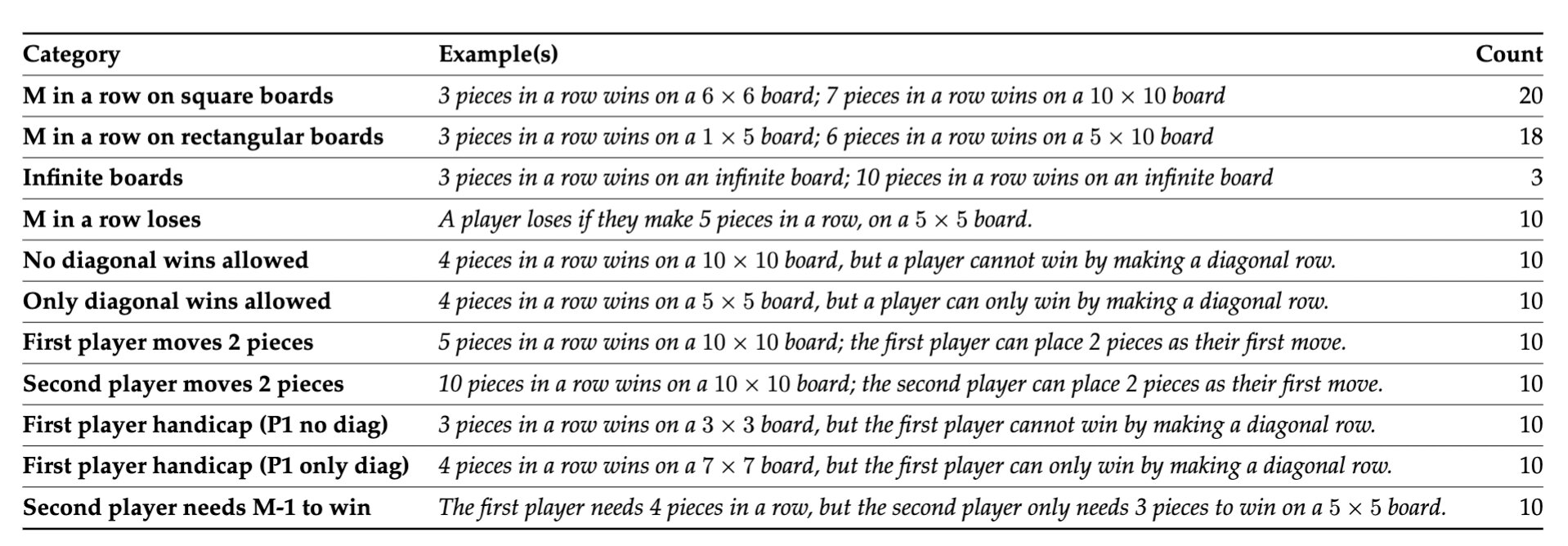}

\caption{\textbf{Diverse set of novel games.} Wide variety of $121$ two-player grid-based strategy games designed to study how people play and think about games when they are novices. Games vary in the board shape (e.g., square versus rectangular), win conditions (e.g., first to $K$ in a row wins versus loses), and game dynamics (e.g., one player can play two pieces on their turn). Counts of games in each category, as well as one representative example game description for each.} 
\label{tab:stimuli}
\end{extfigure}

\begin{extfigure}[h!]
    \centering
    \figimage{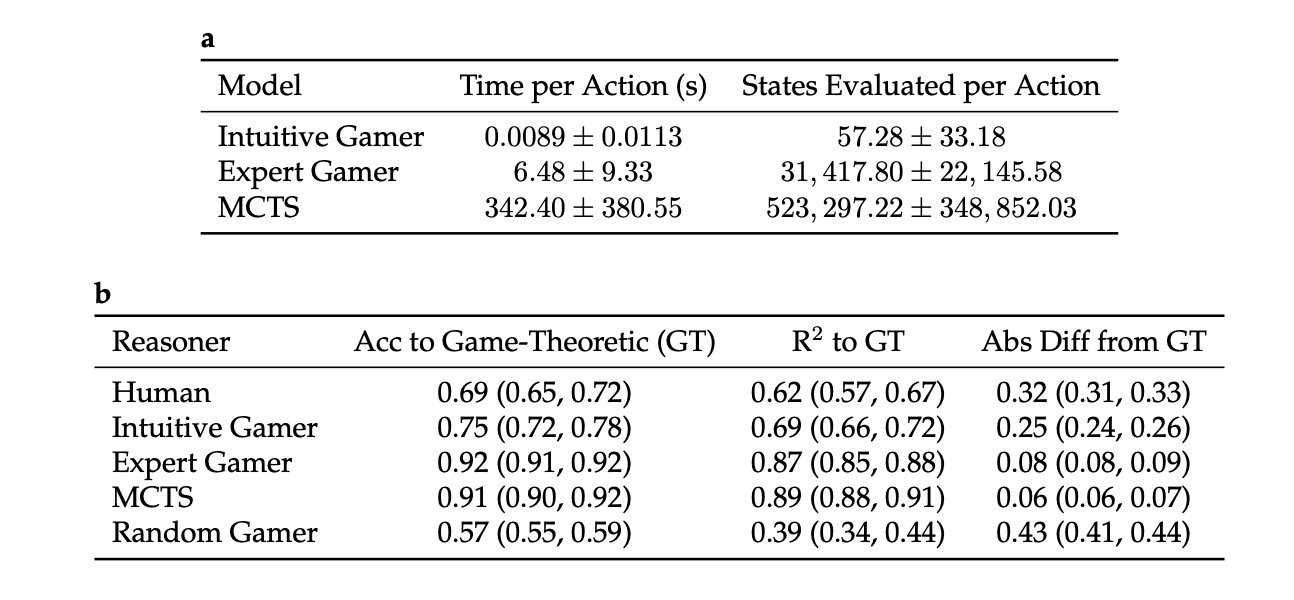}
    \caption{\textbf{Resource-rational game reasoning, comparing compute efficiency and similarity of payoff estimates to estimates of game-theoretic optimal payoff.} \textbf{a}, The Intuitive Gamer is more efficient than more sophisticated game reasoners, both in terms of average time to evaluate before deciding what action to select and average number of game states to evaluate before deciding what action to select. represent mean $\pm$ standard deviation over all moves when simulating evaluations for the game reasoning payoff predictions, for a subset of $41$ of the $121$ games. Additional details on cost comparisons are included in Supplementary Information 3. \textbf{b}, Human and model judgments are compared to the $78$ of the $121$ games where the game-theoretic optimal payoff is estimatable. Accuracy between predicted payoff and the approximate game-theoretic optimal is computed by taking the predicted payoff as ``correct'' if the game reasoner predicted a payoff $> 0.5$ and the expected payoff is $1$; correct if the predicted payoff is $< -0.5$ and the expected payoff is $-1$; correct if the predicted payoff is between $-0.5$ and $0.5$ and the game is expected to end in a draw. $R^2$ correlation is computed between the raw predicted payoffs and the game-theoretic optimal values, as well as the average absolute difference between the expected predicted payoff and approximate game-theoretic payoff (lower is closer to ``correct''). Bootstrap 95\% confidence intervals (CIs) are shown in parentheses, where bootstraps are over bootstrapped samples of participants (with replacement) for people or model simulations.}
    \label{tab:game_resourcerational_reasoning}
\end{extfigure}

\begin{extfigure}[h!]
    \centering
    \figimage{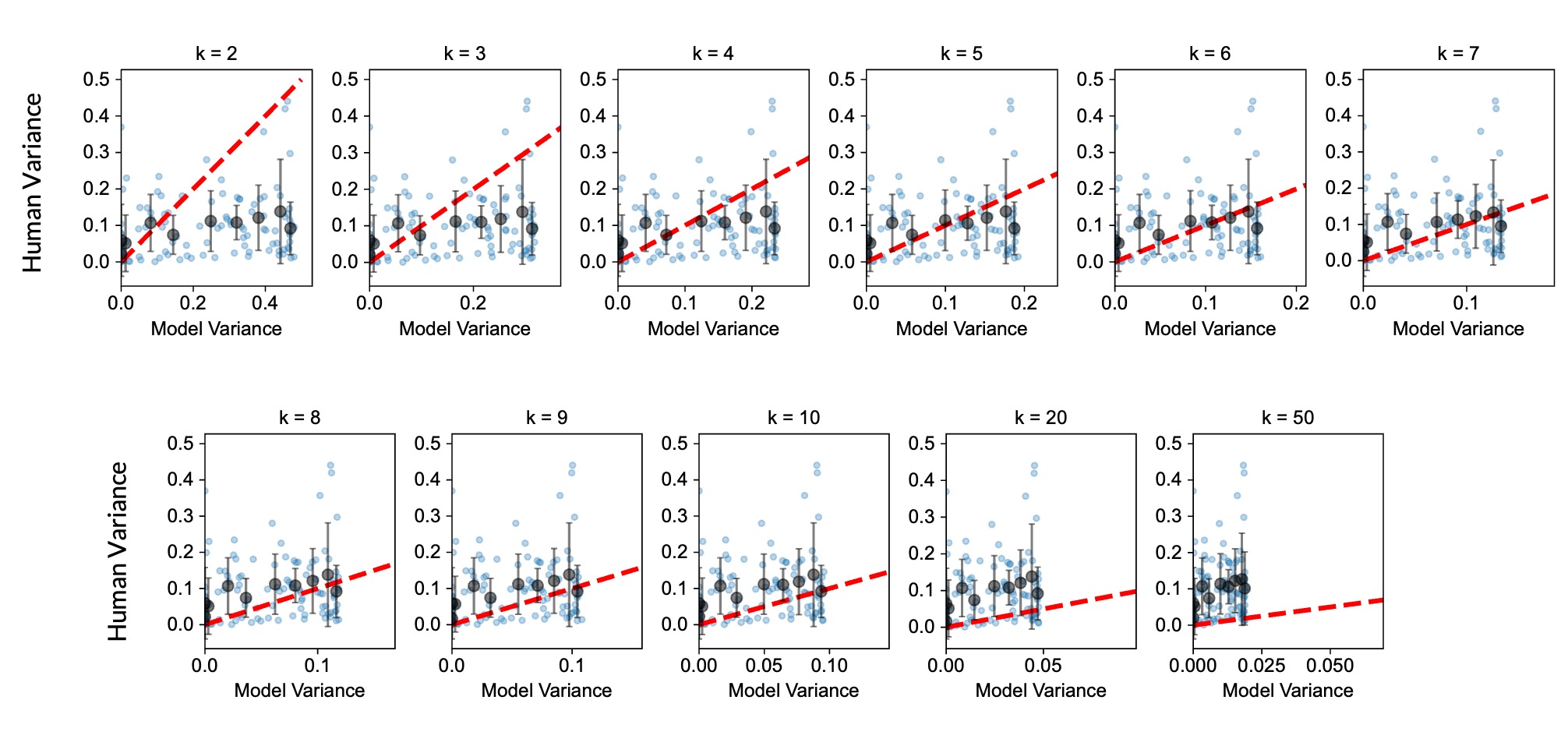}
    \caption{\textbf{Sample complexity analysis, based on the variance of human and model payoff predictions with varied compute budget (number of simulations, $k$) for the Intuitive Gamer model).} The variance over each approximately $20$ human participant payoff judgments per game and compare against the average variance under simulated sets of $N=20$ simulated participants each drawing $k$ simulations from the model. Model variance is binned into quintiles along the horizontal axis and the average human variance is computed for points in that bin (the black dot overlaid on the scatterplot). Error bars are standard deviation of the variance for points in that bin.}
    \label{fig:vary-think-k-dev}
\end{extfigure}

\begin{extfigure}[h!]
    \centering
    \figimage{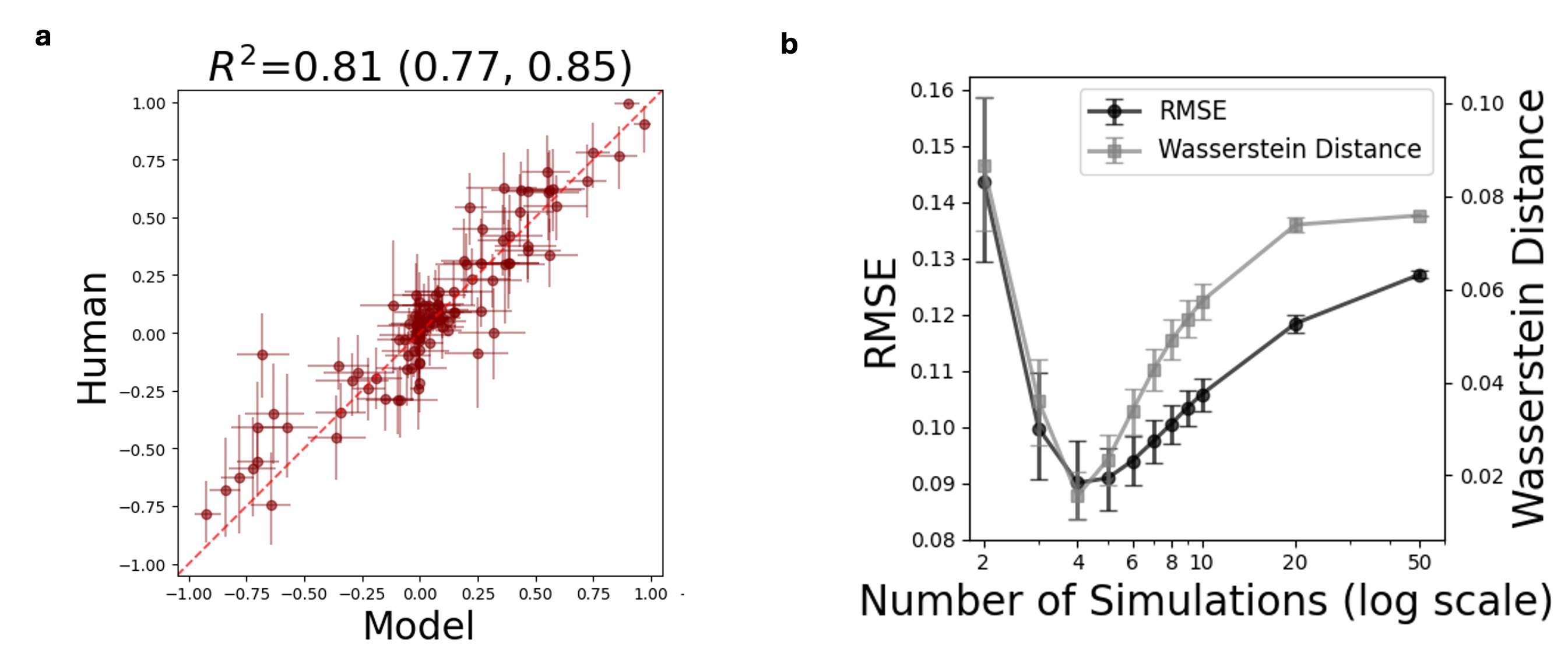}
    \caption{\textbf{Partial game simulations in the Intuitive Gamer reasoning module.} All main results are reported under the assumption that the game reasoning module runs simulations to the end. While most games do not involve a large number of moves, under the Intuitive Gamer player module, for the game to end---it is possible that people are engaging even cheaper partial simulations when evaluating novel games for the first time. \textbf{a,} Exploratory analyses into running the Intuitive Gamer reasoning module under a probabilistic stopping rule equivalently captures the majority of the explainable variance in human payoff judgments as running the Intuitive Gamer reasoning module under full simulations (see Figure~\ref{fig:splash-judge}a). For direct comparison with the full simulation model, $k=6$ partial simulations were run to estimate game payoff. The partial simulations were computed as follows. For each simulated match of a game, a stopping time was sampled from a uniform distribution over board size and terminated early if the game did not end before that time. If the game terminated before a winner or draw was called, the game counted as a draw. \textbf{b,} To assess the impact of compute budget in the number of simulations used to estimate payoff, the variance of payoff under different number of partial simulations $k$ was again compared with the variance in people's payoff evaluations (as in Figure~\ref{fig:splash-judge}c and Extended Data Figure~\ref{fig:vary-think-k-dev}). The best fitting number of samples ($k$) is similarly greater than one and less than ten when using partial simulations, and perhaps even a bit less (e.g., $k=4$) compared to using full game simulations (see Figure~\ref{fig:splash-judge}c).} 
    \label{fig:partial}
\end{extfigure}

\begin{extfigure}[h!]
    \centering
    \figimage{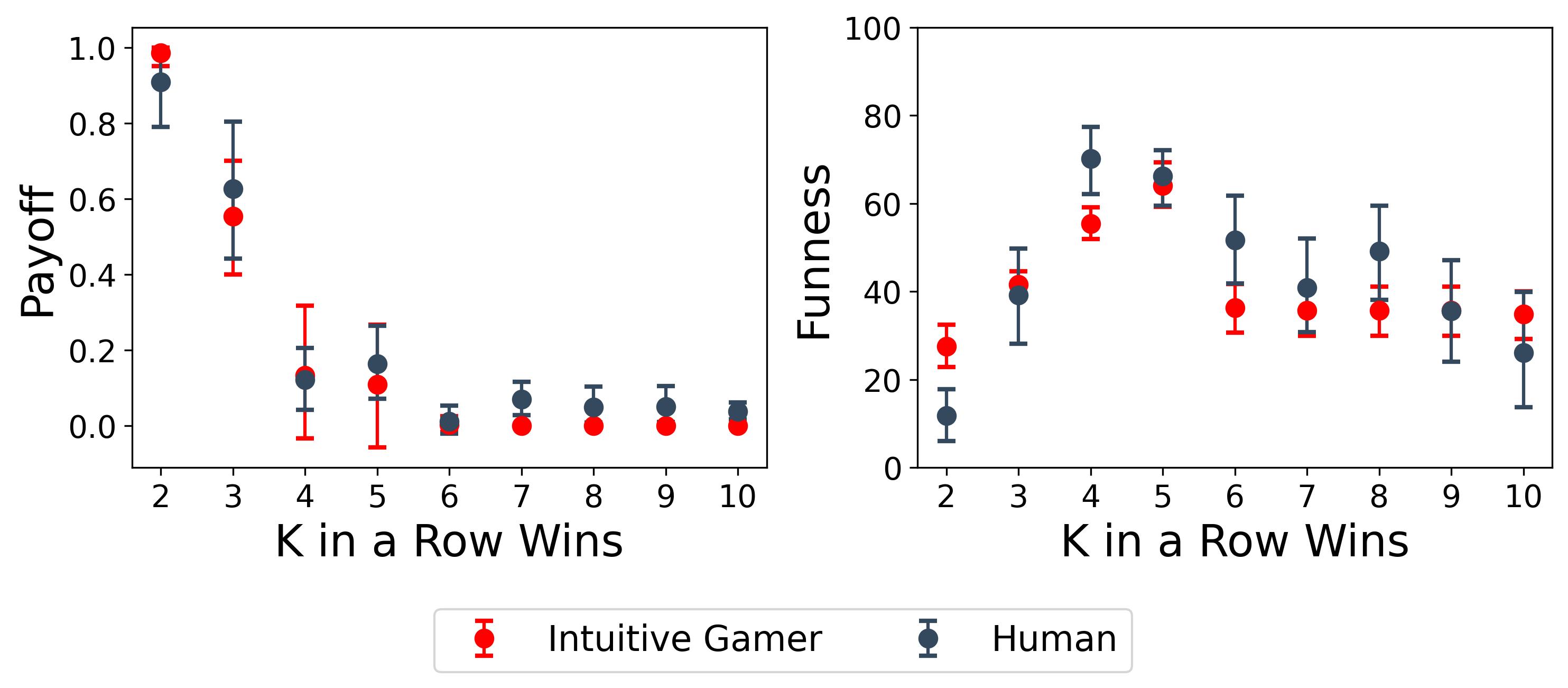} 
    \caption{\textbf{Systematic variation in human and model predictions with change in game rules.} 
    Human and model judgments follow similar qualitative trends in \textbf{a,} payoff and \textbf{b,} funness predictions as $K$ varies for a $10 \times 10$ board. Human and model average predictions are shown. Error bars depict $95\%$ CIs over the mean ($1000$ bootstraps). Funness model error bars (\textbf{b}) are computed over repeated model fits to $50\%$ splits of the games, holding out the subset of $10 \times 10$ games of varied $K (\in \{2,...,10\})$ from each train split. These data reveal systematic and interpretable patterns in the human data (e.g., estimated payoff plateauing towards zero with increased $K$; funness peaking at a intermediate values of $K$). In particular, when $K$ is too small ($3$ or less) the game is unbalanced, as the first player can easily win and the Intuitive Gamer simulates this easily.  For too large values of $K$, the game is hard for either player to win and there is little reward for thinking; the Intuitive Gamer also simulates this. } 
    \label{fig:vary-10x10}
\end{extfigure}

\begin{extfigure}[h!]
    \centering
    \figimage{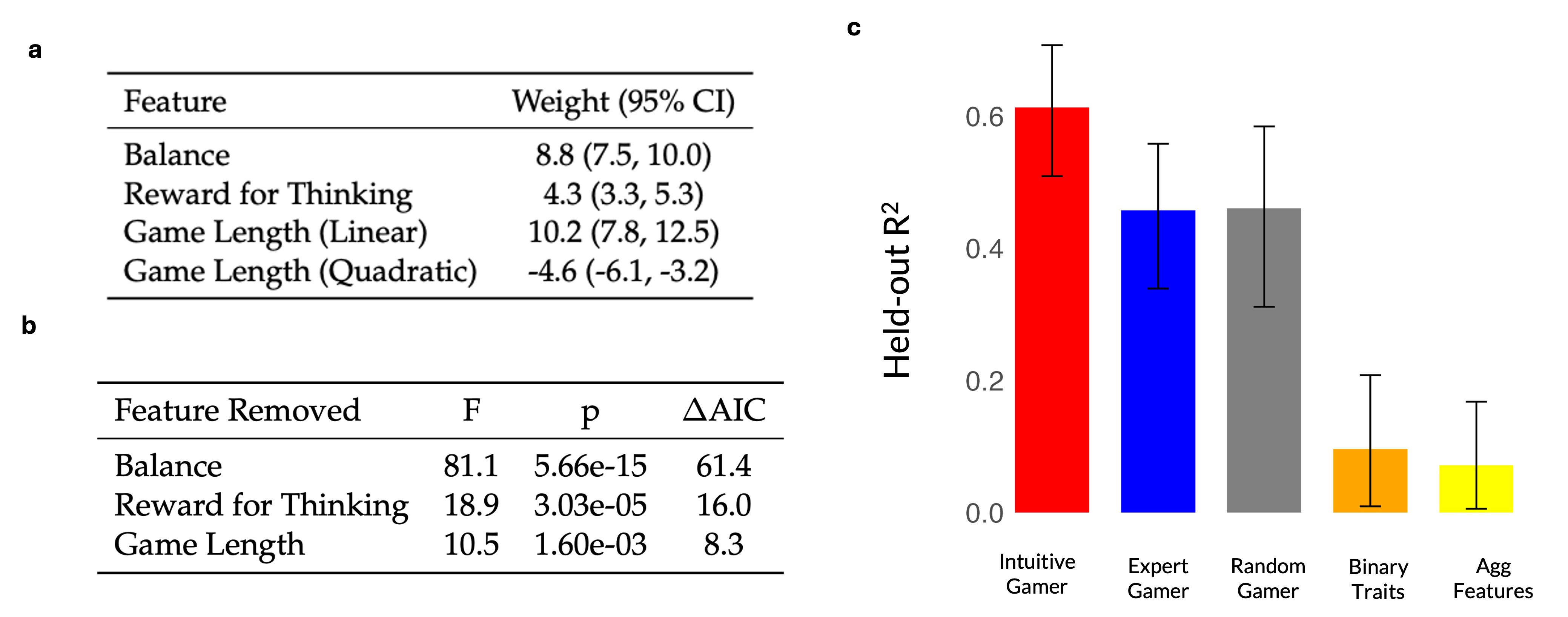} 
    \caption{\textbf{Funness model feature importances and generalization test comparing funness under alternate Gamer agents and non-simulation based funness models.} \textbf{a,} Funness model feature weights reveal that games that are more balanced are generally considered more fun, as are games that are more rewarding to think about (though they contribute less per unit to the funness than balance). Game length has an inverted-U shaped relationship to fun (as evidenced by the positive linear weight and negative quadratic weight; games that are too short or too long are less fun). Weights of the features over $1000$ models fits to bootstrapped mean human funness predictions, over all games. Coefficients are over normalized features (see Methods). \textbf{b,} Ablating each linear feature in the primary funness model reveals that all features matter. Higher $F$-statistic means that ablating the feature explains less of the variance in the human data as the full model. $\Delta$AIC is AIC$_\text{ablated}$ - AIC$_\text{full}$ (a higher value indicates that ablating the feature impairs fit to human data, relative to the full model). \textbf{c,} Simulations under the Intuitive Gamer model better capture human judgments than simulations under models varying the computational sophistication (Random and Expert Gamer) in a generalization test, where the parameters of each model are fit on $50\%$ of the games and tested on the other $50\%$ of games. Funness models based on explicit simulation better capture human judgments than models based only on non-simulation based linguistic features (e.g., binary game traits---whether the game has asymmetric win conditions, non-square boards, etc) and aggregate features based binary traits (counts of how many are ``on'' to capture ``novelty'' relative to Tic-Tac-Toe; and board size, encoded as number of rows $\times$ number of columns). Error bars depict 95\% CIs of the held-out $R^2$ over $500$ bootstrapped train/test splits.}
    \label{fig:generalization-fun}
\end{extfigure}

\begin{extfigure}[h!]
    \centering
    \figimage{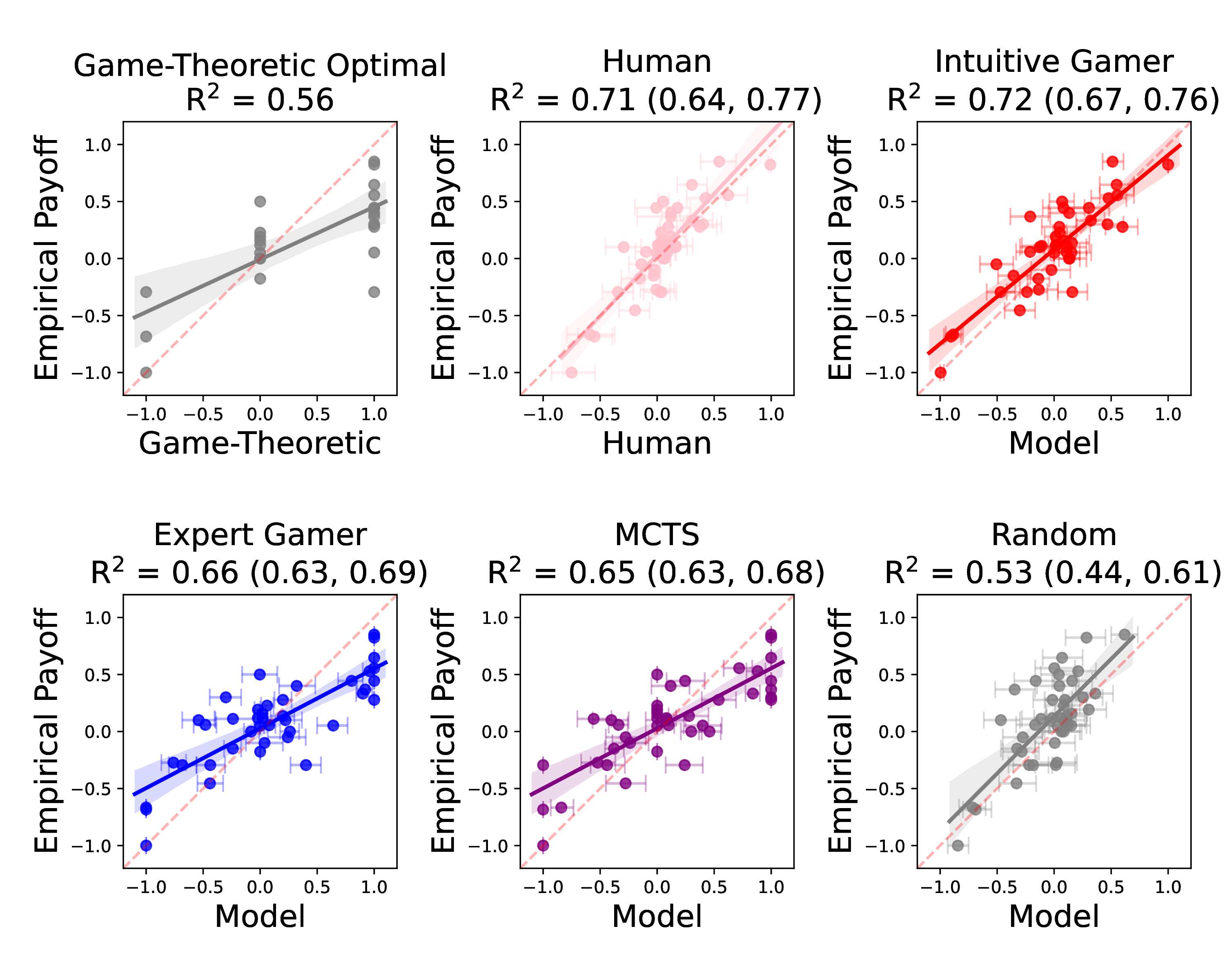}
    \caption{\textbf{Empirical observed payoffs in actual human-human play against human- and model-predicted payoffs.} The empirical payoffs (based on the observed human-human play game outcomes) generally follow the payoffs predicted by people and the Intuitive Gamer model in the ``just think'' experiment (without any live play). The empirical payoffs do not align with what one would expect under optimal play, nor play under relatively more sophisticated expert models (MCTS or the Expert Gamer) or game agents that use less compute (random).}
    \label{fig:empirical-obs-payoff-v-think}
\end{extfigure}

\begin{extfigure}[h!]
    \centering
    \figimage{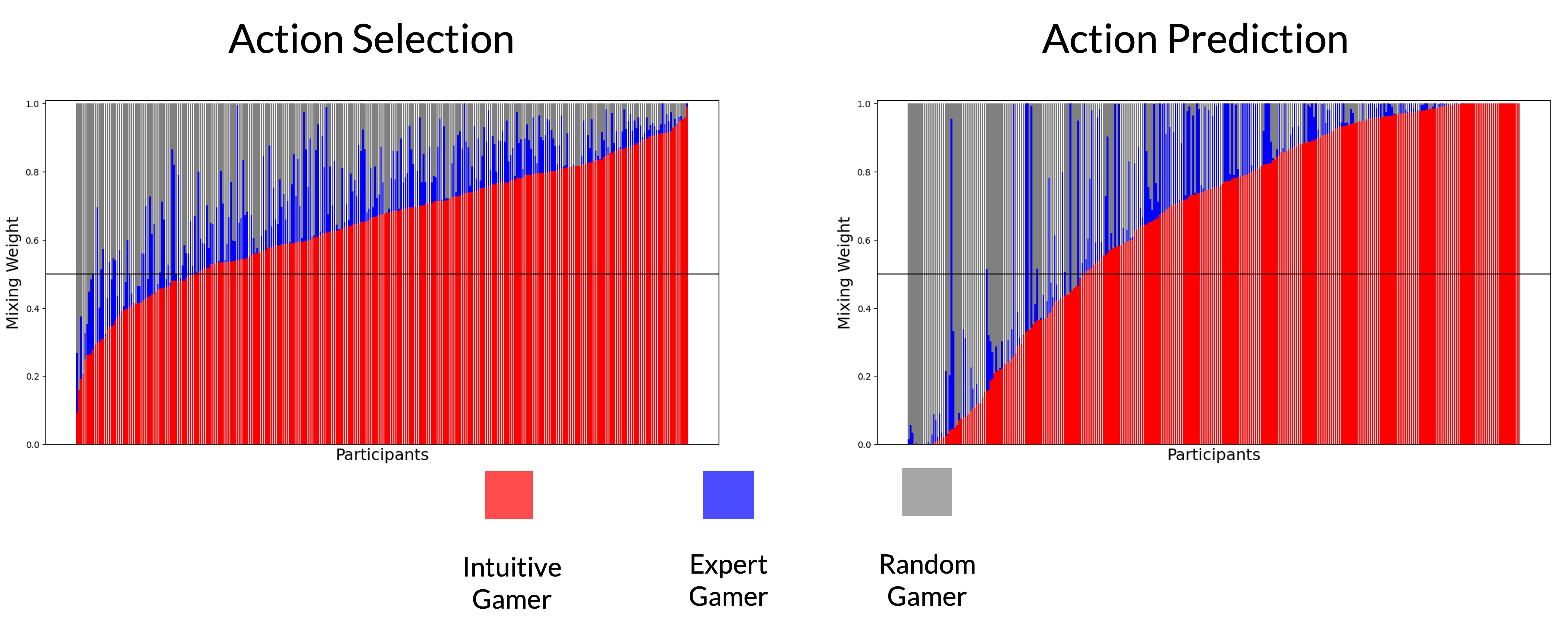} 
    \caption{\textbf{Modeling individual game players.} A probabilistic mixture model (admixture) is fit for each individual participants' action choices (left) and action predictions (right). Bar heights represent the estimated contribution of each model in the probabilistic mixture. The Intuitive Gamer is the dominant component for most participants; however, there is variability across individual participants' behavior---some participants make action choices more aligned with a Random Gamer or more sophisticated Expert Gamer model.}
    \label{fig:individ-play}
\end{extfigure}

\begin{extfigure}[h!]
    \centering
    \figimage{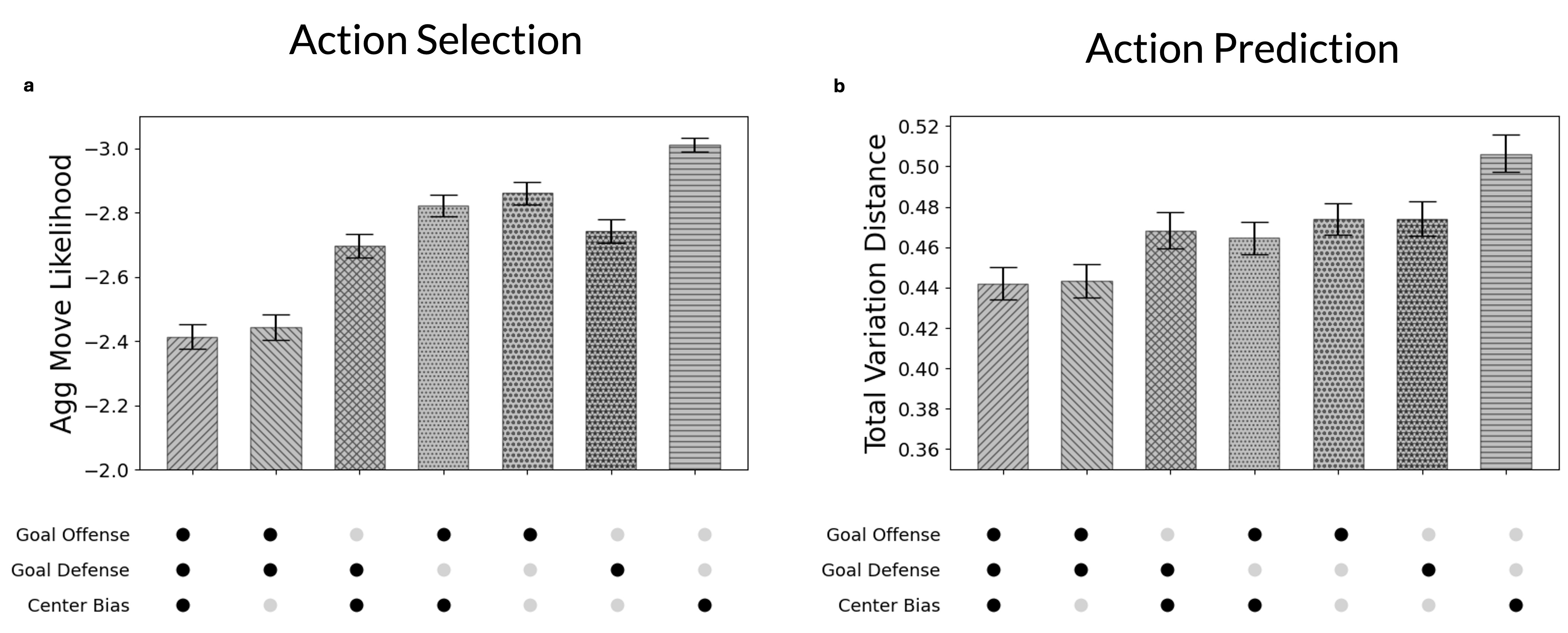} 
    \caption{\textbf{Heuristic value function ablations.} To assess the importance of components of the Intuitive Gamer model to fitting participant data, key features of the full Intuitive Gamer value function $\tilde{\mathcal{V}}$ are lesioned by setting the component weight to zero and leaving the others fixed at one. \textbf{a,} Aggregate log probability under each variant of the human played moves from the ``play'' (action selection) experiment; \textbf{b,}  aggregate TVD comparing the model and human predictor distributions per game stage made in the ``watch-and-predict'' experiment. Error bars depict bootstrapped 95\% CIs over the mean measure per setting. Ablating the connect weight or the block weight has major effects on how well the model explains human data in all three experimental conditions. However, ablating the center weight only has effects in the play experiment. The lack of impact of the center bias in the ``watch-and-predict'' data may arise from the center bias only mattering in the first moves (and participants are only asked to provide judgments over the third move onwards for the ``watch-and-predict'' task).}
    \label{fig:ablate-value-comp}
\end{extfigure}

\begin{extfigure}[h!]
    \centering
    \figimage{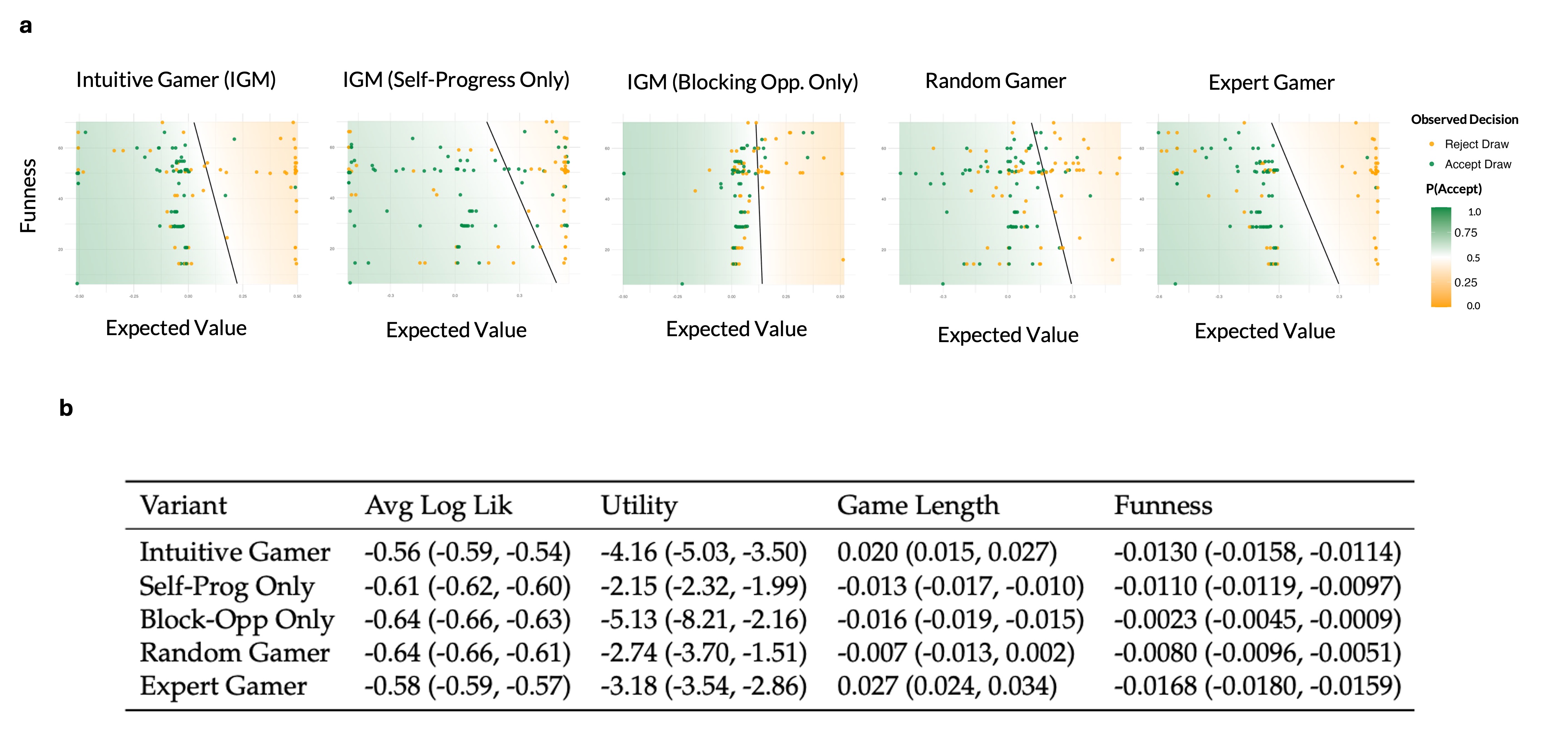} 
    \caption{\textbf{Modeling people's decisions about whether to keep playing when they receive a draw request.} \textbf{a,} Logistic regression models predict whether a player would accept or reject a draw request if received are fit to the expected value of continuing to play under the respective player module variants, and the expected funness of games (as predicted by people in the just-think experiments). Decision surfaces are plotted, averaged over $100$ bootstrapped samples of expected value calculations over $k=6$ simulations. The Intuitive Gamer and Expert Gamer qualitatively encode the intuition that games that have higher expected value from continuing or are expected to be fun are games where players are more open to continuing to play. \textbf{b,} Coefficients for draw request models, depicting $95\%$ CIs over bootstrapped subsamples of $k=6$ game simulations. The actions taken by a player (whether to accept or reject a draw) are assessed under the models by computing the log likelihood of doing the action, given the features computed over the particular board. The model was trained to predict whether a draw request would be accepted; therefore, negative weights for features mean that a higher value of that feature is more predictive of rejecting a draw (i.e., wanting to continue a game). Log likelihood is computed as the averaged likelihood assigned to each decision and averaged over bootstrapped model fits. Higher likelihood means better fit. Using the expected value of continuing as computed under Intuitive Gamer and deeper Expert Gamer simulations similarly explain people's decisions of whether to accept or reject a draw. Lesioning either or both of the goal-directed components of the Intuitive Gamer value function substantially impairs fit to people's decisions, particularly when lesioning the offensive component (i.e., the component specifically driving a player to make progress towards their own goal). This may be because, upon receiving a draw request, a player decides whether to accept or reject from a more offensive stance.}
    \label{fig:accept-reject}
\end{extfigure}

\clearpage

\begin{bibunit}[naturemag]

\renewcommand{\contentsname}{Supplementary Information}
\tableofcontents
\newpage

\section{Summary of experimental variations on full Intuitive Gamer model}

We next summarize how the experiments varying model components show the necessity of all the different aspects of the Intuitive Gamer model (that it is fast, flat, goal-directed, and probabilistic). 


\textbf{That simulations are flat:} by ``flat,'' we mean the Intuitive Gamer player module is depth-limited. Our comparisons to the Expert Gamer (which is approximately depth-$5$) in Figure~\ref{fig:splash-judge}a, as well as our comparisons to a more intermediate depth gameplayer (depth-$3$) in Figure ~\ref{fig:splash-judge}b and Supplementary Information~\ref{fig:depth-3-predicted-payoff} show distinctly lower $R^2$ relative to human payoff predictions. We also find that computing the funness model features under a deeper gameplay agent (Extended Data Figure ~\ref{fig:splash-judge-fun}) does not capture human funness judgements as well as a depth-1 Intuitive Gamer agent. Similarly, we find that a depth-limited player module better captures people’s prediction actions in-game and their judgments about others’ likely play (see Figure~\ref{fig:splash-live-choices-play-watch}, comparison between Intuitive Gamer and Expert Gamer). 

\textbf{That simulations are fast:} by ``fast'', we mean that the Intuitive Gamer's reasoning is fast. This is in part because simulations are shallow, of course, but also because only a small number of simulations need to be run to make each judgment at the level of resolution or confidence we see in people's judgments. The model's judgments are always stochastic, and by decreasing or increasing the number of simulations that each run of the reasoning module conducts per trial, we can obtain more or less variable judgments about game fairness and payoff over different runs of the model. By comparing the variance in judgments across Intuitive Gamer model runs (each run simulating a single human participant) with the variance in judgments across a sample of actual human participants, and equating for number of actual and simulated participants, we can estimate the effective number of simulations people are running (to the extent they are well modeled by an Intuitive Gamer model). We find that using only around $5-7$ simulations from the Intuitive Gamer reasoning module best captures the variability in human judgments, and show this result in Figure ~\ref{fig:splash-judge}c and Extended Data Figure~\ref{fig:vary-think-k-dev}. 

\textbf{That simulations are goal-directed:} by ``goal-directed,'' we mean that the Intuitive Gamer player module chooses actions in a motivated fashion, towards its own goal---and assumes the other player is also motivated to meet their goal (and hence, the Intuitive Gamer tries to block the opponent’s progress). Ablating either or both of these two features---progress towards one’s own goal — is necessary to capture people’s payoff judgments (see Figure ~\ref{fig:splash-judge}b), as well as their actual in-game moves and suggested moves for others (Extended Data Figure ~\ref{fig:ablate-value-comp}). Goal-directedness is also critical for capturing people’s decisions about whether or not to accept a draw request (Extended Data Figure~\ref{fig:accept-reject}).  

\textbf{That simulations are probabilistic:} by ``probabilistic,'' we mean that actions are selected by the Boltzmann rule, sampled from a softmax distribution over the estimated value of actions (based on the above goal-directed heuristic). We also consider a deterministic action selection policy (that effectively takes the best action on each move), which we find neither qualitatively nor quantitatively captures the variance in people’s payoff predictions (see Figure ~\ref{fig:splash-judge}a and ~\ref{fig:splash-judge}b). We conduct a wider range of temperature analyses below.

\section{Model pseudocode}
We include pseudocode for our primary and alternate models.

\subsection{Intuitive Gamer reasoning module}

We provide example pseudocode for the reasoning module applied to estimating the payoff of matches. Critically, the reasoning module simulates $k$ games; based on these game playouts various queries could be computed (e.g., estimating the payoff of the game, or other properties like the length of the match).

\begin{algorithm}[h!]
\caption{Reasoning Module}\label{alg:reasoning-module}
\begin{algorithmic}
\State $k \gets $ number of simulated matches 
\State $\textit{outcomes} \gets []$ 
\For{$i = 1$ \textbf{to} $k$}
    \State $o \gets \Call{PlayerModule}{}$\Comment{Simulate one match (self-play)}
    \State $\textit{outcomes}.\text{append}(o)$ \Comment{Store outcome}
\EndFor
\State \Return $\Call{Payoff}{\textit{outcomes}}$ \Comment{Estimate expected payoff from outcomes}
\end{algorithmic}
\end{algorithm}

\subsection{Intuitive Gamer player module}

We next provide pseudocode on the player modules.

\begin{algorithm}[h!]
\caption{Intuitive Gamer: MakeMove($s$)}\label{alg:approximate-expert}
\begin{algorithmic}
\State $n_1 \gets $length of longest \textit{live} connection extended 
\If{$n_1$ is winning length}
\State $n_1 = n_1 + 1$
\EndIf
\State $n_2 \gets $length of longest \textit{live} opponent connection blocked $-$ $0.5$
\If{$n_2$ is winning length}
\State $n_2 = n_2 + 0.5$
\EndIf
\State $ \epsilon \gets $ Euclidean distance to center of board, normalized so $d=1$ for the corner pieces
\State $\tilde{\mathcal{V}}(a, s) = 2^{(1-\epsilon(a))} + 2^{n_1(a)} + 2^{n_2(a)}$
\State \Return $\textrm{Sample} (\textrm{Softmax} (\{a: \mathcal{\tilde{V}}(a, s)\textrm{ for }a\in s.actions\})$
\end{algorithmic}
\end{algorithm}

As explained in our main text, our Intuitive Gamer player module is flat, goal-directed, and probabilistic. It incorporates shallow search and considers only a constrained set of heuristics ($\tilde{\mathcal{V}}$) for measuring intermediate game states. Actions are evaluated based on their proximity to the center of the board $d$, the length of the live connections they extend $n_1$, and the length of the opponent's live connections that they block. By \textit{live connections} we mean connected pieces that have enough open positions on the
corresponding direction such that they can be extended to
form a winning configuration. Actions are then sampled from a softmax distribution over open-board states, weighted by the heuristic value computations.

\subsection{Expert Gamer player module}
We provide pseudocode for the Expert Gamer model (designed to generalize the 4-in-a-row model of human gameplay from \citep{van2023expertise}). Its algorithm repeats three sub-procedures, SelectNode, ExpandNode, and Backpropagate, within the main procedure MakeMove, which ultimately makes a move by sampling from a softmax distribution.

\begin{algorithm}[h!]
\caption{Expert Gamer: MakeMove}\label{alg:approximate-expert-make-move}
\begin{algorithmic}
\State $root \gets \Node(s)$
\For{$i=1$ to $num\_steps$}
    \State $n\gets\SelectNode(root)$
    \State $\ExpandNode(n)$
    \State $\Backpropagate(n, root)$
\EndFor

\State \Return $\textrm{Sample} (\textrm{softmax} (\{c: c.val\textrm{ for }c\in root.children\})$
\end{algorithmic}
\end{algorithm}

The function Node is the constructor for the nodes of the tree search. As described in the ``Methods'' section, in our simulations we repeat the procedure in MakeMove for $num\_steps=636$ iterations, drawing on results from an empirical parameter fit in \cite{van2023expertise} which found that expert human players (playing a specific 4-in-a-row variant) were well modeled by a heuristic search procedure whose number of iterations is geometrically distributed with a mean stopping parameter of $1/636$. Our model does not use a stochastic stopping parameter and just runs deterministically to $num\_steps=636$ iterations for all simulations. In the following, \textit{player\_type} $\in \{X, O\}$ and \textit{move\_number} is the index of the current move over the entire game (to account for variants which specify that a given player goes twice as their opening move).

\begin{algorithm}[h!]
\caption{Expert Gamer: $\SelectNode(n)$}\label{alg:ae-select-node}
\begin{algorithmic}
\State $root\gets n$
\While{$n.children\neq\emptyset$}
    \If{rand()$<\epsilon$}
        \State $ n= \textrm{unirandom}(\{c: c.val \textrm{ for } c \in n.children\})$
    \Else
        \If{$n.player\_type=root.player\_type$}
            \State $n=\textrm{rand-argmax}(\{c: c.val\textrm{ for }c\in n.children\})$
        \Else
            \State $n=\textrm{rand-argmin} (\{c: -c.val\textrm{ for }c\in n.children\})$
        \EndIf
    \EndIf
\EndWhile
\State \Return $n$
\end{algorithmic}
\end{algorithm}

\begin{algorithm}[h!]
\caption{Expert Gamer: $\ExpandNode(n)$}\label{alg:ae-expand-node}
\begin{algorithmic}
\State $s\gets n.state$
\ForAll{$m\in \LegalMoves(s)$}
    \State $b\gets \PlacePiece(s,m,n.player\_type)$
    \State $i\gets n.move\_number$
    \State $n.children.append(\Node(b,i+1, \tilde{\mathcal{V}}^\text{EG}(s))$
\EndFor
\end{algorithmic}
\end{algorithm}

Within the procedure ExpandNode, the function $\tilde{\mathcal{V}}^\text{EG}$ is an extension of our Intuitive Gamer heuristic function $\tilde{\mathcal{V}}$ defined above, and it represents a more sophisticated heuristic. To set a frame of reference, suppose player $X$ is evaluating the utility of board state $b$. The heuristic $\tilde{\mathcal{V}}^\text{EG}\textbf{(b)}$ is evaluated as 
\begin{equation}
    \label{eq:expert-val}
    \begin{split}
        \tilde{\mathcal{V}}^\text{EG}(s)=\sum_i (2\times \mathbb{I} \{\textrm{the }i\textrm{'th move is done by }X\}-1)\tilde{\mathcal{V}}(p_i,s_i).
    \end{split}
\end{equation}
where $b_i$ is the $i$'th board observed in the game and $p_i$ is the $i$'th move played in the game. The value function is defined symmetrically when $O$ acts. We make this modification so that $\tilde{\mathcal{V}}^\text{EG}$ incorporates information from all potential future actions on the board, unlike $\tilde{\mathcal{V}}$ which only incorporates local information about an action-state pair $(p, b)$.

\begin{algorithm}[h!]
\caption{Expert Gamer: $\Backpropagate(n, root)$}\label{alg:ae-backpropagate}
\begin{algorithmic}
\If{$n.player\_type=root.player\_type$}
    \State $n.val \gets \max_{c\in n.children}c.val$
\Else
    \State $n.val\gets\min_{c\in n.children}c.val$
\EndIf
\end{algorithmic}
\end{algorithm}

\subsection{Monte Carlo Tree Search}
Similar to the ``Expert Gamer'' model, Monte Carlo Tree Search (MCTS) repeats four sub-procedures: SelectNode, ExpandNode, DepthCharge and Backpropagate, but the SelectNode, ExpandNode, and Backpropagate sub-procedures are different from their corresponding counterparts in the human expert model. We run MCTS for $10,000$ steps, which we find balances a close approximation of the optimal for many games while keeping the runtime of any single game trial to less than two days.

\begin{algorithm}[h!]
\caption{MCTS: MakeMove}\label{alg:mcts}
\begin{algorithmic}
\State $root \gets \Node(s)$
\For{$i=1$ to $num\_steps$}
    \State $n\gets\SelectNode(root)$
    \State $\ExpandNode(n)$
    \State $game\_outcome = \DepthCharge(n, root)$
    \State $\Backpropagate(n, root, game\_outcome)$
\EndFor
\State \Return $\argmax_{c\in root.children}c.val$
\end{algorithmic}
\end{algorithm}

\begin{algorithm}[h!]
\caption{MCTS: $\SelectNode(n)$}\label{alg:mcts-select-node}
\begin{algorithmic}
\While{$n.children\neq\emptyset$}
    \State $n=\argmax_{c\in n.children} c.UCB$
\EndWhile
\State \Return $n$
\end{algorithmic}
\end{algorithm}
To balance exploring and exploiting in the MCTS SelectNode sub-procedure, we use a standard upper confidence bound (UCB) \citep{browne2012survey}. 
\begin{align*}
c.UCB = &\argmax_{c\in root.children}\left(\frac{c.val}{c.visits}+\sqrt{2\frac{\log(c.parent.visits)}{c.visits}}\right). 
\end{align*}
\begin{algorithm}[h!]
\caption{MCTS: $\ExpandNode(n, root)$}\label{alg:mcts-expand-node}
\begin{algorithmic}
\State $s\gets n.state$
\ForAll{$m\in \LegalMoves(s)$}
    \State $b\gets \PlacePiece(s,m,n.player\_type)$
    \State $i\gets n.move\_number$
    \State $n.children.append(\Node(b,i+1))$
\EndFor
\end{algorithmic}
\end{algorithm}

\begin{algorithm}[h!]
\caption{MCTS: $\Backpropagate(n, root, game\_outcome)$}\label{alg:mcts-backpropagate}
\begin{algorithmic}
\State $n.visits \gets n.visits+1$
\If{n.parent}
    \If{n.parent.player\_type == root.player\_type}
        \State $n.val \gets n.val+game\_outcome$
    \Else
        \State $n.val \gets n.val+(1-game\_outcome)$
    \EndIf
    \State $\Backpropagate(n.parent, root, game\_outcome)$
\EndIf
\end{algorithmic}
\end{algorithm}

\begin{algorithm}[h!]
\caption{MCTS: $\DepthCharge(n)$}\label{alg:mcts-depthcharge}
\begin{algorithmic}
\State $s\gets n.state$
\State $i\gets n.move\_number$
\While{not $\HasTerminated(s)$}
    \State $player\_type\gets n.player\_sequence[i]$
    \State $m\gets \RandMove(s)$
    \State $i\gets i+1$
    \State $s\gets\PlacePiece(s,m,player\_type)$
\EndWhile
\If{$\IsWin(s, root.player\_type;game\_rules)$}
    \State \Return 1
\ElsIf{$\IsDraw(s; game\_rules)$}
    \State \Return 1/2
\Else
    \State \Return 0
\EndIf
\end{algorithmic}
\end{algorithm}

Finally, as noted above, our implementations of the Expert Gamer and MCTS algorithms always parameterize simulation with a formal specification of the rules of any given game (e.g., restrictions on whether a given player can win only in certain directions on the board). We implement $\mathcal{\tilde{V}}$ and DepthCharge as described in the main text so that intermediate value functions and win conditions are calculated with respect to the specific game dynamics and win conditions for each game variant. \textit{MakeMove} similarly takes in the current \textit{move\_number} move index to account for games in which player order is not strictly alternating, such as games in which the first or second player goes twice as their opening move.

\section{Additional model details}

We report additional details on the models, including compute usage, model-model simulations, and other design choices informing the Expert Gamer model\footnote{One of the $121$ games did not complete and save for the infinite board where $10$ in a row is needed to win; we filter out this game from the ``just think'' game fairness comparisons for the Expert Gamer model. Infinite boards are not used in the funness, play, nor watch and predict experiments.}.

\subsection{Computational cost of models}

We assess the relative compute demands of different gameplay models on two measures: (1) the time it takes each model to select an action and (2) how many board states have their value assessed in the process of the model deciding what action to take. We measure the first metric as the average time per move counts the time, in seconds, that each model takes selecting an action of where to move. This quantity is then averaged across all games, game trials, and actions within a game. Table \ref{tab:time_per_move} shows the average wall-clock time per move for the three models broken down by game. We note that wall-clock time is potentially limited by the exact implementation and hardware used. To reduce some of these confounds, all models utilize a shared codebase and are run on the same hardware. We measure the second quantity by counting the number of nodes in the constructed tree search for each move. This quantity is then averaged across game trials and moves within a game trial. Table \ref{tab:nodes_per_move} shows the average number of nodes whose heuristic values are evaluated in the search tree for each move for the Intuitive Gamer, Expert Gamer, and MCTS models. Each row in the table corresponds to one of the $41$ games used in the ``human-human play'' experiment.  All game simulations involve self-play with that same model. 

\begin{table}[h!]
    \centering
    \small
    \begin{tabular}{lccc}
        \toprule
        Game & Intuitive Gamer & Expert Gamer & MCTS \\
        \midrule
        3x3 3 P1/2 P2 & $7.71 \pm 1.09$ & $340.79 \pm 421.19$ & $6,946.28 \pm 7,373.94$ \\
        3x3 3 & $6.72 \pm 2.00$ & $489.21 \pm 477.21$ & $11,769.89 \pm 13,908.48$ \\
        3x3 3 (P2 2p) & $6.41 \pm 2.23$ & $923.50 \pm 1,062.07$ & $12,168.06 \pm 13,930.55$ \\
        3x3 3 (P1 2p) & $6.73 \pm 1.99$ & $152.83 \pm 73.01$ & $13,445.12 \pm 14,038.39$ \\
        3x3 3 L & $6.46 \pm 2.17$ & $380.69 \pm 547.15$ & $13,618.89 \pm 15,778.72$ \\
        2x5 3 & $7.04 \pm 2.43$ & $218.78 \pm 172.61$ & $18,019.79 \pm 18,641.68$ \\
        1x10 3 & $6.93 \pm 2.50$ & $259.58 \pm 219.57$ & $18,860.72 \pm 19,800.13$ \\
        4x4 3 L & $11.63 \pm 3.43$ & $3,618.14 \pm 2,540.70$ & $62,150.06 \pm 44,828.95$ \\
        4x4 3 D & $11.45 \pm 3.70$ & $3,367.23 \pm 2,507.85$ & $64,164.73 \pm 41,121.09$ \\
        4x4 3 HV & $12.87 \pm 2.90$ & $3,171.31 \pm 3,045.33$ & $74,368.59 \pm 41,785.08$ \\
        5x5 2 & $24.02 \pm 0.81$ & $606.18 \pm 470.70$ & $75,410.55 \pm 56,977.89$ \\
        5x5 3 L & $18.23 \pm 5.03$ & $7,547.74 \pm 3,772.27$ & $124,690.63 \pm 71,792.33$ \\
        4x6 5 & $16.41 \pm 5.70$ & $6,569.44 \pm 3,984.28$ & $127,823.20 \pm 62,653.84$ \\
        4x6 4 & $16.75 \pm 5.56$ & $6,310.25 \pm 4,502.73$ & $129,860.14 \pm 64,085.15$ \\
        5x5 4 D & $17.20 \pm 5.83$ & $7,599.20 \pm 4,259.91$ & $134,135.65 \pm 64,384.30$ \\
        5x5 4 HV & $17.93 \pm 5.46$ & $6,464.76 \pm 4,240.71$ & $136,459.28 \pm 64,148.43$ \\
        5x5 4 (P2 2p) & $17.89 \pm 5.60$ & $7,103.19 \pm 5,034.45$ & $136,702.71 \pm 64,837.41$ \\
        5x5 4 & $18.05 \pm 5.36$ & $6,618.10 \pm 4,702.53$ & $144,846.18 \pm 62,310.97$ \\
        5x5 3 (P1 HV) & $22.57 \pm 1.86$ & $6,168.12 \pm 6,553.56$ & $153,006.93 \pm 71,992.53$ \\
        5x5 3 (P1 D) & $22.18 \pm 2.22$ & $6,355.83 \pm 6,316.61$ & $153,930.07 \pm 71,405.38$ \\
        5x5 4 (P1 2p) & $19.14 \pm 4.72$ & $6,984.52 \pm 5,777.07$ & $155,149.51 \pm 57,176.56$ \\
        5x5 3 & $22.54 \pm 1.93$ & $6,167.83 \pm 6,553.92$ & $157,519.28 \pm 71,077.72$ \\
        5x5 4 P1/3 P2 & $21.94 \pm 2.29$ & $7,475.19 \pm 6,392.12$ & $176,543.76 \pm 45,694.49$ \\
        3x10 3 & $27.25 \pm 2.24$ & $7,697.46 \pm 7,894.15$ & $204,352.35 \pm 82,146.78$ \\
        4x9 4 & $26.77 \pm 7.70$ & $13,698.40 \pm 7,066.52$ & $240,662.55 \pm 85,097.34$ \\
        7x7 4 L & $33.72 \pm 11.05$ & $17,258.99 \pm 7,808.52$ & $301,968.11 \pm 119,135.63$ \\
        5x10 5 & $34.18 \pm 11.61$ & $17,794.77 \pm 8,438.64$ & $313,634.43 \pm 120,188.42$ \\
        7x7 4 HV & $38.99 \pm 9.33$ & $17,756.62 \pm 8,252.03$ & $377,773.04 \pm 101,370.85$ \\
        7x7 4 D & $36.39 \pm 10.42$ & $18,122.45 \pm 8,075.41$ & $378,955.54 \pm 111,335.85$ \\
        7x7 4 (P1 D) & $42.87 \pm 4.78$ & $20,734.93 \pm 10,316.27$ & $387,077.75 \pm 89,789.16$ \\
        7x7 4 & $42.49 \pm 5.05$ & $21,326.28 \pm 10,600.34$ & $387,483.47 \pm 110,703.17$ \\
        7x7 4 (P1 HV) & $41.06 \pm 6.16$ & $19,509.22 \pm 11,271.34$ & $387,637.26 \pm 100,578.80$ \\
        7x7 4 (P2 2p) & $42.86 \pm 5.22$ & $20,767.11 \pm 10,223.01$ & $388,333.98 \pm 109,289.19$ \\
        7x7 4 (P1 2p) & $43.00 \pm 4.69$ & $17,274.72 \pm 12,929.50$ & $397,684.27 \pm 92,618.19$ \\
        5x10 4 & $42.84 \pm 5.23$ & $21,163.57 \pm 10,751.05$ & $397,818.92 \pm 109,009.24$ \\
        7x7 4 P1/3 P2 & $45.42 \pm 2.84$ & $16,165.24 \pm 13,332.40$ & $399,809.50 \pm 91,272.35$ \\
        10x10 5 (P1 D) & $82.90 \pm 13.98$ & $44,812.99 \pm 16,198.57$ & $865,140.51 \pm 103,573.52$ \\
        10x10 3 & $97.43 \pm 1.99$ & $33,495.23 \pm 25,525.04$ & $872,437.53 \pm 177,821.03$ \\
        10x10 5 (P1 HV) & $77.77 \pm 17.70$ & $44,346.05 \pm 16,328.17$ & $874,752.68 \pm 111,721.08$ \\
        10x10 5 & $83.64 \pm 15.43$ & $48,697.30 \pm 16,778.13$ & $885,028.95 \pm 117,608.47$ \\
        10x10 4 & $92.99 \pm 6.52$ & $43,112.32 \pm 24,438.41$ & $897,954.52 \pm 136,325.31$ \\
        \bottomrule
    \end{tabular}
    \caption{\textbf{Efficiency, measured by number of states evaluated, across game reasoning modules.} Average game board states explored per move across all game configurations. Values represent mean $\pm$ standard deviation.}
    \label{tab:nodes_per_move}
\end{table}

\begin{table}[h!]
    \centering
    \small
    \begin{tabular}{lccc}
        \toprule
      Game & Intuitive Gamer & Expert Gamer & MCTS \\
        \midrule
        3x3 3 P1/2 P2 & $4.11 \times 10^{-4} \pm 1.02 \times 10^{-4}$ & $0.05 \pm 0.02$ & $16.25 \pm 4.00$ \\
        3x3 3 (P1 2p) & $4.10 \times 10^{-4} \pm 1.46 \times 10^{-4}$ & $0.03 \pm 0.01$ & $17.61 \pm 5.01$ \\
        3x3 3 & $4.25 \times 10^{-4} \pm 1.73 \times 10^{-4}$ & $0.08 \pm 0.03$ & $18.85 \pm 6.30$ \\
        3x3 3 (P2 2p) & $4.02 \times 10^{-4} \pm 1.47 \times 10^{-4}$ & $0.09 \pm 0.05$ & $19.00 \pm 6.41$ \\
        3x3 3 L & $4.08 \times 10^{-4} \pm 1.47 \times 10^{-4}$ & $0.08 \pm 0.03$ & $19.08 \pm 6.07$ \\
        2x5 3 & $3.17 \times 10^{-4} \pm 1.36 \times 10^{-4}$ & $0.04 \pm 0.02$ & $19.13 \pm 6.33$ \\
        1x10 3 & $3.38 \times 10^{-4} \pm 1.40 \times 10^{-4}$ & $0.05 \pm 0.02$ & $19.39 \pm 6.11$ \\
        4x4 3 HV & $4.37 \times 10^{-4} \pm 1.19 \times 10^{-4}$ & $0.21 \pm 0.10$ & $38.03 \pm 6.55$ \\
        4x4 3 L & $7.27 \times 10^{-4} \pm 2.67 \times 10^{-4}$ & $0.45 \pm 0.20$ & $38.34 \pm 12.44$ \\
        4x4 3 D & $5.54 \times 10^{-4} \pm 1.79 \times 10^{-4}$ & $0.22 \pm 0.07$ & $38.95 \pm 11.64$ \\
        5x5 2 & $1.85 \times 10^{-3} \pm 4.62 \times 10^{-4}$ & $0.13 \pm 0.03$ & $51.45 \pm 5.11$ \\
        4x6 5 & $1.11 \times 10^{-3} \pm 3.40 \times 10^{-4}$ & $0.39 \pm 0.12$ & $54.35 \pm 16.23$ \\
        5x5 4 HV & $7.20 \times 10^{-4} \pm 2.58 \times 10^{-4}$ & $0.46 \pm 0.15$ & $68.63 \pm 19.58$ \\
        5x5 4 D & $9.78 \times 10^{-4} \pm 3.28 \times 10^{-4}$ & $0.83 \pm 0.27$ & $70.84 \pm 21.44$ \\
        4x6 4 & $1.17 \times 10^{-3} \pm 3.44 \times 10^{-4}$ & $0.61 \pm 0.20$ & $71.58 \pm 21.57$ \\
        5x5 3 L & $1.58 \times 10^{-3} \pm 4.87 \times 10^{-4}$ & $1.30 \pm 0.55$ & $72.12 \pm 24.70$ \\
        5x5 4 & $1.55 \times 10^{-3} \pm 4.84 \times 10^{-4}$ & $0.84 \pm 0.36$ & $76.41 \pm 21.90$ \\
        5x5 4 (P2 2p) & $1.43 \times 10^{-3} \pm 4.42 \times 10^{-4}$ & $0.82 \pm 0.25$ & $77.91 \pm 22.59$ \\
        5x5 3 (P1 HV) & $1.18 \times 10^{-3} \pm 2.00 \times 10^{-4}$ & $0.60 \pm 0.18$ & $78.96 \pm 12.05$ \\
        5x5 3 (P1 D) & $1.51 \times 10^{-3} \pm 3.69 \times 10^{-4}$ & $0.71 \pm 0.22$ & $79.25 \pm 15.16$ \\
        5x5 3 & $1.81 \times 10^{-3} \pm 4.57 \times 10^{-4}$ & $0.83 \pm 0.27$ & $80.16 \pm 10.07$ \\
        5x5 4 (P1 2p) & $1.45 \times 10^{-3} \pm 3.56 \times 10^{-4}$ & $0.74 \pm 0.28$ & $85.30 \pm 19.61$ \\
        3x10 3 & $1.91 \times 10^{-3} \pm 4.44 \times 10^{-4}$ & $0.87 \pm 0.22$ & $89.06 \pm 13.60$ \\
        5x5 4 P1/3 P2 & $1.73 \times 10^{-3} \pm 4.55 \times 10^{-4}$ & $0.84 \pm 0.22$ & $89.83 \pm 13.46$ \\
        4x9 4 & $2.30 \times 10^{-3} \pm 6.36 \times 10^{-4}$ & $1.59 \pm 0.33$ & $129.93 \pm 35.15$ \\
        7x7 4 HV & $2.37 \times 10^{-3} \pm 6.76 \times 10^{-4}$ & $2.02 \pm 0.46$ & $189.08 \pm 43.82$ \\
        5x10 5 & $3.81 \times 10^{-3} \pm 1.20 \times 10^{-3}$ & $2.73 \pm 0.70$ & $211.20 \pm 68.08$ \\
        7x7 4 L & $4.58 \times 10^{-3} \pm 1.42 \times 10^{-3}$ & $4.41 \pm 1.25$ & $231.47 \pm 75.59$ \\
        7x7 4 (P1 HV) & $4.02 \times 10^{-3} \pm 9.90 \times 10^{-4}$ & $3.00 \pm 0.58$ & $235.27 \pm 46.61$ \\
        7x7 4 P1/3 P2 & $6.14 \times 10^{-3} \pm 1.62 \times 10^{-3}$ & $3.67 \pm 0.65$ & $239.62 \pm 40.60$ \\
        7x7 4 D & $3.24 \times 10^{-3} \pm 9.26 \times 10^{-4}$ & $3.56 \pm 0.99$ & $243.65 \pm 47.14$ \\
        7x7 4 (P1 D) & $4.75 \times 10^{-3} \pm 1.11 \times 10^{-3}$ & $3.68 \pm 0.58$ & $261.60 \pm 49.34$ \\
        7x7 4 & $6.53 \times 10^{-3} \pm 2.22 \times 10^{-3}$ & $4.30 \pm 0.56$ & $261.65 \pm 43.19$ \\
        5x10 4 & $5.74 \times 10^{-3} \pm 1.44 \times 10^{-3}$ & $3.98 \pm 0.81$ & $266.12 \pm 35.31$ \\
        7x7 4 (P2 2p) & $6.45 \times 10^{-3} \pm 1.57 \times 10^{-3}$ & $4.68 \pm 0.92$ & $270.42 \pm 41.26$ \\
        7x7 4 (P1 2p) & $6.29 \times 10^{-3} \pm 1.87 \times 10^{-3}$ & $3.65 \pm 0.90$ & $291.32 \pm 46.20$ \\
        10x10 3 & $2.78 \times 10^{-2} \pm 6.36 \times 10^{-3}$ & $17.34 \pm 2.94$ & $862.42 \pm 66.75$ \\
        10x10 5 (P1 HV) & $1.49 \times 10^{-2} \pm 3.54 \times 10^{-3}$ & $15.06 \pm 2.14$ & $953.86 \pm 263.76$ \\
        10x10 5 (P1 D) & $1.97 \times 10^{-2} \pm 4.86 \times 10^{-3}$ & $17.58 \pm 2.61$ & $987.14 \pm 265.40$ \\
        10x10 4 & $2.50 \times 10^{-2} \pm 5.11 \times 10^{-3}$ & $20.28 \pm 4.42$ & $1015.55 \pm 124.69$ \\
        10x10 5 & $2.13 \times 10^{-2} \pm 3.28 \times 10^{-3}$ & $21.28 \pm 2.43$ & $1151.12 \pm 169.73$ \\
        \bottomrule
    \end{tabular}
    \caption{\textbf{Efficiency, measured by wall-clock time per action, across game reasoning modules.} Average time per move (seconds) across all game configurations. Values represent mean $\pm$ standard deviation. The values are sorted by the average time per move of the MCTS model.}
    \label{tab:time_per_move}
\end{table}

\begin{table}[h!]
    \centering
    \begin{tabular}{c|c|c|c}
        Matchup & Wins (\%) & Losses (\%) & Draws (\%) \\
        \hline
        Expert vs Ablated Expert & 43.0 & 16.4 & 40.5 \\
        Ablated Expert vs Expert & 19.3 & 41.4 & 39.3 \\
    \end{tabular}
    \caption{\textbf{Comparing the Expert Gamer with inheritance in the value function against a lesioned variant.} Game outcome percentages for Expert Gamer versus an Ablated Expert Gamer, which does not incorporate the entire global state matchups in is computation of the heuristic value of each state. Results are based on $40$ match simulations over the subset of $41$ games used in the play experiment, showing win, loss, and draw rates of Player 1 averaged over all simulations and games. Rows are formatted as Player 1 model vs. Player 2 model.}
    \label{tab:expert_ablated_results}
\end{table}

\begin{table}[h!]
    \centering
    \begin{tabular}{c|c}
        $\epsilon$ & Expected payoff for the Expert Gamer \\
        \hline
        0.0 & $0.0831 \pm 0.863$ \\
        1e-4 & $0.0845 \pm 0.861$ \\
        1e-3 & $0.0683 \pm 0.866$ \\
        1e-2 & $0.0651 \pm 0.865$ \\
        0.1 & $0.0328 \pm 0.879$ \\
        0.2 & $-0.00396 \pm 0.887$ \\
    \end{tabular}
    \caption{\textbf{Selection of $\epsilon$ in the Expert Gamer based on payoff in cross-model gameplay.} Expected payoff and standard deviation for the Expert Gamer model, averaged across opponent and whether the expert played first or second.}
    \label{tab:expvnovicedepth}
\end{table}

\subsection{Heuristic quality of Expert Gamer model}
The Expert Gamer model is more sophisticated than the Intuitive Gamer across two axes---search depth and heuristic sophistication. While it is easy to verify that the Expert Gamer model has a deeper search depth than the Intuitive Gamer, it is difficult to compare the heuristic sophistication of the two models without knowing the objective value of each board state of each game. To this end, we simulate game trials of the Expert Gamer model against a faithful implementation of the Intuitive Gamer model that incorporates search depth. We observe that the Expert Gamer model dominates regardless of whether or not it plays as Player 1 or 2. Our results are reported in Table~\ref{tab:expert_ablated_results}. The Ablated Expert removes the mechanism that incorporates features of the global board state instead of the purely local features. In particular, the value function of the ablated expert differs from the value function of our expert model as given by Equation~\ref{eq:expert-val}. The value function of the ablated expert is given by:

\begin{equation}
    \label{eq:ablated-expert}
    \begin{split}
        \mathcal{V}^\textrm{EG, abl}(s')=(2\times \mathbb{I} \{\textrm{the }i\textrm{'th move is }X\}-1)\mathcal{V}(a,s),
    \end{split}
\end{equation}
where $s$ is the board state that immediately preceded $s'$ on the game tree and $a$ is the action that takes board state $a$ to state $s'$. The value function of the ablated expert is clearly very myopic as it depends only on the most recent action of the board state.

We incorporate an exploration component to our Expert Gamer model by setting a probability $\epsilon$ with which we explore a uniformly random child node. We have the Expert Gamer model play against the Intuitive Gamer (flat and depth-$3$ variants, as described at the end of the Supplement) on the subset of 41 games used in the play experiment with different values of $\epsilon'$ and set $\epsilon$ to be the $\epsilon'$ that yields the greatest expected advantage over the two models. Results are reported in Table~\ref{tab:expvnovicedepth}, which shows the expected payoff averaged across the expert's opponent. Choice of $\epsilon$ has a minimal impact (Table~\ref{tab:expvnovicedepth}); still, we take the maximum shows, this value turns out to be $\epsilon=0.0001$.

\subsection{Intuitive Gamer parameters}

\begin{figure}[h!]
\includegraphics[width=1.0\linewidth]{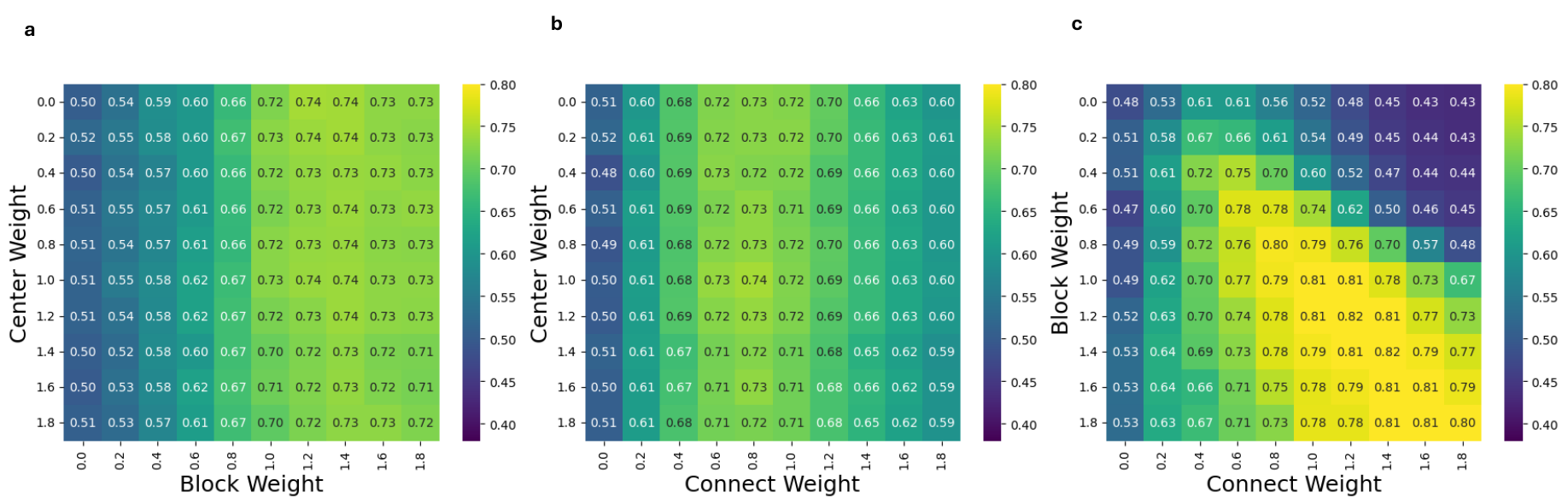}
    \caption{\textbf{Parameter sensitivity analyses.} Analyzing the impact of the goal progress (offensive ``connect'' component), goal blocking (defensive ``block'' component), and center weight on fit to people's judgments in the ``just think'' experiment, with subset of simulations for different settings of the heuristic value function weights. We run $100$ bootstrapped subsamples of $20$ participants from the ``just think'' experiment per game with $k=6$, and compute $R^2$ between the average human payoff from the subset versus the average model payoff. Each cell averages over the held-out parameter (e.g., averaging over all settings of the connect weight in the leftmost plot). Fit to participants is substantially more impacted by the components controlling offense and defense, rather than center; interestingly, the defensive blocking component generally requires a value as high or more than the corresponding offensive component.} 
    \label{fig:param-sweeps-think}
\end{figure}

We additionally assess the sensitivity of the Intuitive Gamer model to choice of parameters ($w$) for the heuristic value function ($\tilde{\mathcal{V}}$). We fix the temperature $\tau = 1$ and vary $w_\text{connect}$, $w_\text{block}$, and $w_\text{center}$. We sweep over $w_\text{connect}$, $w_\text{block}$, and $w_\text{center}$ between $0$ and $2$ in steps of $0.2$ respectively, and run $50$ simulations of our model for each setting (which we bootstrap subsample from $100$ times, simulating $20$ simulated participants running $k=6$ mental simulations each). We then compute the $R^2$ under each variants' predictions relative to the human predicted payoffs. We observe in Figure~\ref{fig:param-sweeps-think} higher sensitivity to the parameterization of the goal-progress and goal-blocking components, highlight the importance of ``goal-directedness'' in the Intuitive Gamer model and appropriately balancing offense and defense. The sweeps also underscore that our selection of simply weighting all values with $1$ is a reasonable modeling choice. 

\FloatBarrier 

\FloatBarrier 

\section{Example experimental interfaces}

\begin{figure}[h!]
    \centering
    \includegraphics[width=1.0\linewidth]{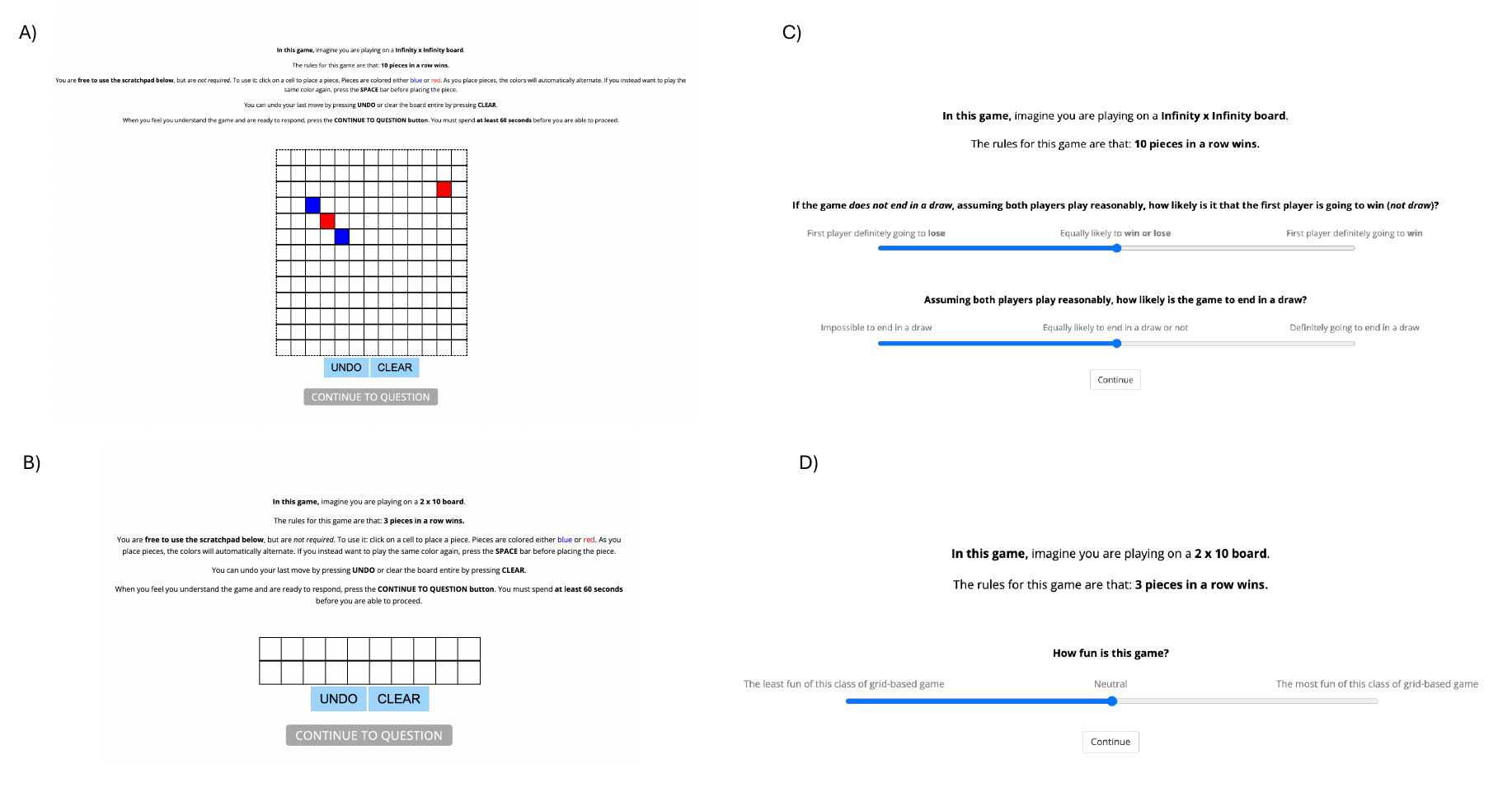}
    \caption{\textbf{``Just think'' evaluating games before any experience example interfaces.} For each game, participants are given a scratchpad which they can optionally use while thinking about their response (\textbf{a-b}). Pieces will automatically change between blue and red upon each click, unless overridden (see Methods). Infinite boards \textbf{(a)} are shown with dashes along the edges. For each game, participants either judge game outcomes \textbf{(c)} or game funness \textbf{(d)} using sliders.}
    \label{fig:just-think-interface}
\end{figure}

\begin{figure}[t!]
    \centering
    \includegraphics[width=1.0\linewidth]{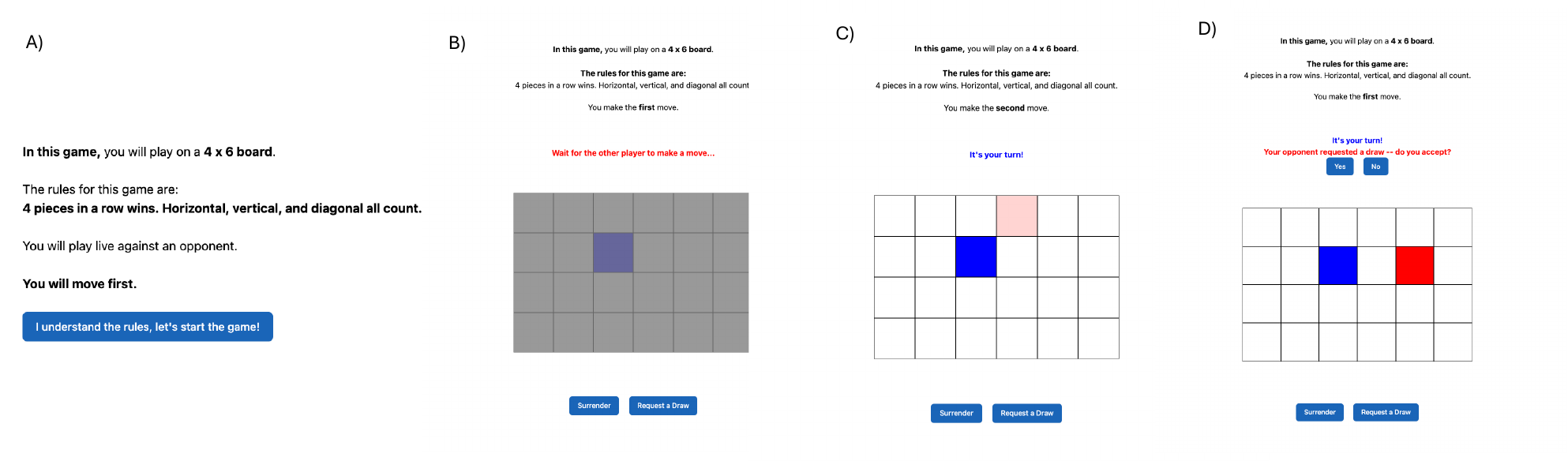}
    \caption{\textbf{Example interfaces from the human-human gameplay study.} \textbf{a,} Participants are first informed of the game rules and whether they will play first or second. \textbf{b,} The board is disabled and greyed-out when it is not their turn. \textbf{c,} When it is their turn, they see a hover (in the color of their piece) before they decide where to play. They decide to play by clicking on the board. \textbf{d,} When a player requests a draw, the other player is informed and allowed to accept or reject the draw.}
    \label{fig:play-interface-play}
\end{figure}

\begin{figure}[t!]
    \centering
    \includegraphics[width=1.0\linewidth]{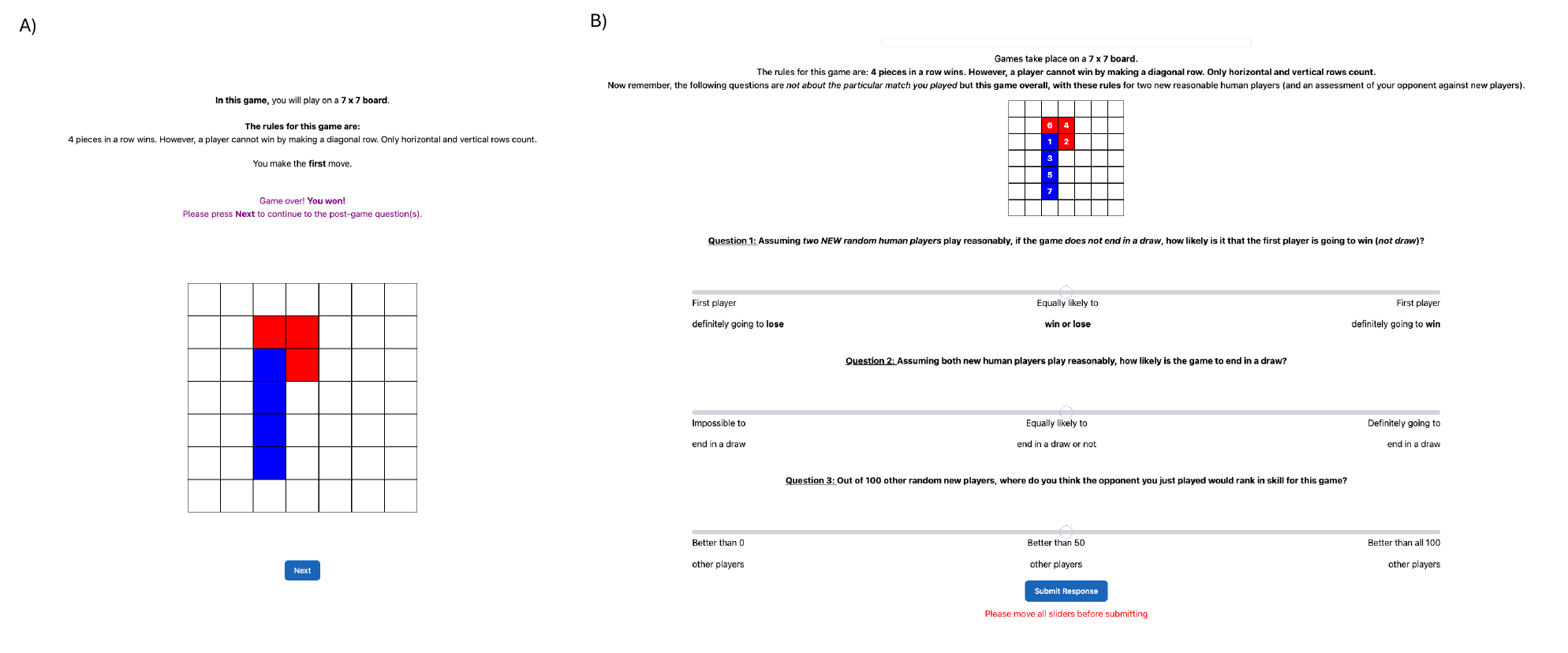}
    \caption{\textbf{Additional example interfaces from the human-human gameplay study.} \textbf{a,} After the game is over, both players are informed of the outcome. \textbf{b,} Each player then makes a series of judgments about the game via sliders and are shown a ``snapshot'' of how their match had unfolded.}
    \label{fig:play-interface-judge}
\end{figure}

\begin{figure}[t!]
    \centering
    \includegraphics[width=1.0\linewidth]{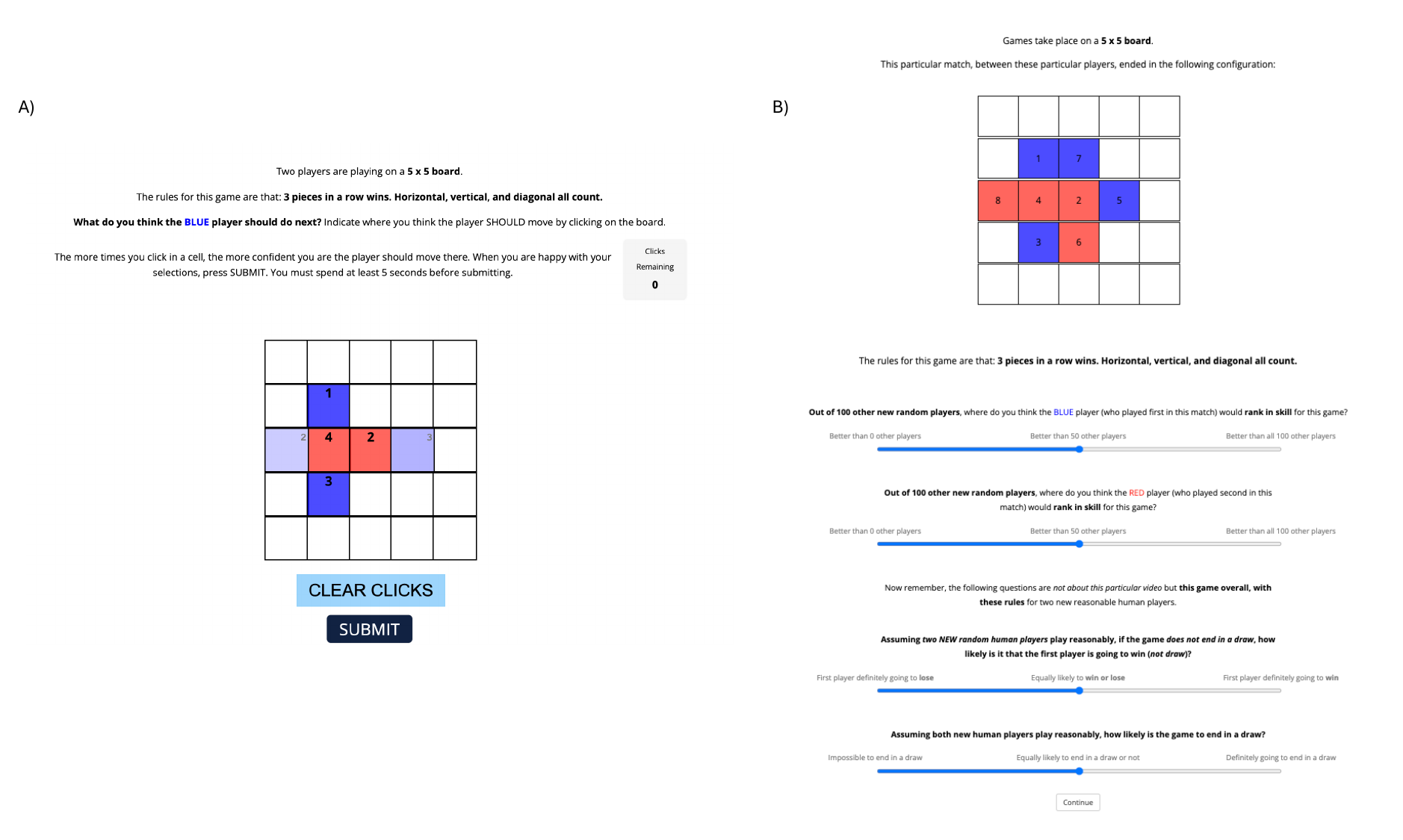}
    \caption{\textbf{Example interfaces from the people watching others play and predicting moves study.} Participants watch gameplay, which is frozen at three timepoints. At each timepoint, participants are tasked with predicting where a player should play by spreading $5$ clicks over the board. \textbf{a,} The cells are colored in proportion to the number of clicks made by the participant, and the number of clicks remaining (out of 5) are shown to the participant. \textbf{b,} At the end of watching a full match, participants are shown a snapshot of the played match, and asked to make a series of judgments about the game via sliders.}
    \label{fig:watch-interface}
\end{figure}

We include example interfaces for each of our main behavioral experiments. Figure~\ref{fig:just-think-interface} depicts example interfaces from the ``just think'' judging game outcome and game funness experiments. Figures ~\ref{fig:play-interface-play} and ~\ref{fig:play-interface-judge} depict example interfaces from the human-human gameplay experiment. Figure~\ref{fig:watch-interface} depicts examples from the indirect human watching and predicting play experiment.

\section{Additional analyses into human and model game evaluation}

\subsection{Participant scratchpad usage}

\begin{figure}
    \centering
    \includegraphics[width=1.0\linewidth]{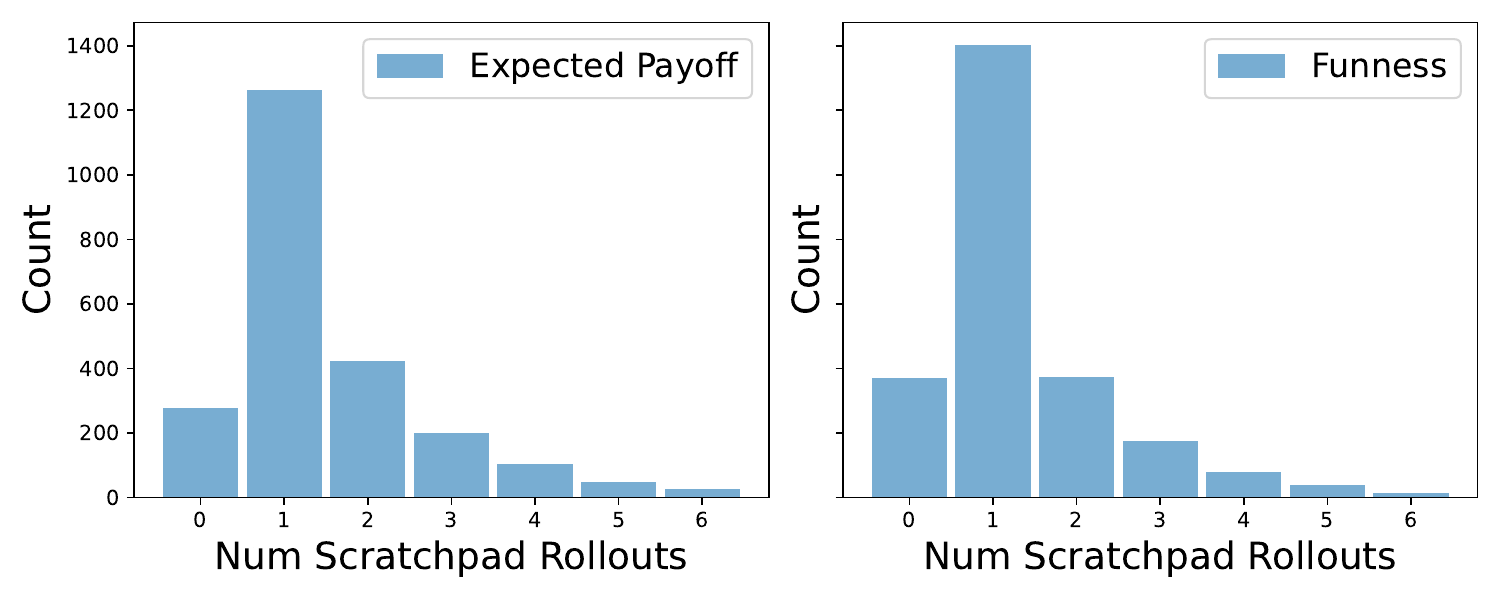}
    \caption{\textbf{Scratchpad usage.} Explicit scratchpad ``rollouts'' per game, per participant. A ``rollout'' is counted as any interaction with the scratchpad, involving at least one click. $2$ rollouts means the participant pressed ``RESET'' once. Most participants engage at least once with any game, both when evaluating \textbf{a,} fairness and \textbf{b} funness; few conduct more than three explicit rollouts on the scratchpad.}
    \label{fig:scratchpad-rollouts}
\end{figure}

Before making their judgments, participants had the option to interact with an interactive version of the board to simulate self-play (a ``scratchpad'') to minimize demands on spatial working memory. Participants made on average $1.60 \pm 1.46$ SD rollouts for the payoff task and $1.36 \pm 1.15$ SD rollouts for the funness task. We define a rollout as the number of times a participant started interacting on a fresh board (i.e., zero rollouts indicates no interaction with the board; one rollout indicates on usage of the board; two indicates one restart of board clicks). We depict a histogram over rollouts per task in Figure~\ref{fig:scratchpad-rollouts}. We do not draw a direct parallel between scratchpad rollouts and mental simulation (it is highly plausible that participants are doing more mental simulation than they are demonstrating on the scratchpad); we include the scratchpad to reduce the potential demands on spatial working memory of our task such that we can focus on studying reasoning. While from manual inspection, many participants using the scratchpads did appear to simulate play, some participants seemed to just click around without intention.

\subsection{Participant experience}

\begin{figure}
    \centering
    \includegraphics[width=0.9\linewidth]{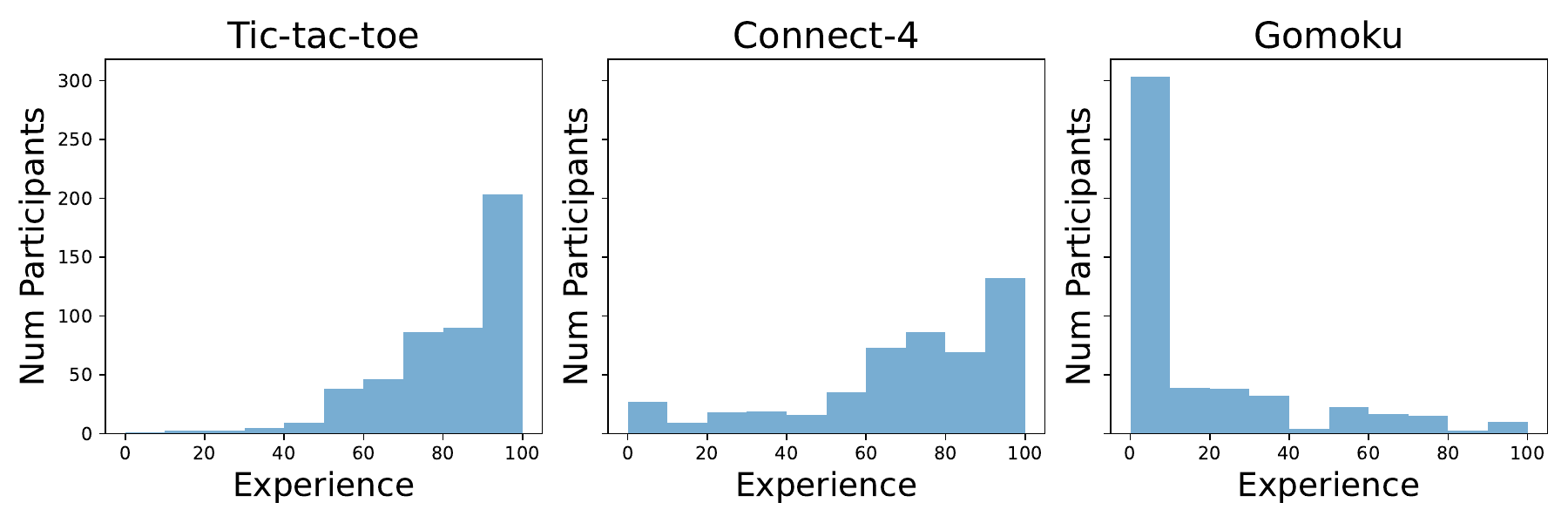}
    \caption{\textbf{Self-reported prior game experience.} Participants in the ``just think'' experiment indicated their prior experience with several existing games, namely, Tic-Tac-Toe, Connect-4, and Gomoku.}
    \label{fig:self-report-experience}
\end{figure}

Participants were also asked at the end of the ``just think'' experiment how much prior experience they had with Tic-Tac-Toe, Connect-4, and Gomoku. Participants self-reported their prior experience on three slider scales; one per game. Sliders ranged from $0$ to $100$, where $0$ = ``No prior experience,'' $50$ = ``Some prior experience playing'', and $100$ = ``Substantial prior experience playing.'' Most participants self-reported experience with Tic-Tac-Toe; Experience with Connect-4 was more variable and few participants had any experience with Gomoku (see Figure~\ref{fig:self-report-experience}). Future work can better investigate the interplay of game-specific experience and pre-play judgments.

\subsection{Simulated game length}

\begin{figure}
    \centering
    \includegraphics[width=0.7\linewidth]{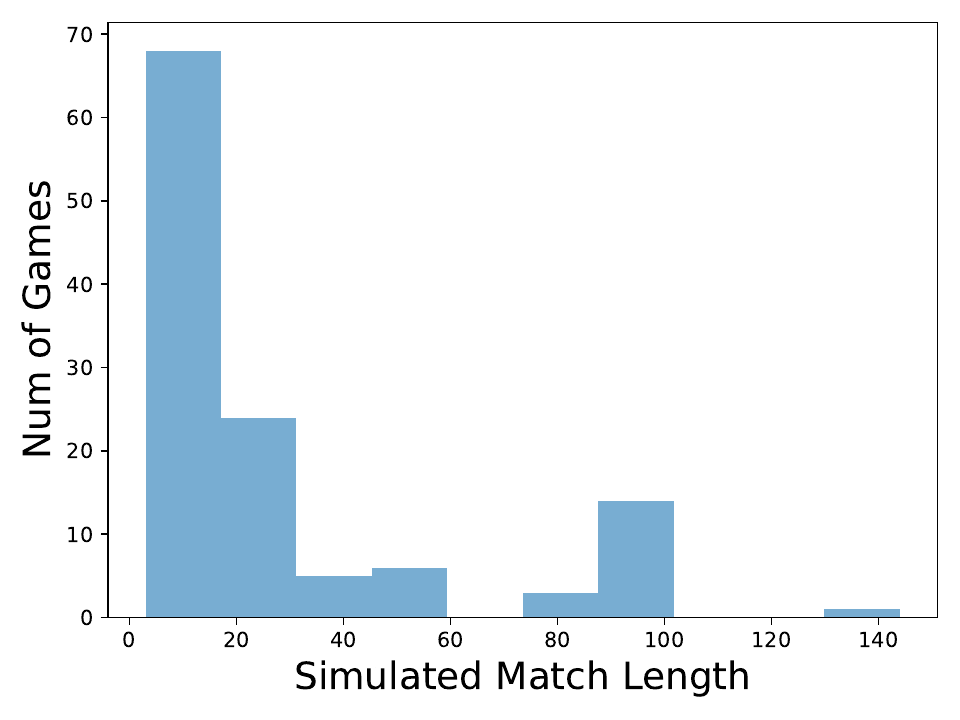}
    \caption{\textbf{Intuitive Gamer simulated match length.} Average simulated match length from the Intuitive Gamer reasoning module, for each of the $121$ games. Simulations were run under the Intuitive Gamer player module.}
    \label{fig:sim-game-len}
\end{figure}

For the primary model, the Intuitive Gamer reasoning module simulates play to the end of the games. The median match length from the $121$ set of games is $14.8$ moves; approximately $75\%$ of the $121$ games ($91$) reach a termination condition (either Player 1 or Player 2 achieves their objective, or the board fills up and ends in a draw) within $30$ moves. Simulated match lengths are shown in Figure~\ref{fig:sim-game-len}. 

\subsection{Decomposed game outcome prediction tasks} 
\label{sec:decomp-scores}
\begin{figure}[t!]
    \centering
    \includegraphics[width=0.9\linewidth]{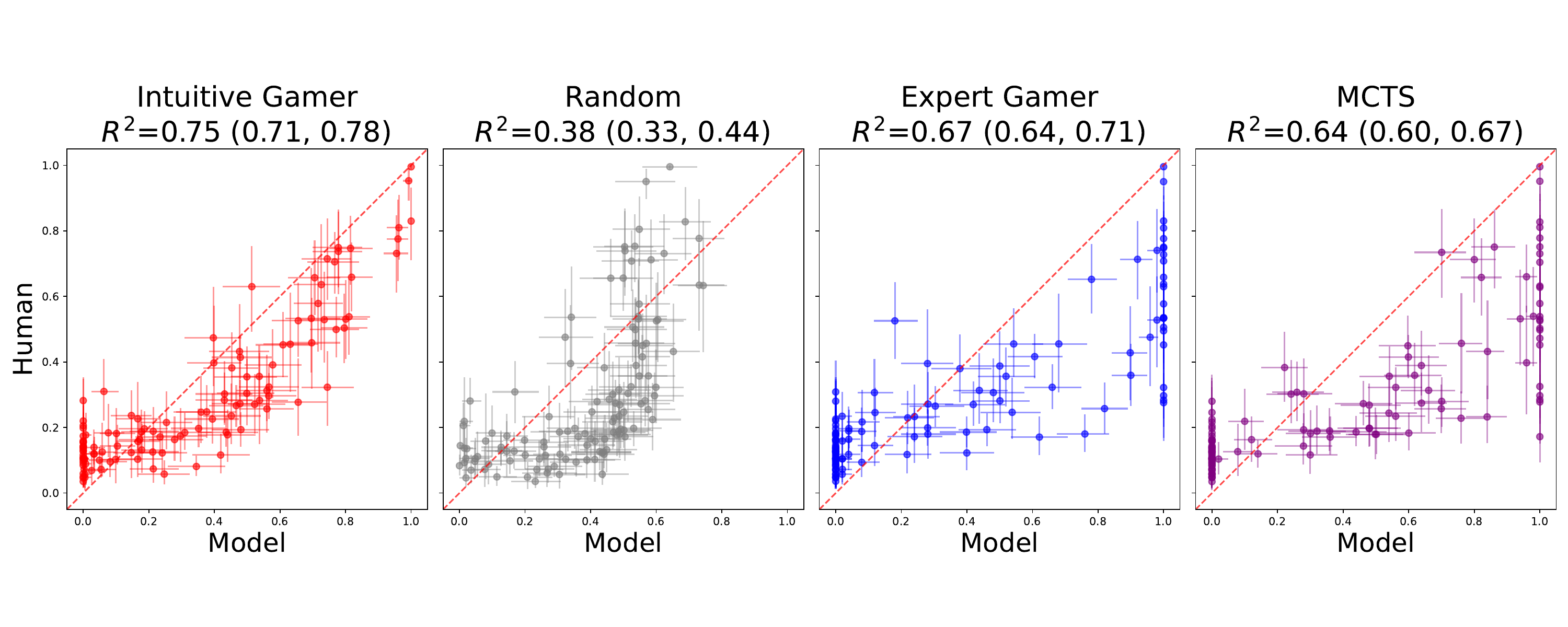}
    \caption{\textbf{Empirical model comparisons for $P(\text{P1 win})$ against people.} Bootstrapped 95\% CIs over participants and samplings of $k=6$ simulations for $N=20$ simulated participants for each model.}
    \label{fig:pwin-models}
\end{figure}

\begin{figure}[t!]
    \centering

    \includegraphics[width=0.9\linewidth]{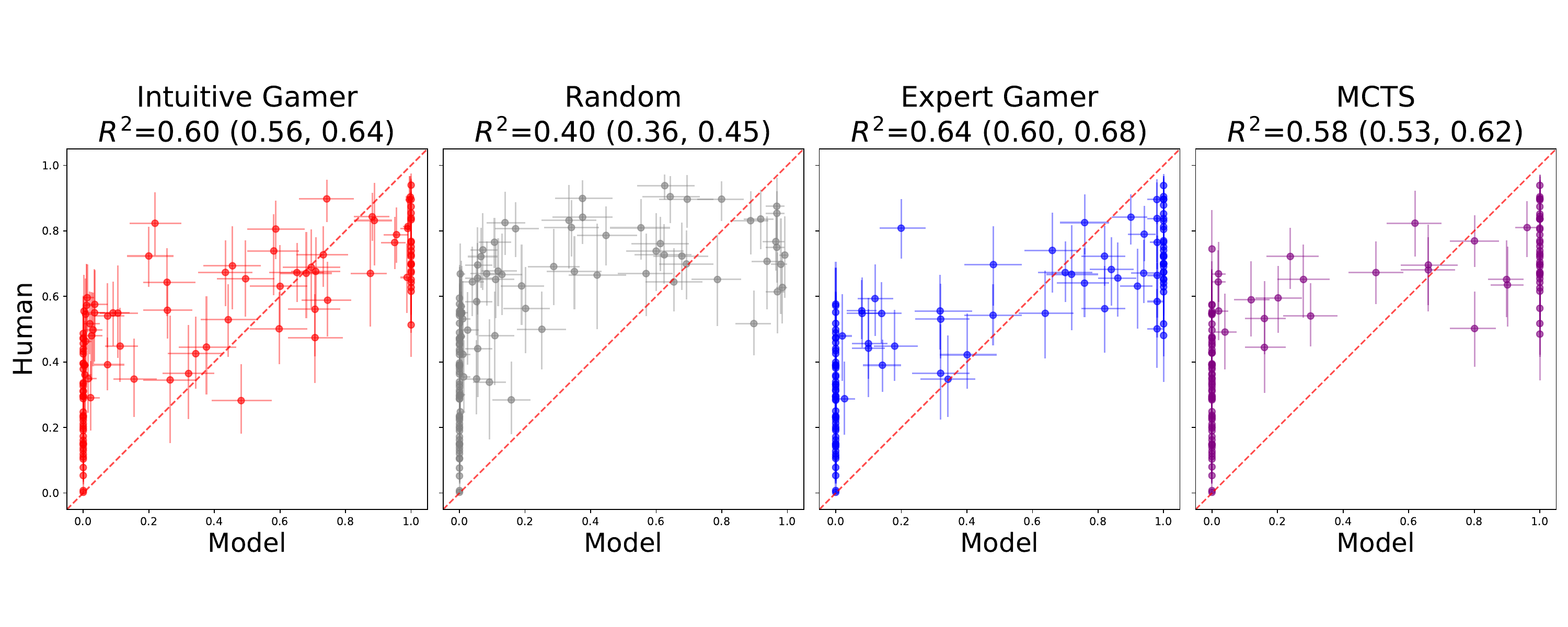}
    \caption{\textbf{Empirical model comparisons for $P(\text{draw})$ against people.} Bootstrapped 95\% CIs over participants and samplings of $k=6$ simulations for $N=20$ simulated participants for each model.} 
    \label{fig:pdraw-models}
\end{figure}

\begin{figure}[t!]
    \centering

    \includegraphics[width=0.9\linewidth]{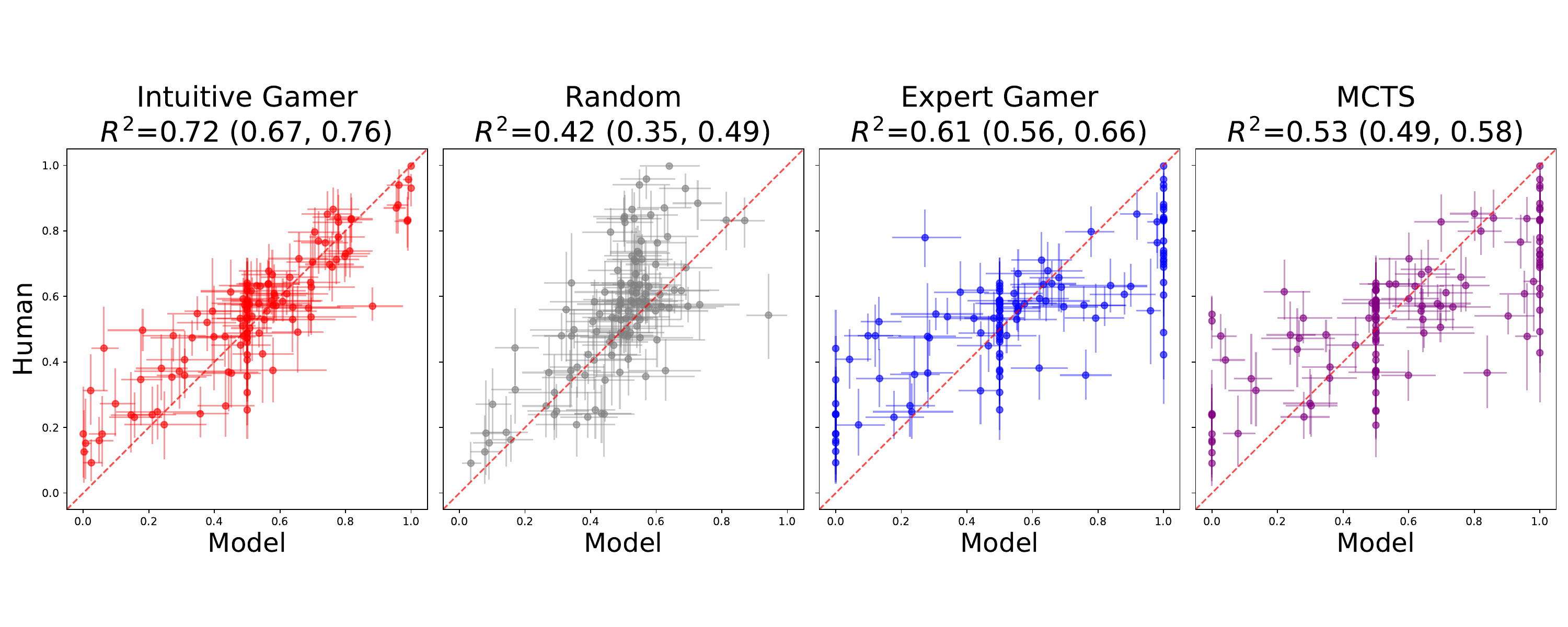}
    \caption{\textbf{Empirical model comparisons for $P(\text{P1 win}|\text{no draw})$ against people.} Games under the model that only have draws are imputed with $0.5$. Bootstrapped 95\% CIs over participants and samplings of $k=6$ simulations for $N=20$ simulated participants for each model.}
    \label{fig:pwin-no-draw-models}
\end{figure}

\begin{figure}[t!]
    \centering
    \includegraphics[width=0.7\linewidth]{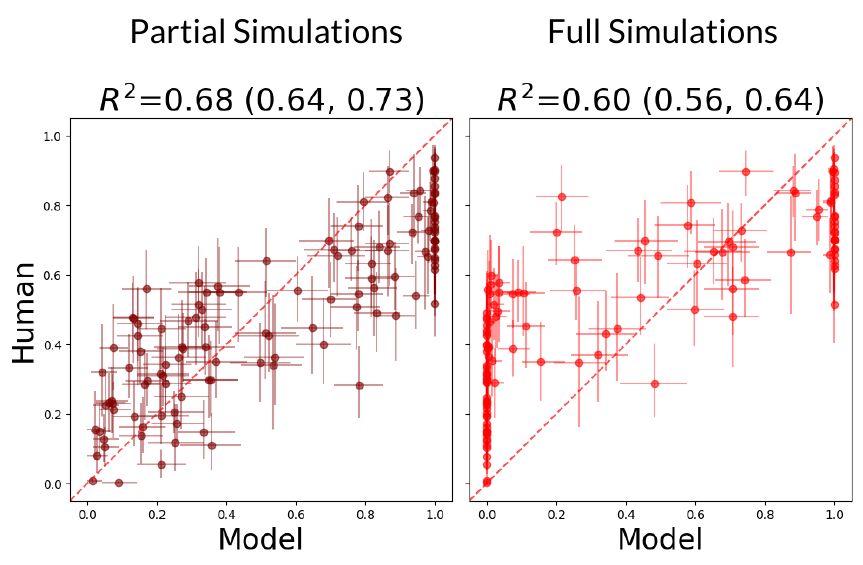}
    \caption{\textbf{Full versus partial simulations to estimate $P(\text{draw})$.} Comparing human and Intuitive Gamer model predicted $P(\text{draw})$ under partial versus full game simulations (both are computed with $k=6$ simulations). Games that end early are deemed draws. Error bars depict 95\% bootstrapped CIs over the average value for each game.}
    \label{fig:early-stop}
\end{figure}

Participants indicated their inferences about likely game outcomes about (1) if the game did not end in a draw, whether the first player was likely to win ($P(\text{P1 win} \mid \text{not draw})$); and (2) how likely the game was to end in a draw ($P(\text{draw})$). We compare the Intuitive Gamer model predictions against human judgments for these decomposed questions in Figure~\ref{fig:pwin-models} and ~\ref{fig:pdraw-models}, respectively. The questions are critical together; e.g., if the participant thinks that the game will definitely end in a draw, the first question is ill-posed. This is accounted for in the payoff, and $P(\text{P1 win})$ when we combine $P(\text{P1 win} \mid \text{not draw}) \times P(\text{draw})$. We report participants' computed $P(\text{P1 win})$ and their reported $P(\text{draw})$ in Figures~\ref{fig:pwin-models} and ~\ref{fig:pdraw-models}. We extract readouts of the corresponding questions to the the Intuitive Gamer model by counting the number of outcomes (in each bootstrapped set of $k=6$ samples) that correspond to each query (e.g., the proportion of draws; or proportion of first player wins) and compare against alternate models using the same empirical outcome frequency in Figures~\ref{fig:pwin-models} and ~\ref{fig:pdraw-models}, respectively. For cases where a game did not have any non-draw simulations, we take $P(\text{P1 win} \mid \text{not draw}) = 0.5$ (which is generally what we notice participants do; see Figure~\ref{fig:pwin-no-draw-models}). In general, the Intuitive Gamer approaches the human split-half $R^2$ for the measures involving $P(\text{win})$ (i.e., split-half R$^2$ for $P(\text{P1 win}) = 0.82$ [$95\%\text{CI}: 0.77, 0.87]$; $P(\text{P1 win} \mid \text{not draw})  = 0.78$ [$95\%\text{CI}: 0.72, 0.83]$). However, the Intuitive Gamer (and alternate models) are all substantially sharper in $P(\text{draw})$ predictions than people and not near the noise ceiling (split-half human $R^2$ for $P(\text{draw}) = 0.78$ [$95\%\text{CI}: 0.72, 0.83]$). It is possible that people compute over a small sample of simulated games in a different way than counting outcomes, which may better correspond to the readouts they provided, or generally have a prior to believe that games are more likely to end in a draw. Our early explorations into the impact over the game reasoning module running only partial simulations, where simulations that stop early are taken to be a draw, may encode such a prior and improves fit to human draw judgments (Figure~\ref{fig:early-stop}). However, we leave a close analysis of partial versus full simulations for future work, as it requires a deeper cross-comparison with alternate models and other stopping rules (which added substantial complexity to this first exploration of an Intuitive Gamer model).

\subsection{Predicting payoff from non-simulation based linguistic features}

\begin{figure}
    \centering
    \includegraphics[width=1.0\linewidth]{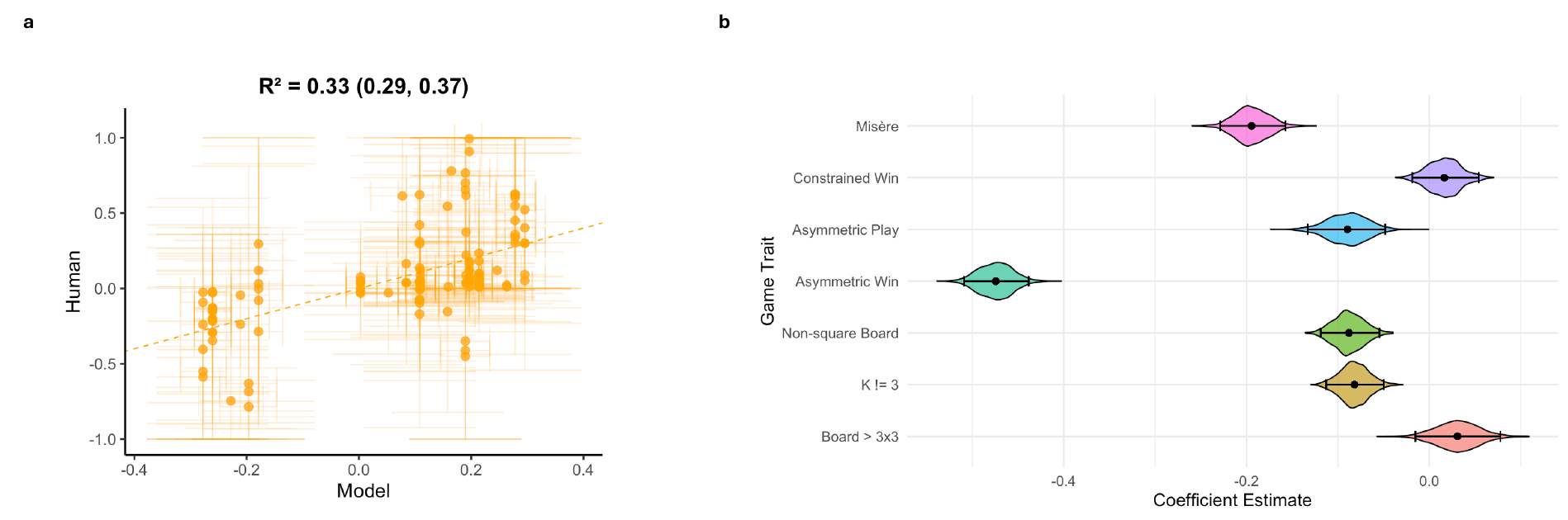}
    \caption{\textbf{Predicting payoff from non-simulation game traits.} Fitting a linear regression model to binary game traits captures some of the variance of payoff judgments. Each point in \textbf{(a)} is a game. The error bars are the 95\% CIs over the human bootstrapped mean payoff and model-predicted mean payoff fit over bootstrapped subsets of 70/30\% of games. \textbf{(b)} 95\% CIs around the parameter fits per game trait.}
    \label{fig:nonsim-payoff-model}
\end{figure}

A key component of our hypothesis is that people assess new problems by drawing on fast probabilistic mental simulations. This hypothesis demands a comparison then against non-simulation-based alternate models. To assess the role of non-simulation-based features, we fit a linear regression model to the binary game traits (as introduced in the Methods). We fit to 70\% of the games and test on the held-out 30\%. We find that the model can capture some variance in human judgments $R^2 = 0.33$ [95\% CI: $0.29, 0.37$] (see Figure~\ref{fig:nonsim-payoff-model}) but comparatively less than models based on explicit simulation, as noted in the main text.

\subsection{Language model payoff and game-theoretic optimal comparisons}

\begin{figure}
    \centering
    \includegraphics[width=1.0\linewidth]{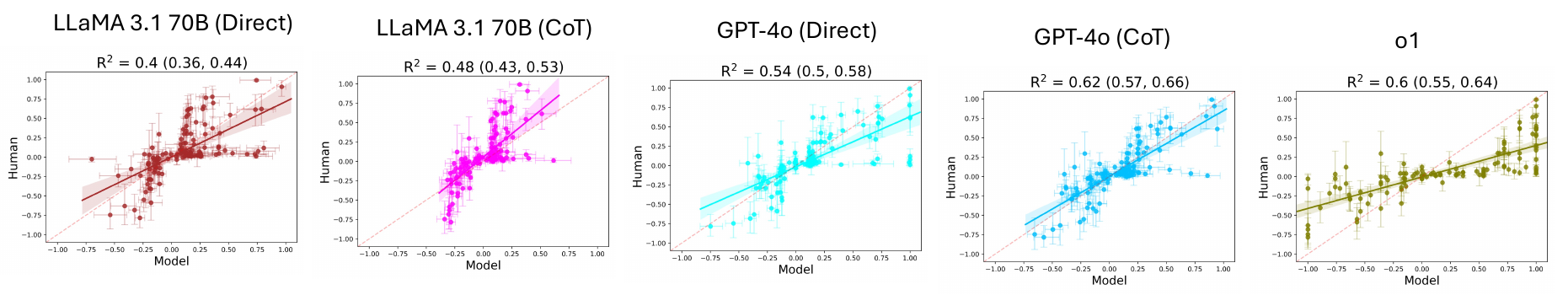}
    \caption{\textbf{Comparing payoff predictions against language models.} Correlations between bootstrapped samples of the human versus language (top) and ``reasoning'' (bottom) model predicted payoffs for the $121$ games. 95\% CIs are over $1000$ bootstrapped subsamples.}
    \label{fig:lm-payoffs}
\end{figure}

We also compare against a series of language models---varying in prompting type (directly asking about game outcomes, or permitting ``chain of thought'' reasoning)---as well as a state-of-the-art reasoning model, o1~\citep{openai2024openaio1card}. We compare predictions under the language models against human predicted payoffs Figure~\ref{fig:lm-payoffs} and against the ``optimal'' game theoretic expected payoffs in Main Text Table~\ref{tab:game_resourcerational_reasoning}b. While o1 is closer to human judgments and a closer fit to the game-theoretic optimal values, it is still a ways off of ``perfect'' highlighting that the games we consider here are non-trivial to reason about.

\subsubsection{Language model prompting}

All language and reasoning models are prompted with a variant of the instructions provided in the human experiment. We prompt models to provide (in a single response) with our two game outcome reasoning questions---how likely the first player will win if the game does not end in a draw, and how likely the game will end in a draw---directly following the questions asked of participants. We provide the models with the same $0-100$ slider labels as in the human study and ask the model to provide a value in the same range per question. We sample $20$ rollouts for each model. We sample all models at the default temperature of $0.7$, except o1, which we sample using its default fixed $1.0$ temperature. 

\subsection{Reasoning about game funness} 

We next provide additional analyses into the funness model and features that matter for capturing the variance in the human data.

\subsubsection{Component ablations}

To assess the impact of the relative contribution of each funness model component, we lesion each component. Analysis of Variance (ANOVA) and Akaike information criterion (AIC) assessments highlight that each component contributes to the fit to human data (see Extended Data in the main text). We also assess the relative benefit for fitting each component with a non-linear spline in Table~\ref{tab:lin-quad-fun}. We see a relative benefit only for the game length component. The balance feature collapses back to a single degree of freedom (linear fit).

\begin{table}[!h]
\centering
\begin{tabular}[t]{lccc}
\toprule
Feature & F & p & $\Delta$AIC\\
\midrule
Balance & 1.242 & 2.67e-01 & -0.7\\
Reward for Thinking & 1.082 & 3.00e-01 & -0.9\\
Game Length & 14.027 & 2.85e-04 & 11.8\\
\bottomrule
\end{tabular}
\caption{\textbf{Impact of quadratic features in funness model.} Comparing the inclusion of quadratic features on each of the simulation-based features in the funness model reveals that only a quadratic term for the game length feature significantly impacts fit to human judgments. $\Delta$AIC is AIC$_\text{lin}$ - AIC$_\text{nonlin}$ (higher indicates better fit from non-linear).}
\label{tab:lin-quad-fun}
\end{table}

\subsubsection{Linguistic features} 
\begin{figure}[t!]
    \centering
    \includegraphics[width=1.0\linewidth]{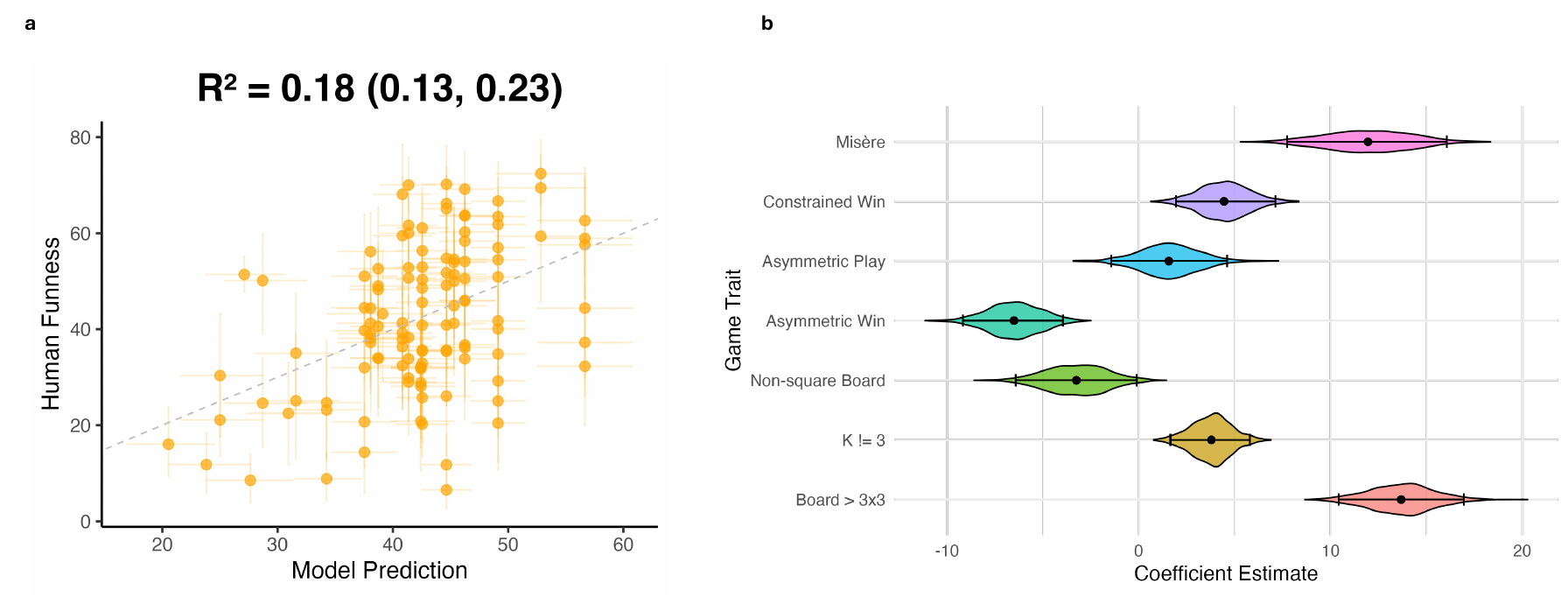}
    \caption{\textbf{Predicting fun from non-simulation game traits.} \textbf{a,} Modeling human funness judgments from non-simulation based binary game traits alone captures some of the variance in human judgments, however substantially less than the simulation-based model. \textbf{b,} Bootstrapped 95\% CIs around the parameter fits from the funness model for each game trait.}
    \label{fig:binary-trait-fun}
\end{figure}

We additionally explore the role of linguistic non-simulation based features, as outlined in the Methods. Fitting a model to the binary traits alone captures a moderate amount of the variance in human funness judgments (Figure ~\ref{fig:binary-trait-fun}a) but comparatively less than the simulation-based models. Some of this may arise from relative correlations between the traits and the simulation-based features (Figure ~\ref{fig:binary-trait-fun}b), e.g., boards that have asymmetric win conditions are generally associated with unfair games (which could connect to the balance component). Adding in the binary traits to the simulation-based funness model does not parsimoniously improve fit to human data. We find minimal evidence for an effect from ``novelty,'' computed as the number of binary traits (or deviations from Tic-Tac-Toe) that are ``active.'' Adding in a linear component does induce a slightly better fit according to ANOVA and AIC tests (Table~\ref{tab:addtl-features-fun}), but has a negligible on our generalization tests (Table~\ref{tab:gen-test-addtl})---that is, held-out fits when fitting to bootstrapped subsamples of 50\% of the games. It is possible that some participants are primarily accounting for linguistic-only (non-simulation based) features when assessing funness; recall, we do see comparatively higher variability in participants' judgments for funness evaluations than payoff predictions. However, at the aggregate population level, accounting for simulation-based features dominates fits. We leave such participant-level modeling of funness for future work.

\begin{figure}[t!]
    \centering
    \includegraphics[width=1.0\linewidth]{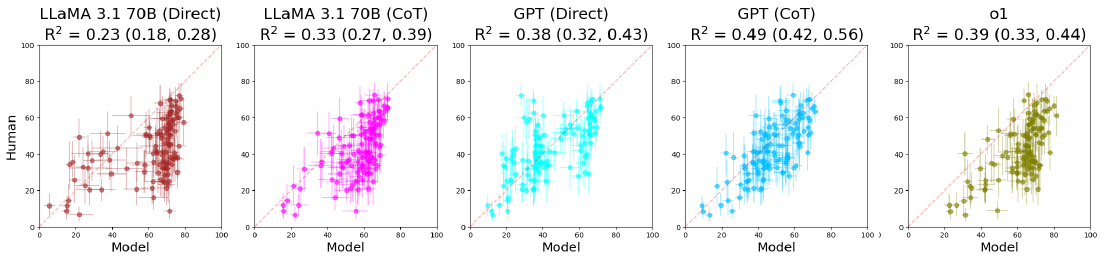}
    \caption{\textbf{Language- and reasoning-model predicted funness per game, compared to people.} Each point is one game. Error bars are bootstrapped 95\% CI over the average human- and model-predicted funness.}
    \label{fig:llm-fun}
\end{figure}

We also compare language- and reasoning-model predicted funness compared against human predicted funness in Figure~\ref{fig:llm-fun}. Models are prompted in the same way as in the payoff predictions described above; that is: using the default temperature of $0.7$ and sampling $20$ rollouts, under a slightly modified version of the full experiment instructions given to people, with the exception of o1 (sampled at its default $1.0$). Most language models yield fits to human data that are comparably worse than the simulation-based features, with some exception to o1. We do not know what kind of simulation, if any, o1 is doing over these games; hence, the model at present does not support the kind of explanatory analyses we are interested in here. However, the comparisons point to potential value in using the human data collected here for benchmarking human-likeness of AI.

\begin{table}[!h]
\centering

\centering
\begin{tabular}[t]{lccc}
\toprule
Added Feature & F & p & $\Delta$AIC\\
\midrule
Board Size & 0.340 & 5.61e-01 & -1.6\\
Approx Novelty & 4.765 & 3.11e-02 & 2.9\\
Binary Traits & 1.224 & 2.96e-01 & -4.8\\
\bottomrule
\end{tabular}
\caption{\textbf{Additions of linguistic non-simulation features to the funness model.} Comparing the inclusion of linguistic non-simulation based features to the funness model reveals only a weak potential effect of incorporating non-simulation based features into the model. $\Delta$AIC is AIC$_\text{sim-only}$ - AIC$_\text{expanded}$ (higher indicates better from including the additional feature).}
\label{tab:addtl-features-fun}
\end{table}

\begin{table}[ht]
\centering
\begin{tabular}{ll}
  \hline
  Variant & Held-Out $R^2$\ \\ 
  \hline
 Simulation Only & 0.62 (0.50, 0.71) \\ 
+ Linear Novelty & 0.62 (0.53, 0.70) \\ 
 + Quadratic Novelty & 0.61 (0.50, 0.69) \\ 
   \hline
\end{tabular}
\caption{\textbf{Generalization test when incorporating non-simulation based features into funness model.} Comparing held-out fits (fit on 50\% of games, test on 50\% of games) when including a ``novelty'' (linear or non-linear feature) ontop of the base regression model featuring only simulation-based features (balance; reward for thinking; quadratic length). This reveals that the additional linguistic feature does not meaningfully impact generalization.
} 
\label{tab:gen-test-addtl}
\end{table}

\section{Additional analyses into human and model action selection and prediction}

\begin{figure}
    \centering
    \includegraphics[width=0.9\linewidth]{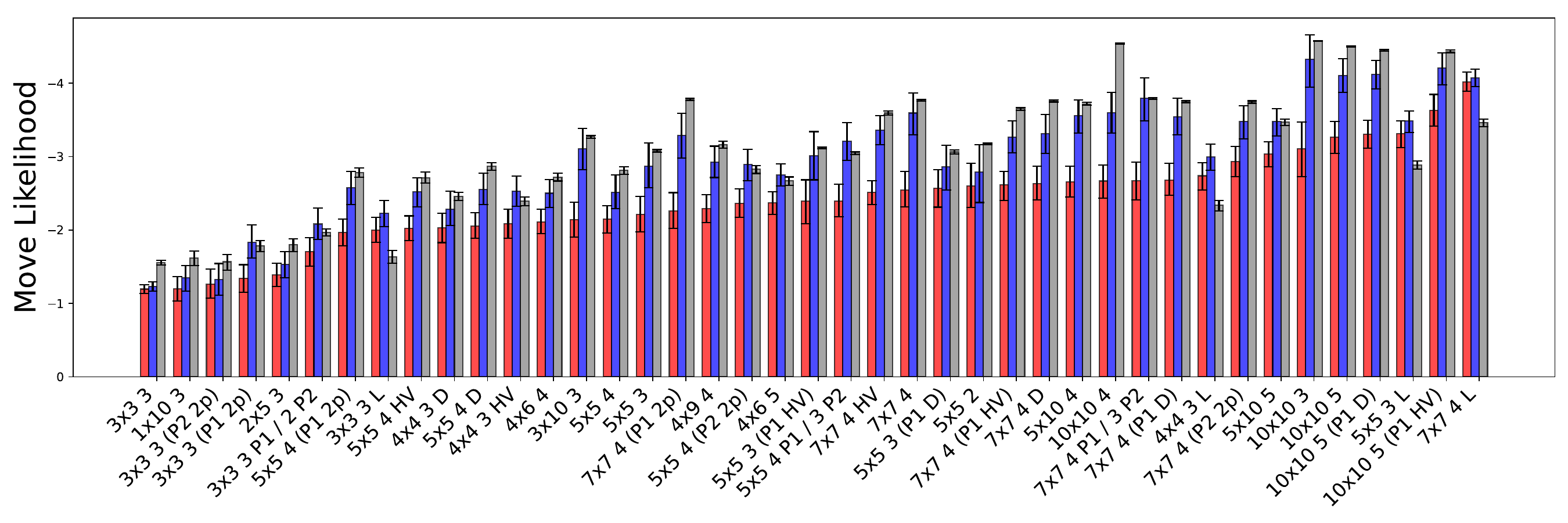}
    \caption{\textbf{Aggregate move likelihood, per game, in the human-human play experiment.} Aggregate move log probability (higher is better) for the moves people played in the human-human gameplay experiment under the Intuitive Gamer (red), Expert Gamer (blue), and random (grey) models, broken up by games. Error bars depict 95\% bootstrapped confidence intervals around the mean log-likelihood over all moves for each game.}
    \label{fig:logprob-per-game}
\end{figure}

We next present additional results from the ``zero-shot human-human play'' and ``watch-and-predict'' experiments. We report the full set of averaged log probabilities for each game in the human gameplay experiment in ~\ref{fig:logprob-per-game}.

\subsection{Action selection accuracy and rank}

\begin{figure}[t!]
    \centering
    \includegraphics[width=1.0\linewidth]{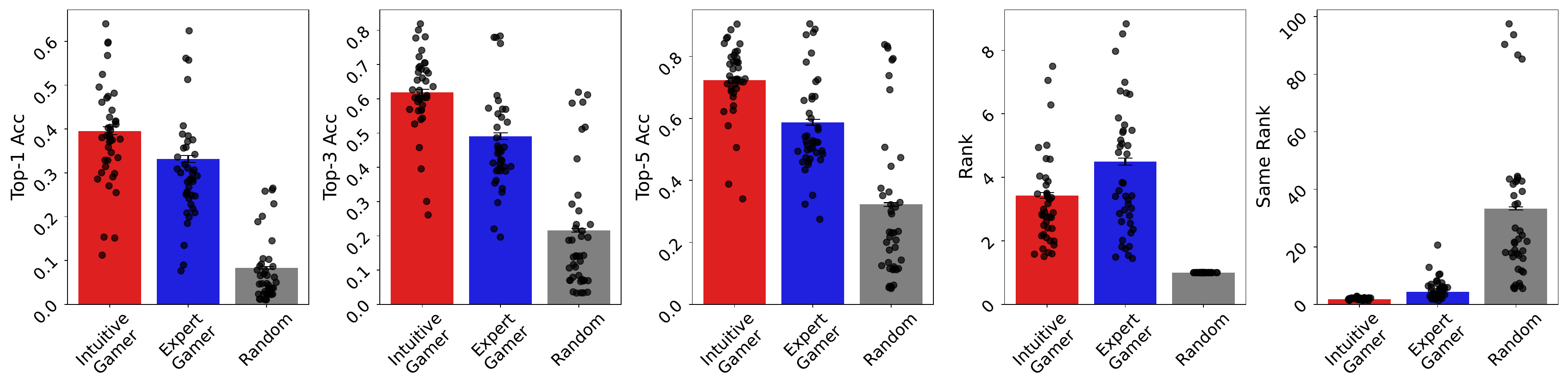}
    \caption{\textbf{Accuracy and rank of human played move in human gameplay experiments.} We additionally include additional statistics comparing the move distributions for the play experiments against observed play, particularly accuracy and the rank of the humans' played move relative to set of possible legal moves (ranks include ties). Each dot corresponds to the average for any one of the $41$ played games.} 
    \label{fig:acc-and-rank-play}
\end{figure}

The Intuitive Gamer model generally is more accurate at capturing peoples' moves according to Top-1, Top-3, Top-5 accuracy of the moves people made ( Figure~\ref{fig:acc-and-rank-play}). Accuracy is averaged over $100$ bootstrapped samples to account for ties; for instance, to compute Top-1 accuracy, if the Intuitive Gamer assigns three moves to the same top probability, then accuracy is computed by sampling from that set. To account for ties, we also compute the average rank of the human played moves (including ties). However, random will appear best under this measure as all moves are assigned the same rank with uniform probability (therefore, appearing as if all played moves are assigned rank $1$). To account for this, we also report the average number of moves assigned the same rank as the move played by people. Together, we see that generally the Intuitive Gamer assigns a relatively low rank to the played move (and not too many other moves to the similarly low rank). 

\subsection{Robustness to choice of distributional measure}

\begin{figure}[t!]
    \centering
    \includegraphics[width=0.9\linewidth]{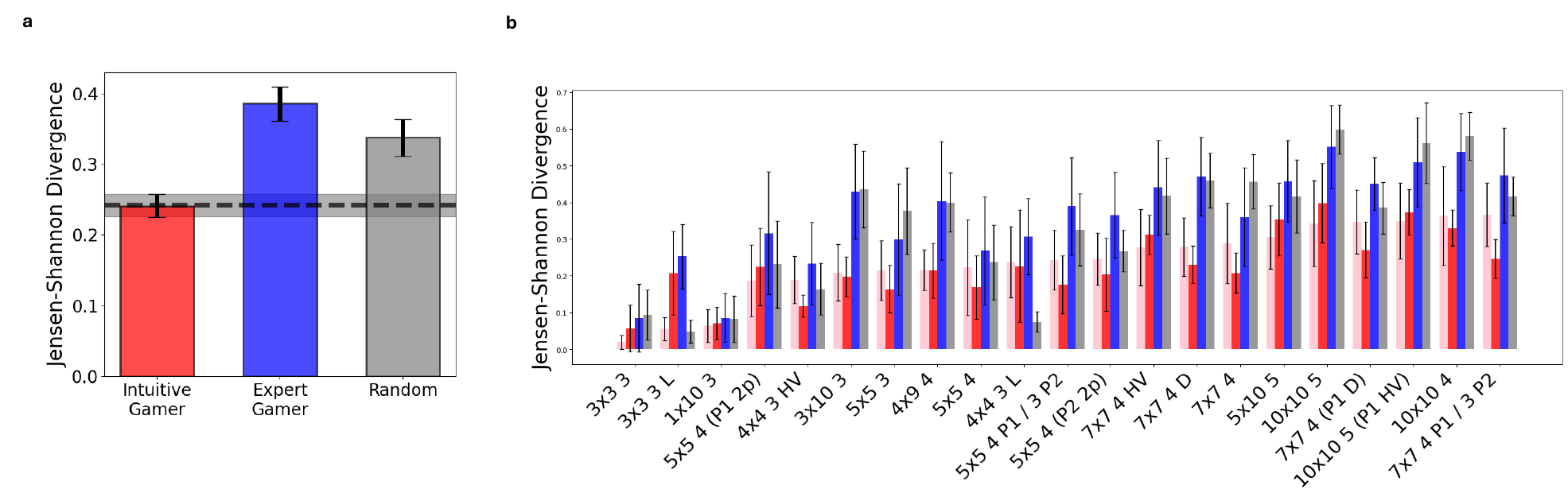}
    \caption{\textbf{``Watch-and-predict'' results are robust to the choice of distributional metric.} Jensen-Shannon Divergence (lower is better) reveals that the Intuitive Gamer distribution are generally better aligned to the human watchers' distributions, over \textbf{a,} all game boards and \textbf{b,} at a per game level. Error bars in \textbf{a} depict bootstrapped 95\%CI around the mean JSD for for all board; error bars in \textbf{b} depict standard deviation over boards per game.}
    \label{fig:jsd-per-game}
\end{figure}

\begin{figure}[t!]
    \centering
    \includegraphics[width=1.0\linewidth]{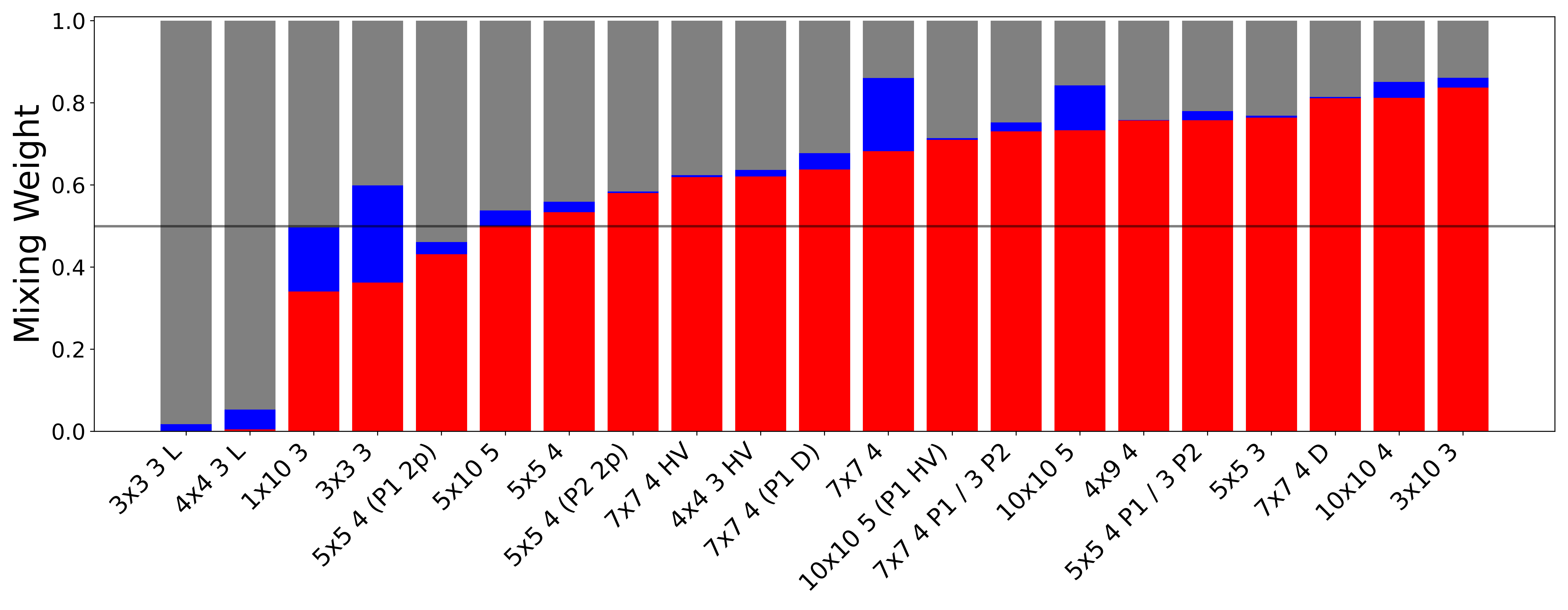}
    \caption{\textbf{Average admxiture weights for each game, fit to the average human watch-and-predict distributions.} The Intuitive Gamer is the dominant component for most games in an admixture fit over the models' distributions to the participants' distributions, optimized against the smoother Jensen-Shannon Divergence. Notable failure modes (e.g., misere games and comparable fits with the Expert Gamer model for Tic-Tac-Toe) align with places of weaker Intuitive Gamer model fits in the play data (see main text). Red means Intuitive Gamer, blue means Expert Gamer, grey means Random Gamer.}
    \label{fig:admixture-jsd}
\end{figure}

To assess the robustness of our ``watch-and-predict'' experiment analyses to the measure of distributional similarity, we repeat our analyses using an alternative distributional measure: the Jensen-Shannon Divergence~\citep{menendez1997jensen}. The Intuitive Gamer again is generally at the human noise ceiling in average and similarly near the split-half noise ceiling for many (though not all, e.g., misere) games (see Figure~\ref{fig:jsd-per-game}). An admixture model over the model distributions to learn a mixing weights to fit the human watcher distributions (optimized using JSD) generally reveals that the Intuitive Gamer model is the dominant component across games---with exceptions for misere games and some other smaller-board games (see Figure~\ref{fig:admixture-jsd}), which warrants further investigation. 

\subsection{Modeling choices with a softmax-based model}

As noted in the main text ``Methods,'' when assessing how models capture the choices people actually made in games, we model participants' moves as some combination of the likelihood under the core model of interest (the Intuitive Gamer or Expert Gamer model) as well as some probability that they play randomly. We sweep over $\alpha$ between $0.5$ to $0.95$ in increments of $0.05$. An alpha of $0.5$ means that any move is equally weighted between the primary model (Intuitive Gamer or Expert Gamer) and random; higher alpha puts more weight on the primary model. The Intuitive Gamer model generally leads to better fits, independent of the choice of $\alpha$. However, both models yield a better fit with human play and human watch distributions when mixed in some proportion with random. This aligns with the admixture results in the main paper, which show that fits for most games and most people are best modeled by some random component. A lower alpha (more weight to random) leads to a comparatively better fit when modeling human play and watch judgments under the Expert Gamer model, further indicating that the Expert Gamer model is likely more sophisticated than how people actually reason in new games for the first time. 

In the admixture analyses, we also repeat admixture fits for different temperatures ($\tau$), ranging in $\tau \in \{0.5, 1.0, 1.5, ...,3.0\}$ where $\tau$ controls the action choice distribution softmax. We refit the admixture at each setting of $\tau$ and apply the same $\tau$ to both the Intuitive Gamer and Expert Gamer model. We do this to ensure that the relative fits in the admixture are as fair as possible to each model as the Expert Gamer model is often relatively sharp. As temperature increases, all models collapse toward random; however, as temperature decreases, we lose the probabilistic nature that is core to our Intuitive Gamer hypothesis. For most settings of $\tau$, the Intuitive Gamer is the dominant mixing component; however, a higher $\tau$ appears to yield a slightly better fit for the watch distributional fits to human watch distributions relative to the played moves, which may be due to the seemingly higher variability in participants' move distributions.

\begin{figure}[t!]
    \centering
    \includegraphics[width=1.0\linewidth]{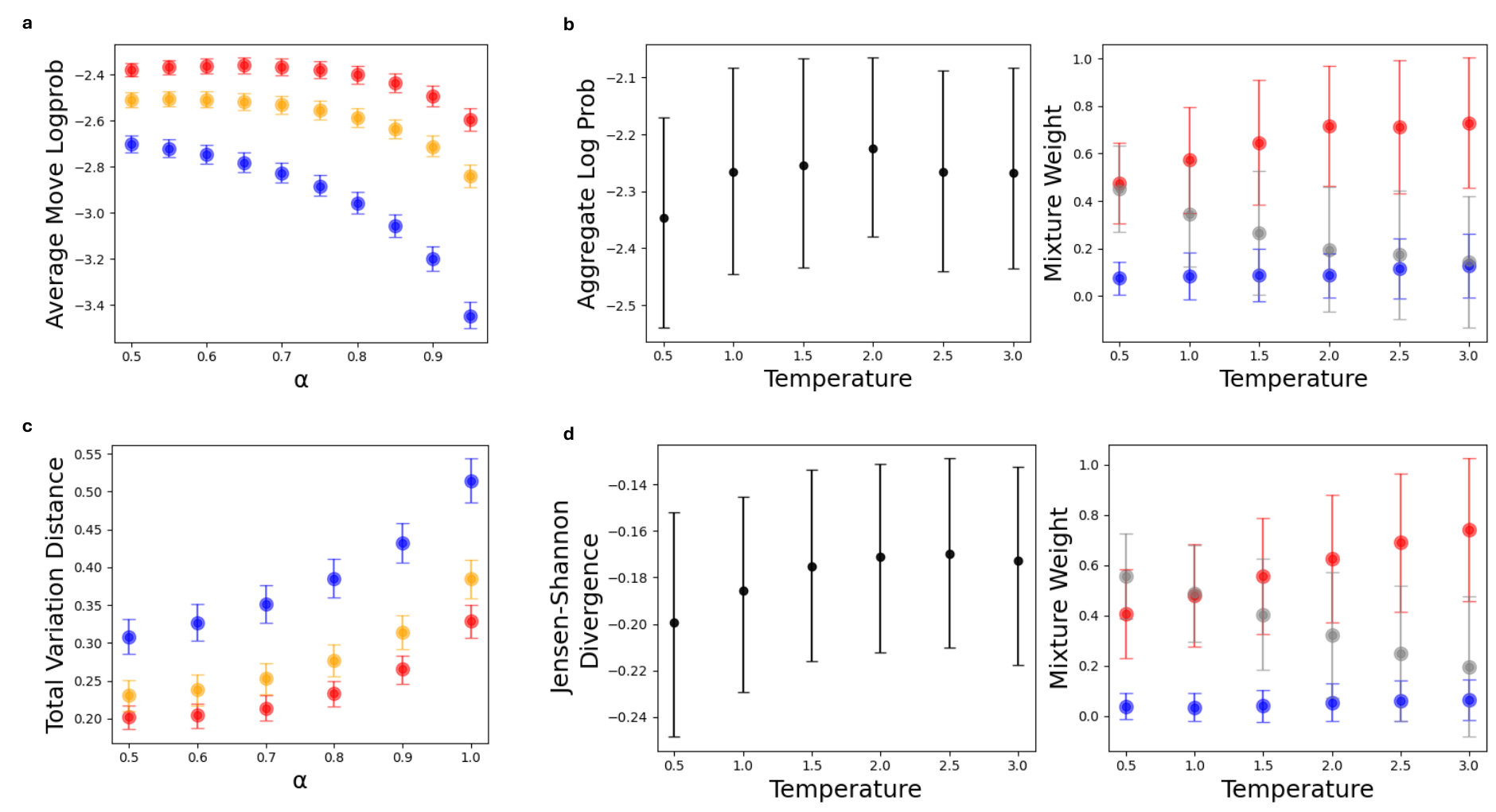}
    \caption{\textbf{Mixture weight and temperature sensitivity in models' fit to people's action selection and prediction data.} Varying the mixture weight $\alpha$ in the $\alpha-$softmax model for the core play \textbf{(a)} and watch \textbf{(c)} experiments. Error bars show 95\% bootstrapped CI over the mean average move log probability (higher is better) and average Total Variation Distance (lower is better) relative to the human watcher move distribution. Red is the Intuitive Gamer model; yellow the depth-3 version of the Intuitive Gamer (non-flat); blue the Expert Gamer model. Admixtures are then fit over the flat Intuitive Gamer (red), Expert Gamer (blue), and random (grey) game agents for varied move distribution temperature for the play \textbf{(b)} and watch \textbf{(d)} experiments, respectively. Admixtures are fit for all data for each game. The left subplot depicts the average log probability of the moves in the resulting admixture for each temperature \textbf{(b)} and the resulting Jensen-Shannon Divergence (as the admixtures are optimized against JSD rather than TVD, given the smoothness of JSD) over the optimized mixture distribution. Error bars depict 95\% CIs over the bootstrapped mean. The right subplot depicts the averaged mixing weights for each temperature value. Error bars depict standard deviation over the games.}
    
    \label{fig:alpha-vary}
\end{figure}

\subsection{Predicted probability of played move relative to human predictions}

\begin{figure}[h!]
    \centering
    \includegraphics[width=1.0\linewidth]{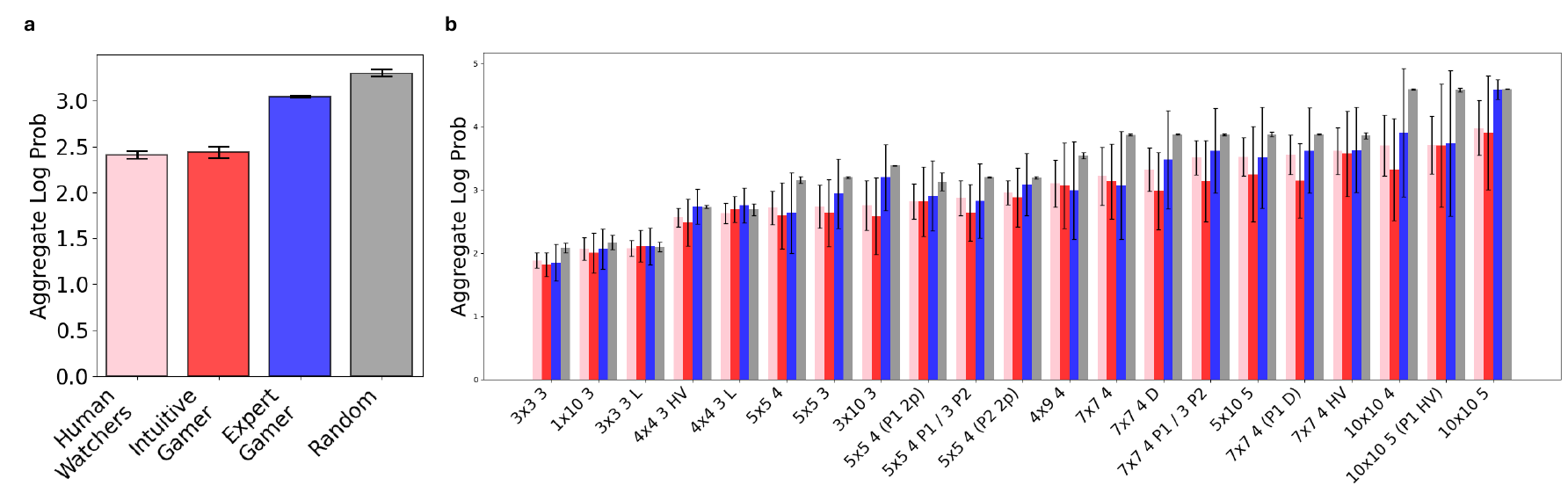}
    \caption{\textbf{People's distribution of predicted moves capture real human player's actions comparably to the Intuitive Gamer model.}  \textbf{a,} From only a single indirect experience with a new game (watching one partial match), participants' distribution over predicted moves (pink) generally place similar probability on the move actually played to that predicted under the Intuitive Gamer model. The suggested moves made by the people predicting from watching videos (pink) to the actual move played by participants in each match and compare the log-likelihood under the human predictors' inferred move to the log-likelihood assigned by the respective models, marginalizing out $\alpha$. Error bars depict bootstrapped 95\% CIs over the averaged log-likelihood over boards, again sweeping over a ``slop'' parameter $\alpha$ (see ``Methods''). \textbf{b,} Predicted move likelihood assigned to the move selected by the live player, broken down by game and ordered by lowest negative log likelihood (lower is better). Error bars depict standard deviation.}
    \label{fig:human-watcher-logprob}
\end{figure}

In the ``watch-and-predict'' experiment, participants were asked to predict where a player should move. To assess whether the distribution of moves people predicted captures how the human player actually played, the same aggregate analysis (as in the main ``play'' experiment), computing the average log likelihood of a player's action under the move distribution predicted by participants watching that match versus the distributions predicted under the Intuitive Gamer model and alternate models along a spectrum of expertise. Human watchers place similar probability on the move people actually made as the Intuitive Gamer model (Figure~\ref{fig:human-watcher-logprob}). 

\subsection{Additional human- and model-predicted distributions}

\begin{figure}
    \centering
    \includegraphics[width=1.0\linewidth]{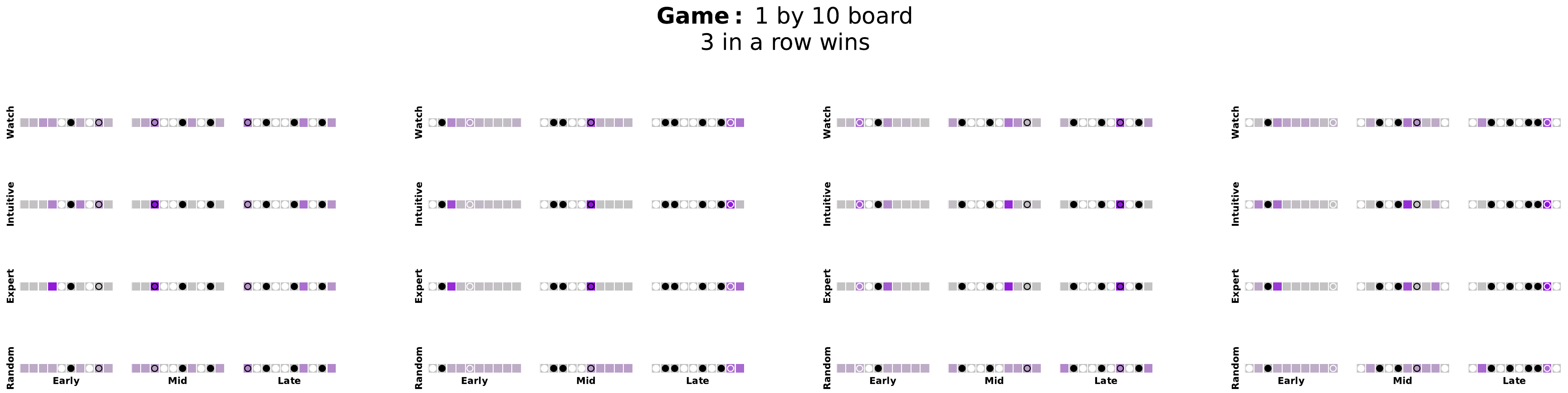}
    \includegraphics[width=1.0\linewidth]{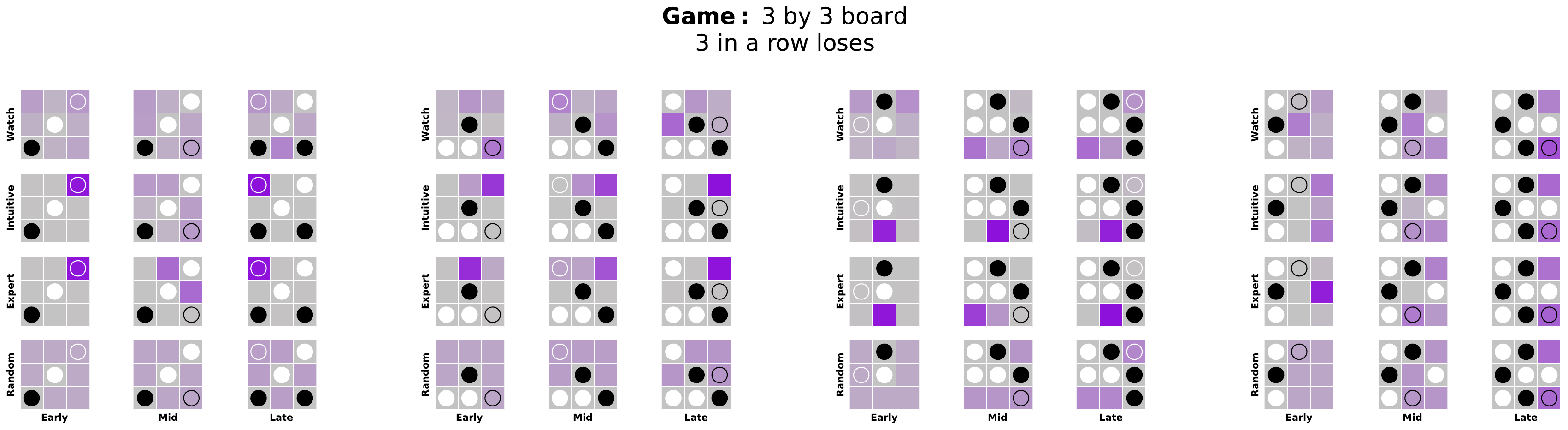}
    \includegraphics[width=1.0\linewidth]{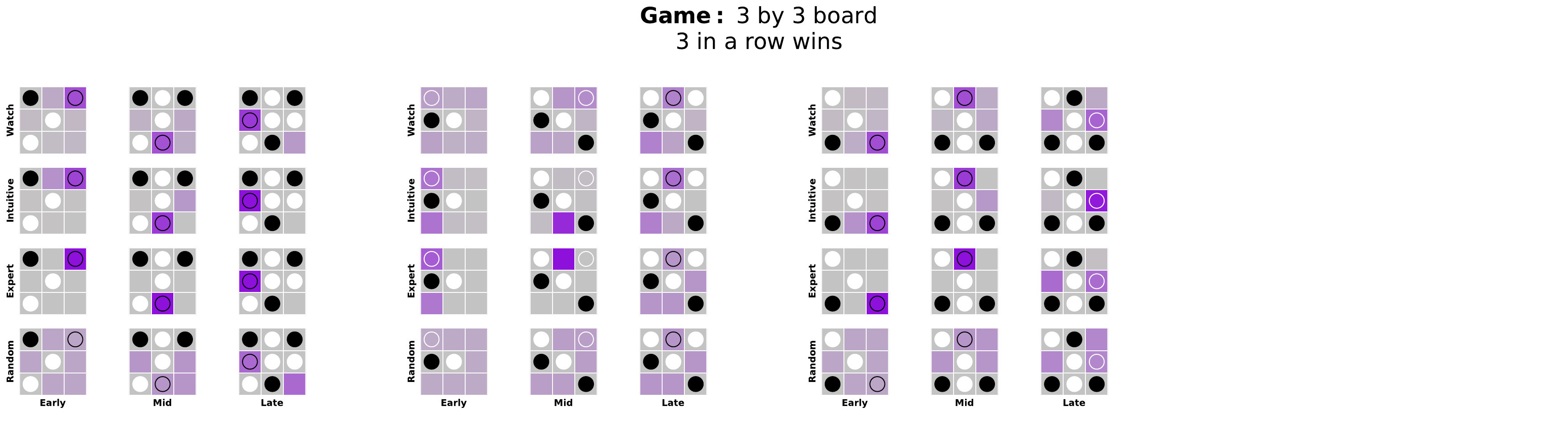}
    \caption{\textbf{Full set of ``watch-and-predict'' distributions per match.} As in the main text, the top entry in each panel depicts the aggregated human-predicted distribution over next moves, as determined by participants in the indirect play (watch-and-predict) study. The circle indicates where a person actually played. The color of each grid cell indicates where the player is predicted to move on that turn, under either the model- or human-predicted distribution; darker purple means a player is more likely to select that grid cell. We show all game boards and human- and model-predictions for each game type. The game board size and rules are written above each set of boards.}
     \label{fig:watch-extra-1}
\end{figure}

\begin{figure}
    \centering
    \includegraphics[width=1.0\linewidth]{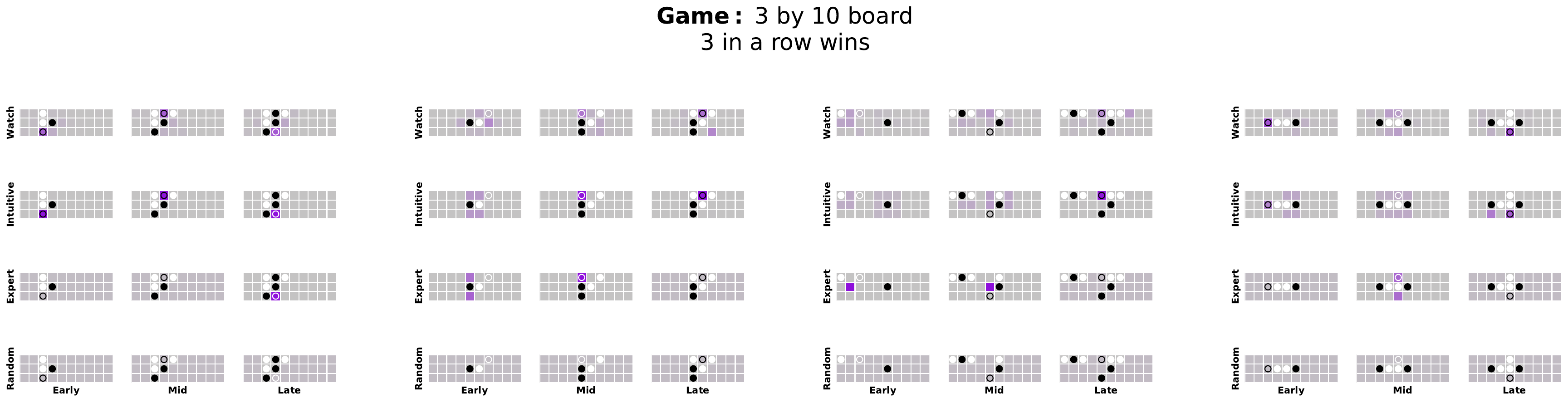}
    \includegraphics[width=1.0\linewidth]{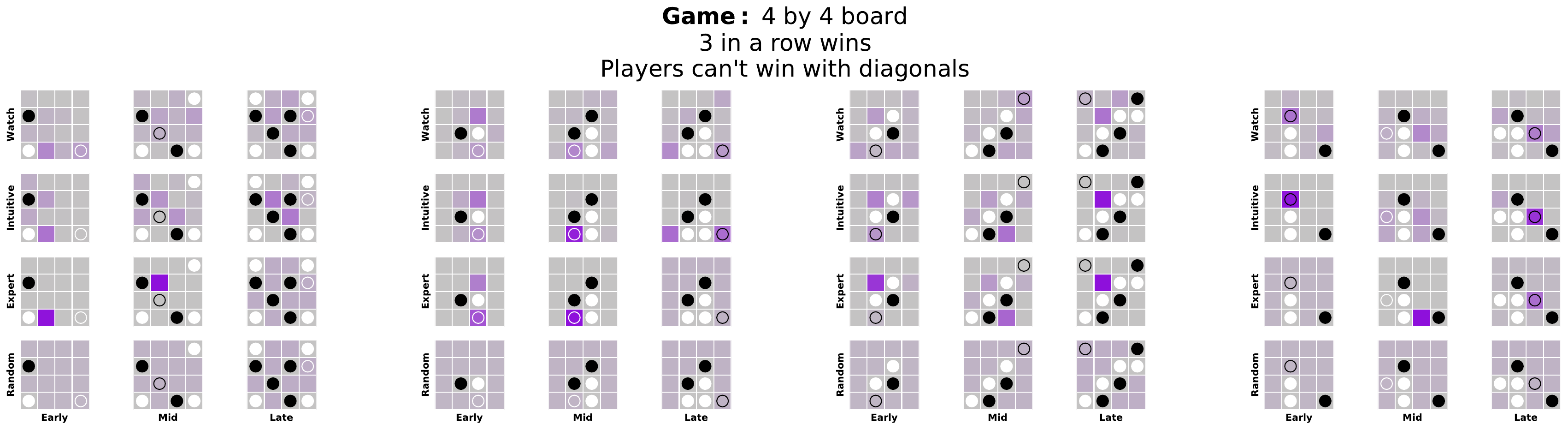}
    \includegraphics[width=1.0\linewidth]{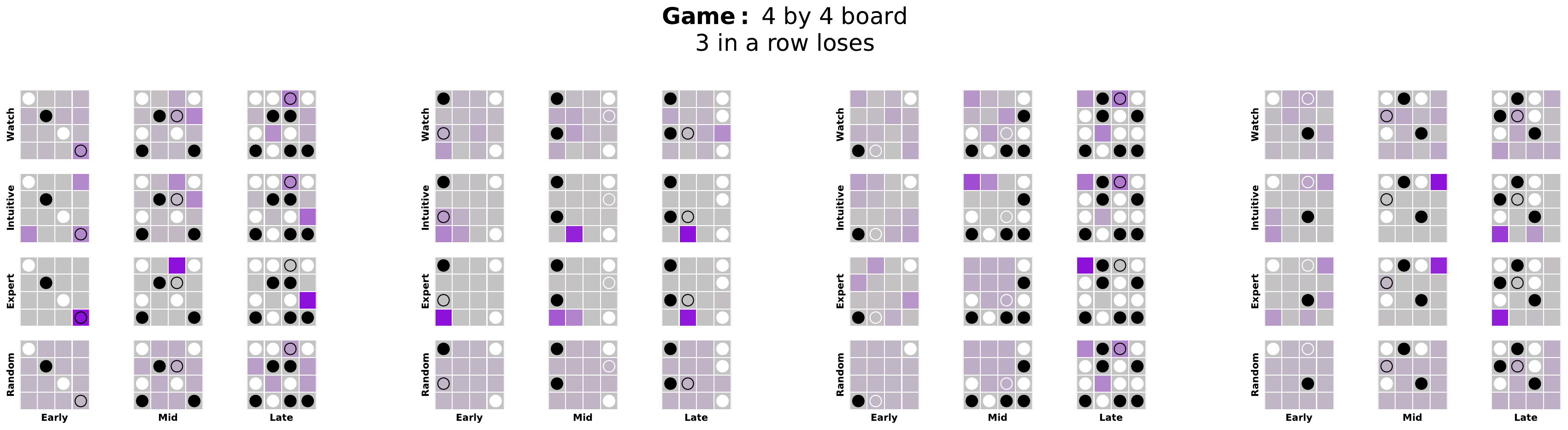}
    \caption{Full set of ``watch-and-predict'' distributions per match (continued).}
    \label{fig:watch-extra-2}
\end{figure}

\begin{figure}
    \centering
    \includegraphics[width=1.0\linewidth]{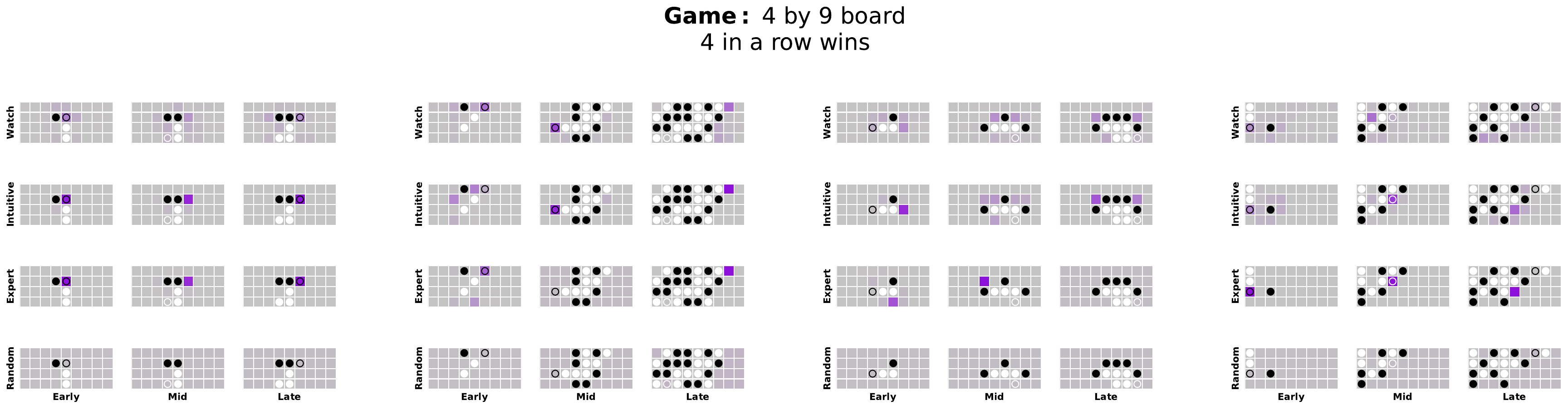}
    \includegraphics[width=1.0\linewidth]{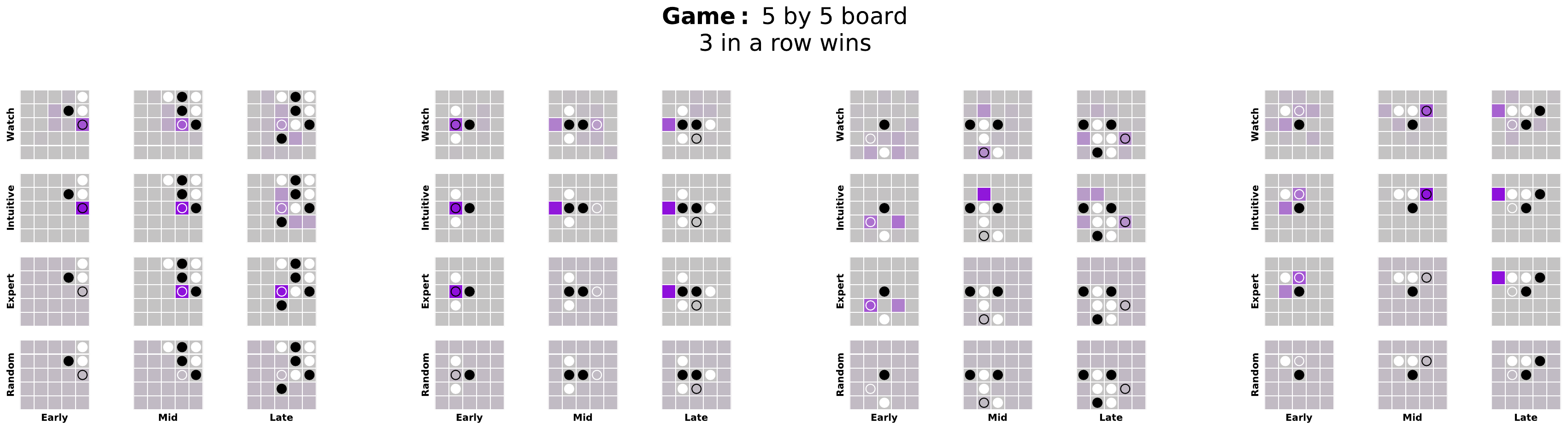}
    \includegraphics[width=1.0\linewidth]{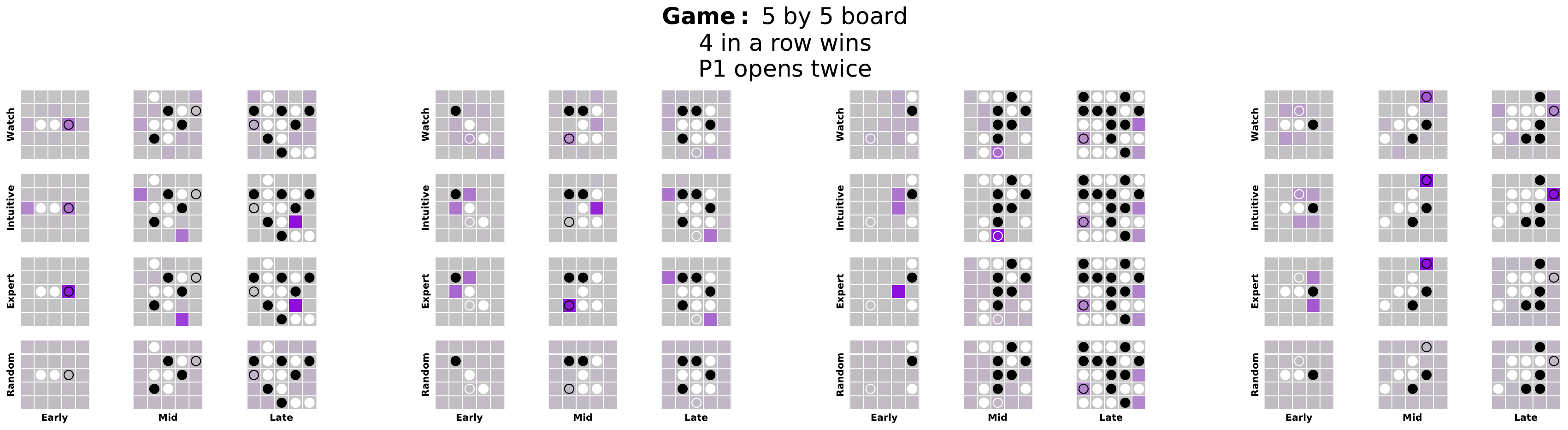}
    \caption{Full set of ``watch-and-predict'' distributions per match (continued).}
    \label{fig:watch-extra-3}
\end{figure}

\begin{figure}
    \centering
    \includegraphics[width=1.0\linewidth]{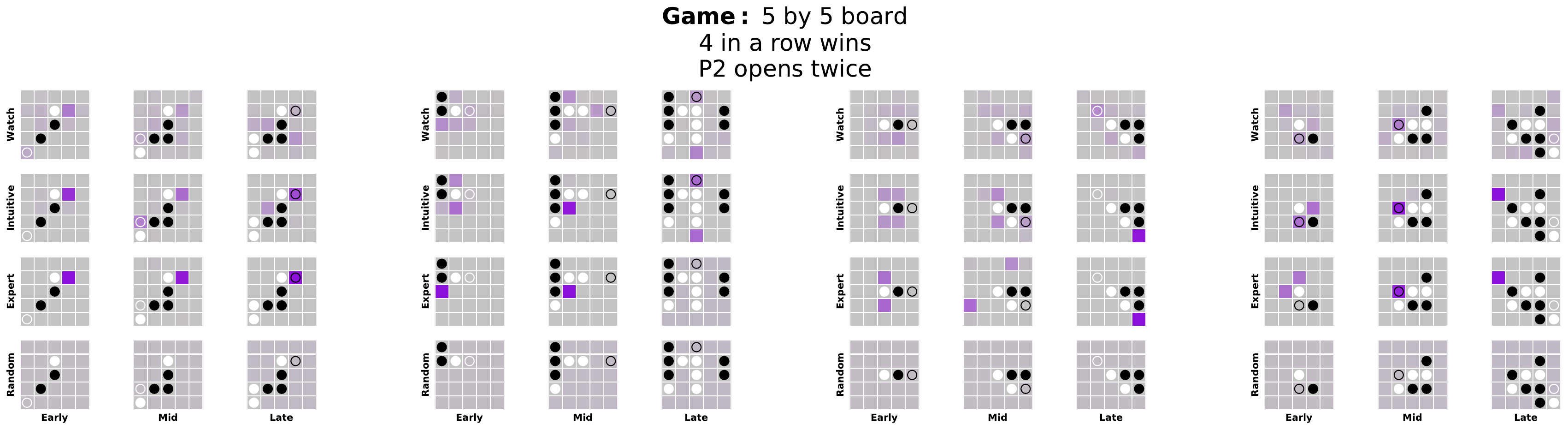}
    \includegraphics[width=1.0\linewidth]{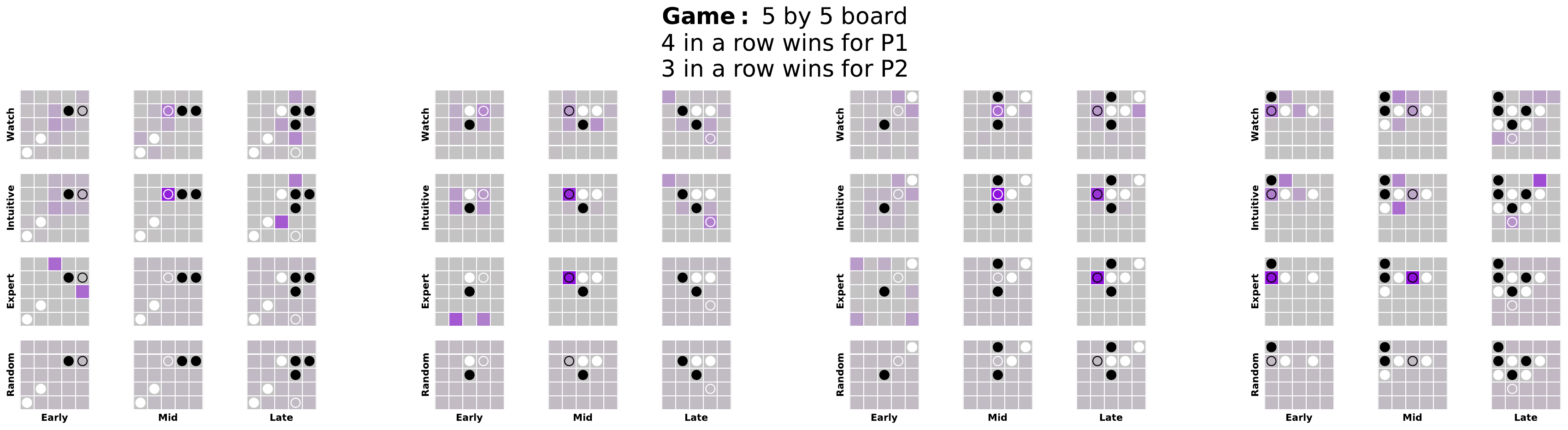}
    \includegraphics[width=1.0\linewidth]{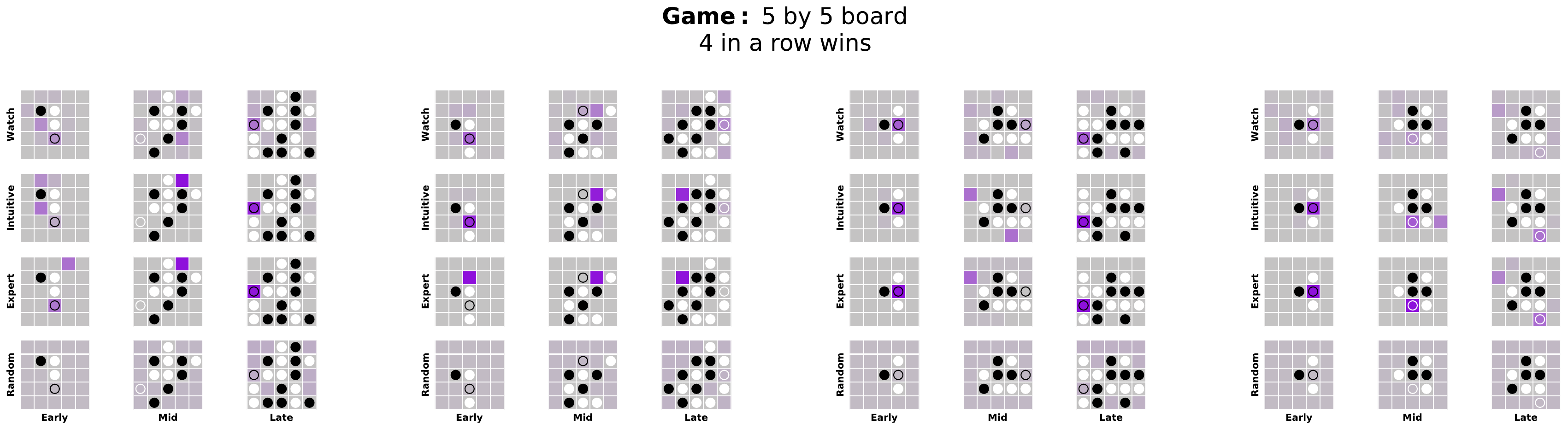}
    \caption{Full set of ``watch-and-predict'' distributions per match (continued).}
    \label{fig:watch-extra-4}
\end{figure}

\begin{figure}
    \centering
    \includegraphics[width=1.0\linewidth]{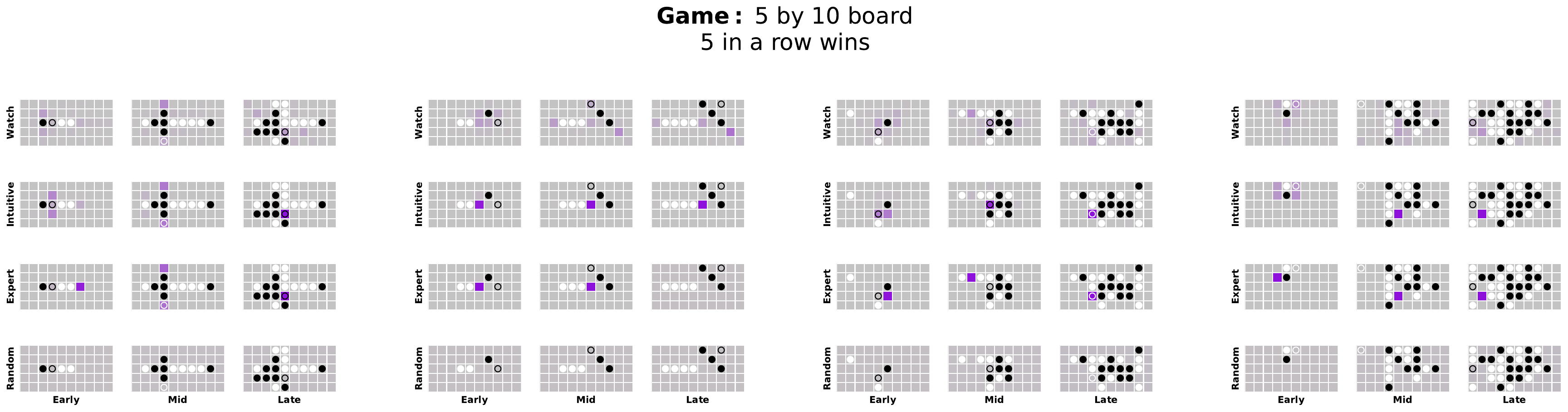}
    \includegraphics[width=1.0\linewidth]{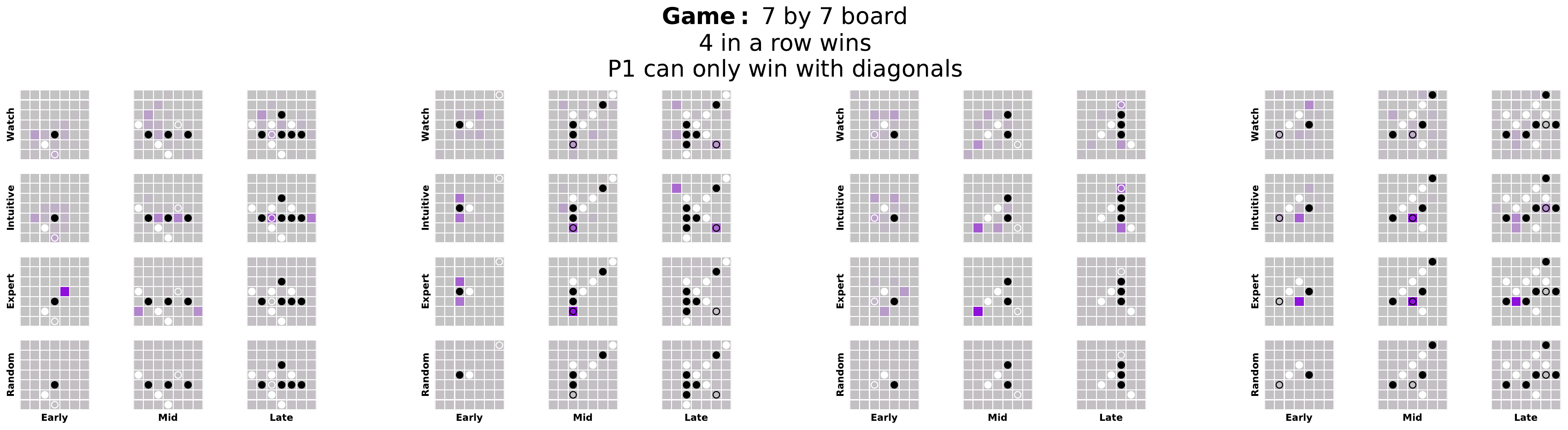}
    \includegraphics[width=1.0\linewidth]{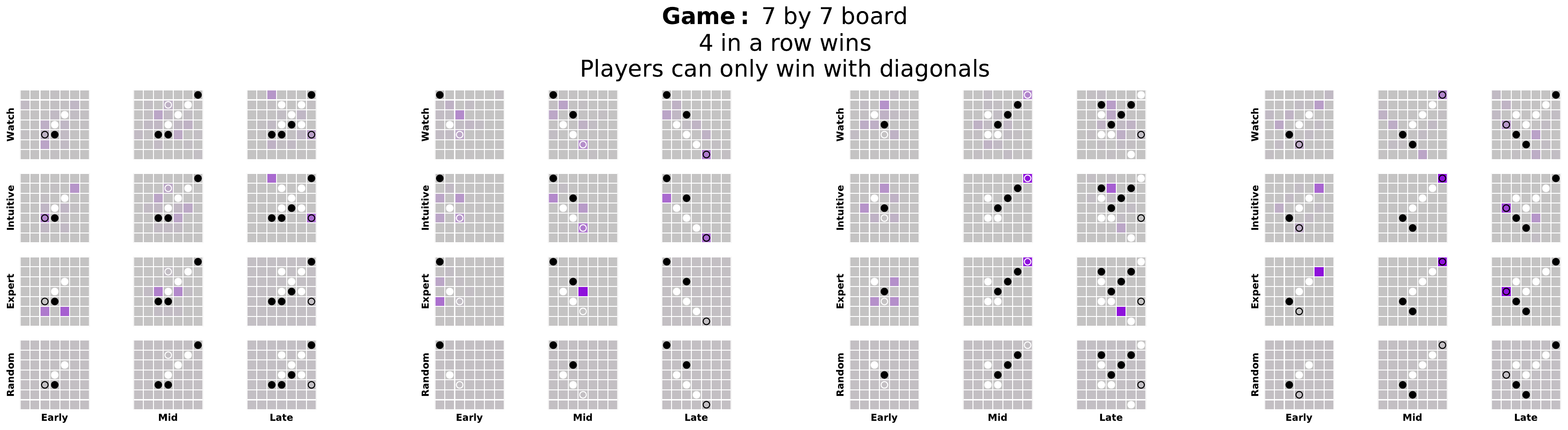}
    \caption{Full set of ``watch-and-predict'' distributions per match (continued).}
    \label{fig:watch-extra-5}
\end{figure}

\begin{figure}
    \centering
    \includegraphics[width=1.0\linewidth]{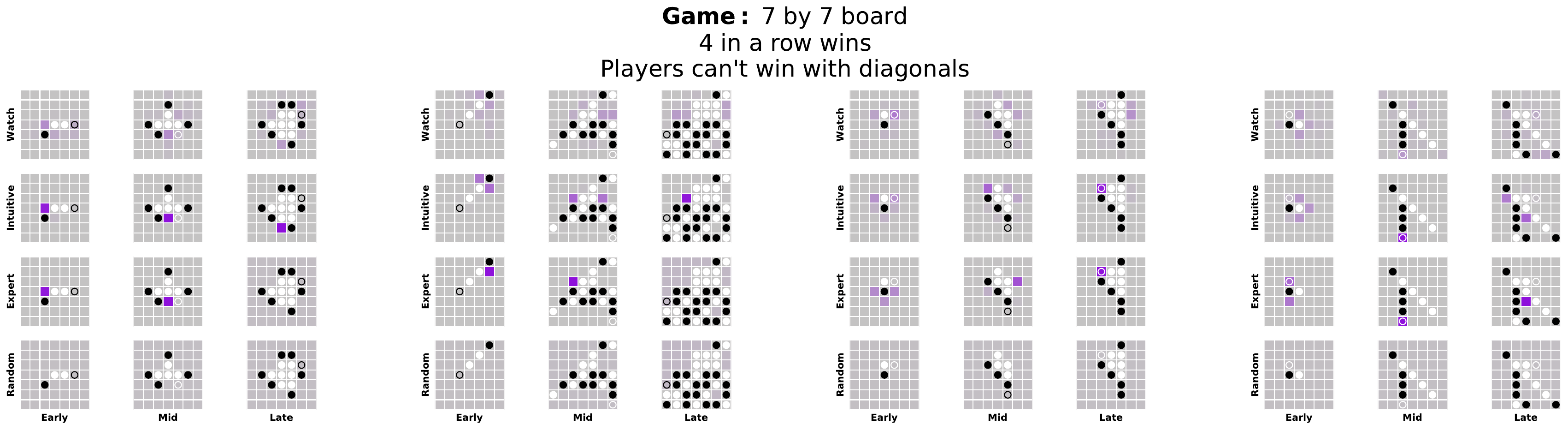}
    \includegraphics[width=1.0\linewidth]{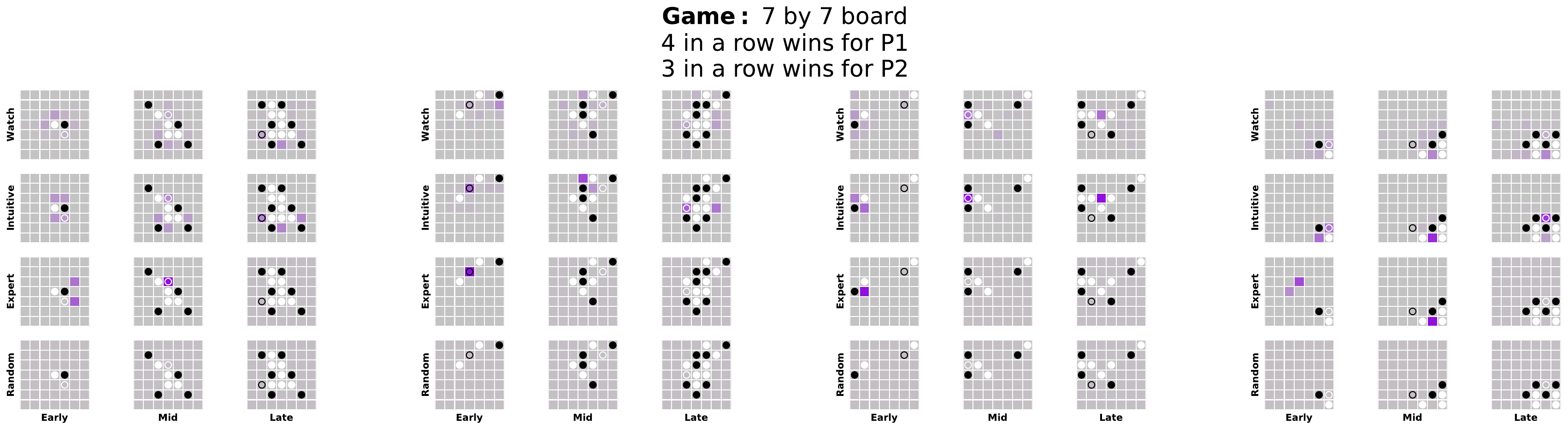}
    \includegraphics[width=1.0\linewidth]{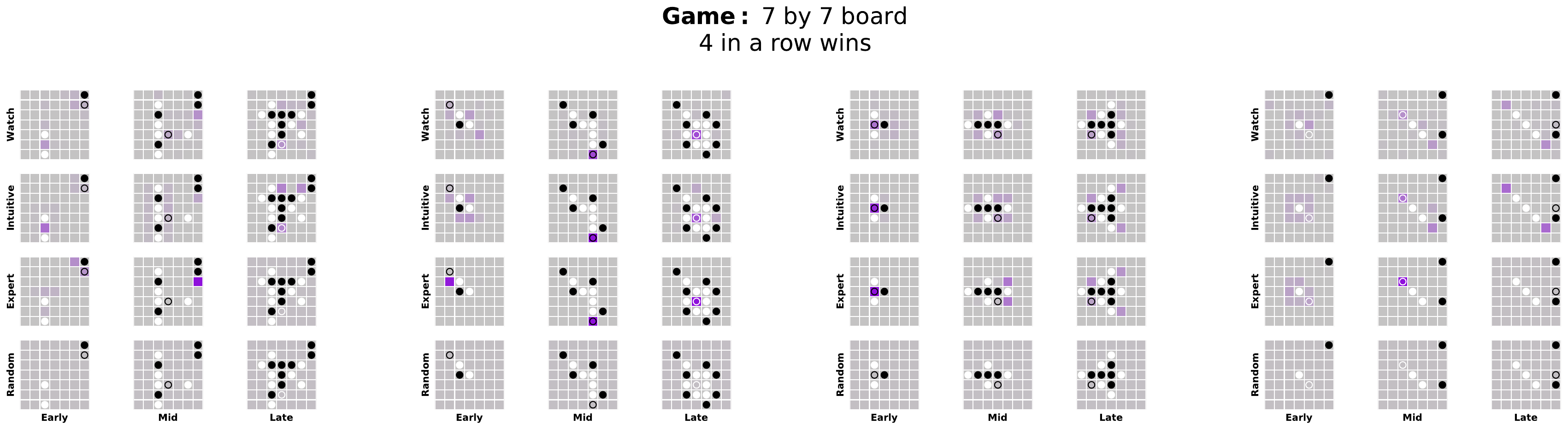}
    \caption{Full set of ``watch-and-predict'' distributions per match (continued).}
    \label{fig:watch-extra-6}
\end{figure}

\begin{figure}
    \centering
    \includegraphics[width=1.0\linewidth]{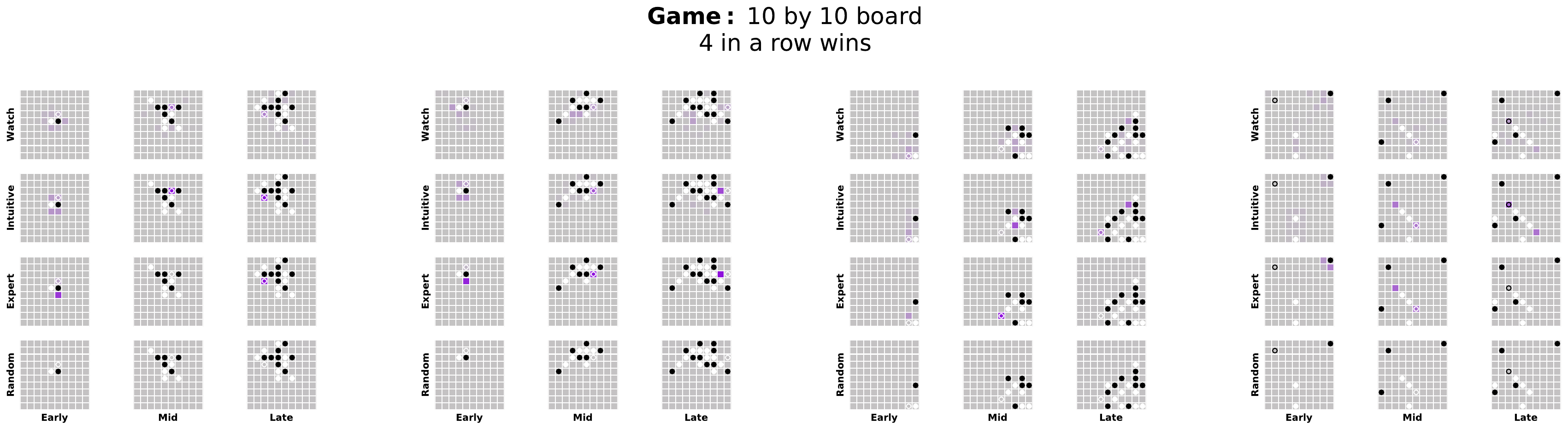}
    \includegraphics[width=1.0\linewidth]{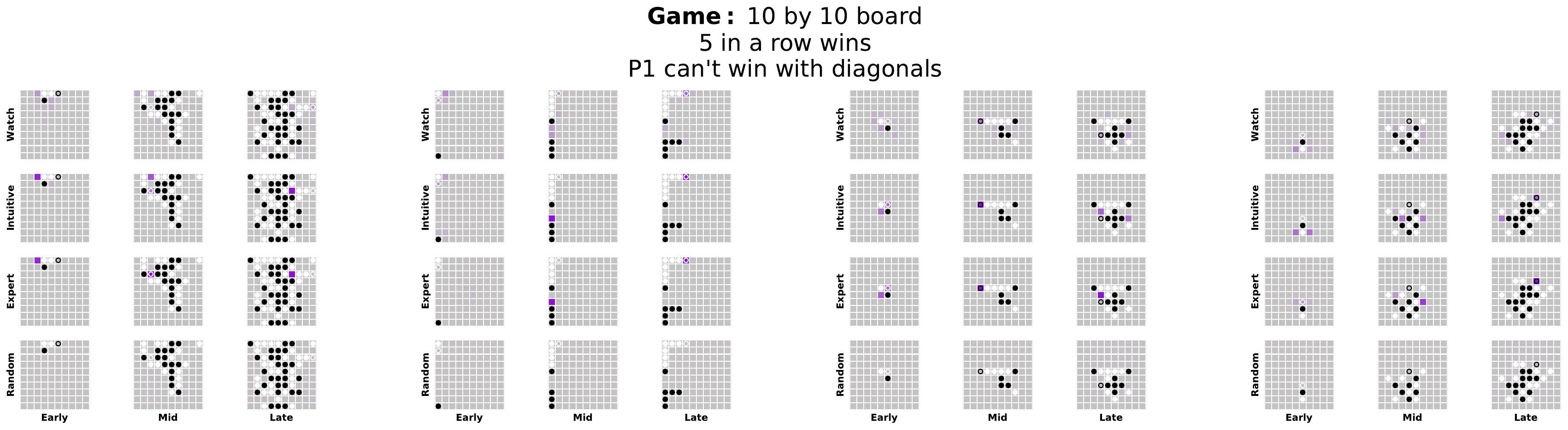}
    \includegraphics[width=1.0\linewidth]{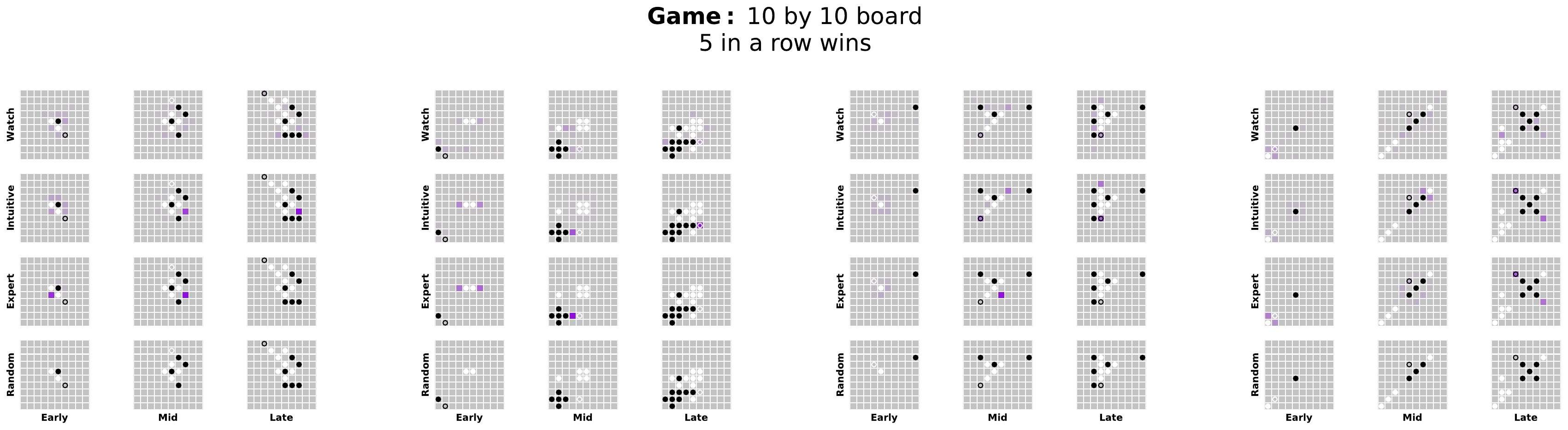}
    \caption{Full set of ``watch-and-predict'' distributions per match (continued).}
    \label{fig:watch-extra-7}
\end{figure}

We depict all aggregate human- and model-predicted distributions from the ``watch-and-predict'' study (see Figures~\ref{fig:watch-extra-1}-~\ref{fig:watch-extra-7}). Four matches, paused at three timepoints (during early-, middle-, and late-stage play) were shown for each game.

\subsection{Game lengths} 

\begin{figure}
    \centering
    \includegraphics[width=1.0\linewidth]{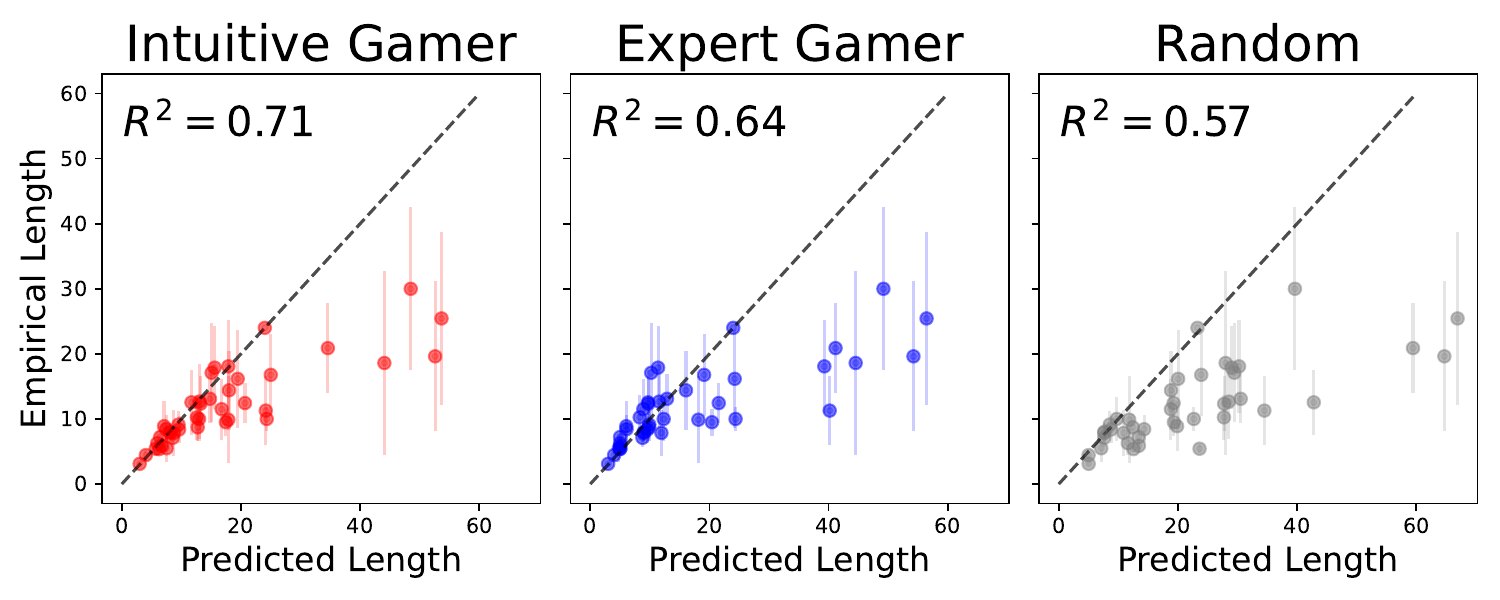}
    \caption{\textbf{Empirical human- vs. model-predicted expected game length}. Comparing the observed game length in human-human played games against the expected game length under simulations from the various game reasoning models. Game simulations are conducted between the same agent type (e.g., Intuitive Gamer against another instanced of itself). Empirical game length is coded as whenever the game ended: either from a player winning, the game ending in a draw. Games that ended early from an accepted draw request or a player surrendering are excluded.} 
    \label{fig:game-length-expected}
\end{figure}

We compare the empirical observed game lengths in the play experiment against the expected game length under model simulations. Intuitive Gamer simulations show the highest correlations with empirical human game length data, while Expert Gamer is more correlated than the random agent (Figure~\ref{fig:game-length-expected}). We observe that the Intuitive Gamer simulations rarely ``overpredict'', and there are several outlier games where it takes all models significantly longer to simulate than humans to finish.

\subsection{Draw requests and surrenders}

\begin{figure}
    \centering
    \includegraphics[width=1.0\linewidth]{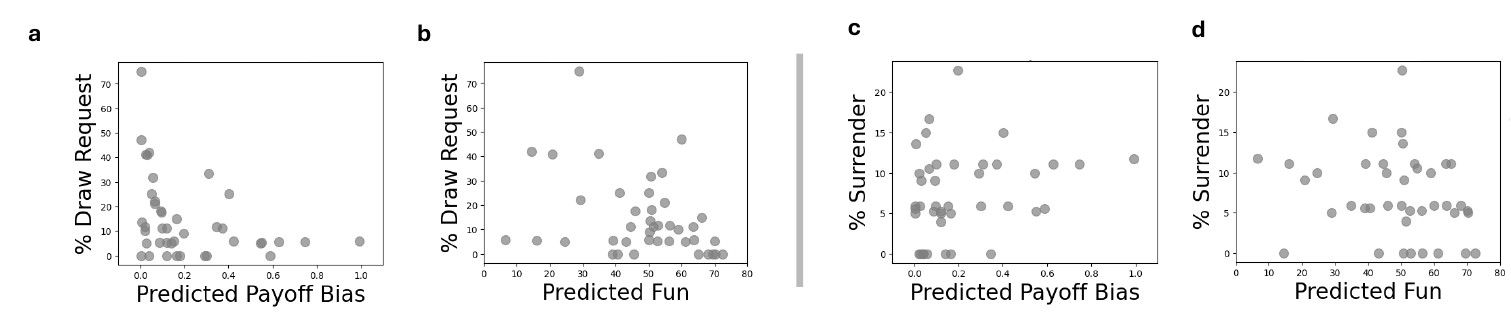}
    \caption{\textbf{Draw and surrender rates and people's ``just think'' game evaluation.} Frequency of draw requests (\textbf{a-b}) and surrenders (\textbf{c-d}) across all played matches per game (each point is a game). \textbf{a,} Games that are more biased (absolute difference in payoff from zero, under the averaged human-predicted payoffs in the ``just think'' study) are generally associated with higher draw rates. \textbf{b,} Games that are more fun (under the people's averaged funness judgments in the ``just think experiment before any play) are generally associated with fewer draw requests. \textbf{c,} There is not a clear relationship between people's evaluation of game fairness (before any play) and surrender rates; \textbf{d,} in contrast, there games that are less fun tend to have somewhat higher surrender rates.}
    \label{fig:draw-surr-payoff-fun}
\end{figure}

We next conduct exploratory analyses into draw requests and surrender decisions in individual matches against model and participant-predicted payoffs. Participants have high draw rates on matches that tend to be less fun and also less biased (see Figure~\ref{fig:draw-surr-payoff-fun}a-b). This suggests that participants are sensitive to other features of games when making decisions about whether the request to end the game. Surrender rates are generally higher for less fun games, though show less of a trend based on game bias (see Figure~\ref{fig:draw-surr-payoff-fun}c-d). Future work can work on richer process models of humans' draw and surrender decisions. The choice to request a draw instead of surrender is particularly worth investigating. If a player wanted to end the game, there should be no reason that they do not immediately surrender, as if they request a draw, the game only ends if the other player agrees to draw. Future modeling could explore other factors that people may consider when evaluating whether a game is worth playing (e.g., related to shame or dignity).

\subsection{Evaluating games after a single exposure}

After participants played a single match or watched and predicted actions in a single match, we asked them to either rate the expected payoff of the game overall or the funness. We also asked participants to rate the skill of their opponent (in the ``play'' condition) or the skill of both players (in the ``watch'' condition). We next conduct an initial exploration into these post-match game evaluations.

\subsubsection{Evaluating games after one round of play}

\begin{figure}
    \centering
    \includegraphics[width=1.0\linewidth]{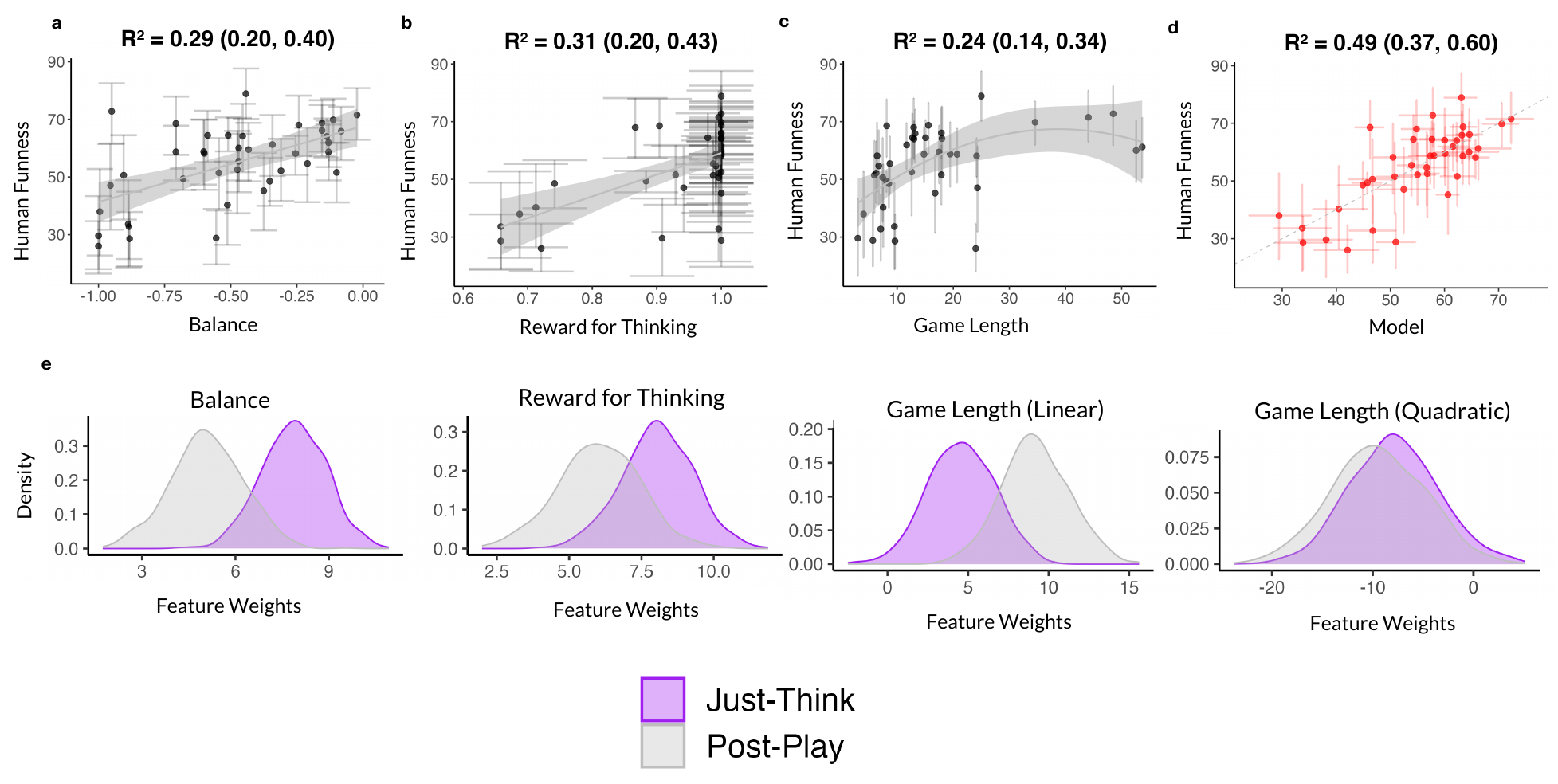}
    \caption{\textbf{Impact of a single instance of play experience on judgments}. \textbf{a,} bootstrapped mean predicted payoff for ``just think'' versus ``post-play'' participant judgments; \textbf{b,} bootstrapped mean predicted funness from ``just thinking'' versus ``post-play''.}
    \label{fig:post-play-v-think}
\end{figure}

\begin{figure}
    \centering
    \includegraphics[width=1.0\linewidth]{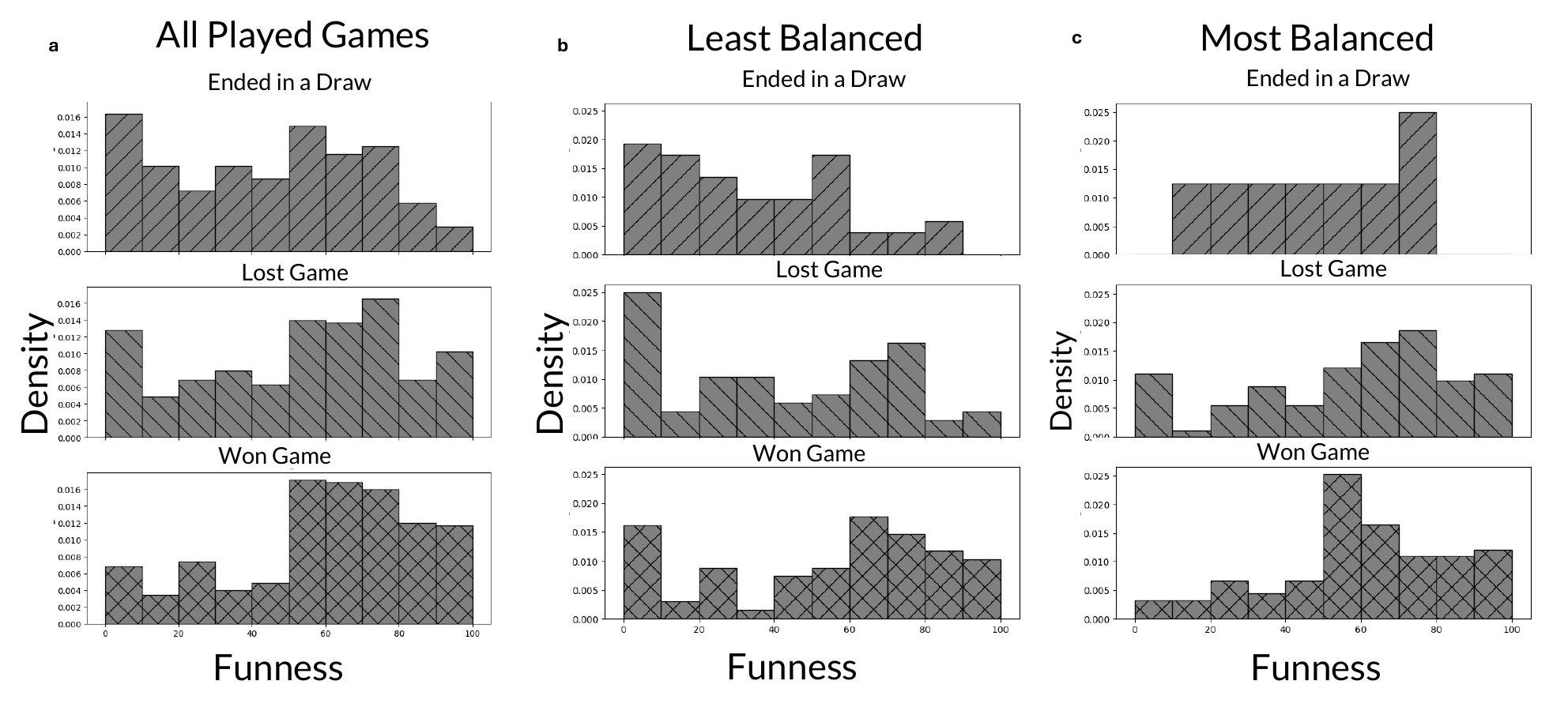}
    \caption{\textbf{Relationship between attained outcome and funness judgments, after a single round of play in a new game.} Funness judgments made by players after each match, broken down by whether they drew (top), lost (middle), or won (bottom). Even in cases where players lost, many still found the game fun. \textbf{a,} Funness for all matches for all played games. \textbf{b-c,} Post-play funness judgments for matches in the least balanced games (according to the Intuitive Gamer ``balance'' feature; the bottom 25 percentile; \textbf{b}) and the most balanced games (upper 75 percentile; \textbf{c}). Winning in an unbalanced game (\textbf{b}) still tends to lead to somewhat higher funness ratings; losing in a balanced game can also still lead to relatively high funness ratings.}
    \label{fig:outcomes-match-funness}
\end{figure}

Participants' payoff predictions are highly correlated with their payoff predictions before any play and well-align with the Intuitive Gamer model (Figure~\ref{fig:post-play-v-think}). Participants' post-play funness judgments are less correlated with pre-play funness judgments (Figure~\ref{fig:post-play-v-think}) and are overall noisier (split-halfs $R^2 = 0.49$ [95\% CI: $0.33, 0.64$] for post-play funness compared to a split-half of $R^2 = 0.69$ [95\% CI: $0.56, 0.79$] for the same subset of $41$ games rated on funness by participants who ``just thought'' about the games without any play). While we asked participants to assess the funness of the game overall---not just with respect to the match they played---it is possible people were biased in some form by the outcome of the game (Figure~\ref{fig:outcomes-match-funness}) or experience playing against the particular other player.

\begin{figure}
    \centering
    \includegraphics[width=1.0\linewidth]{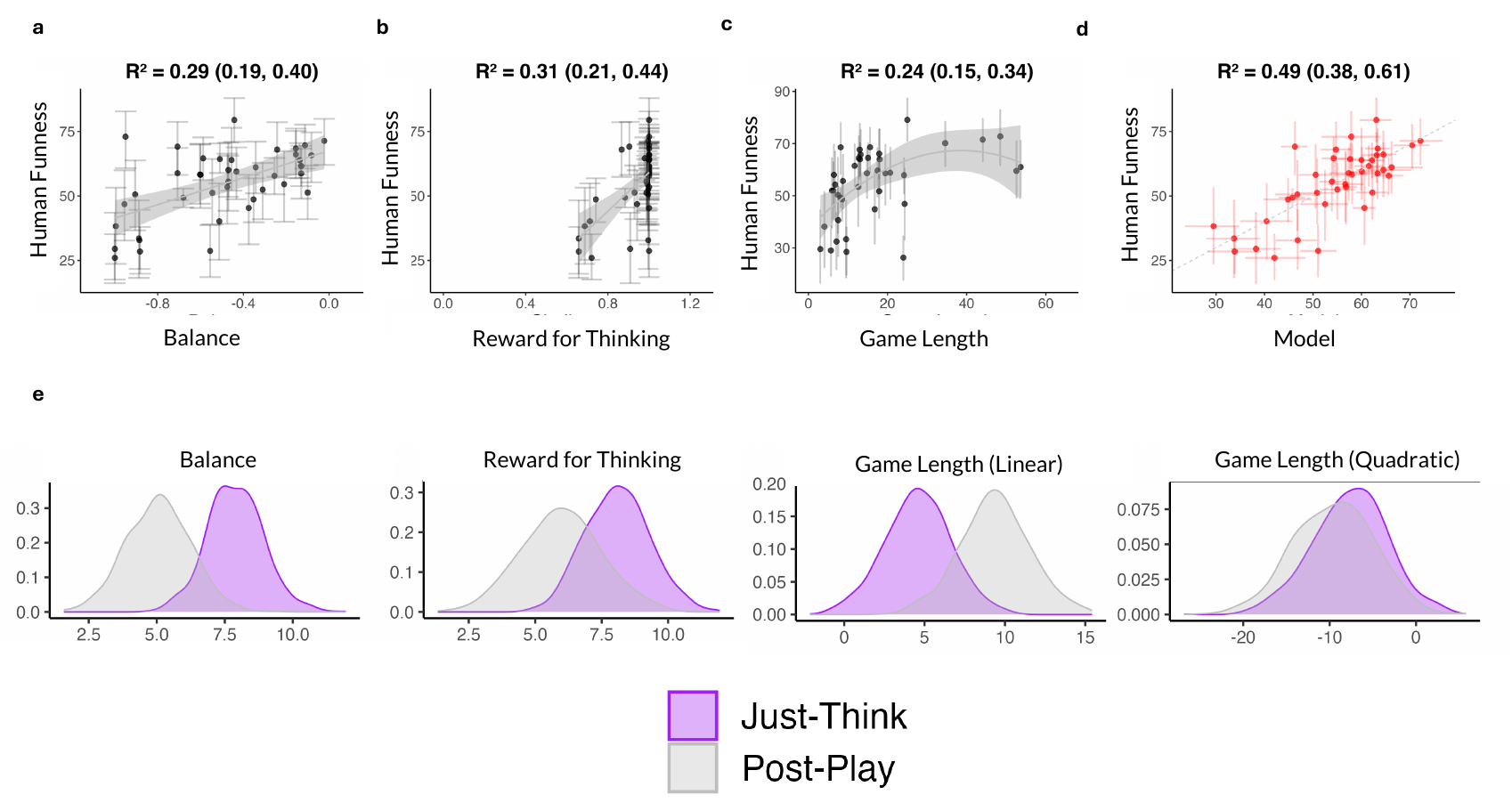}
    \caption{\textbf{Post-play funness modeling}. \textbf{a}-\textbf{c,} Features derived from the Intuitive Gamer model compared against post-play participant funness judgments. \textbf{d,} Regression model fit to the post-play data. 95\% CIs are bootstrapped over participant ratings per games. \textbf{e,} Comparison of bootstrapped funness model parameter fits for the $41$ games using the just think human ratings (purple) versus the ratings provided by participants after one round of play (grey).}
    \label{fig:post-play-fun-model}
\end{figure}

To assess the relative contribution of each simulated feature in our funness model, we rerun the same regression modeling analyses. The refit post-play funness model derived from the Intuitive Gamer features again reaches the split-half noise ceiling, capturing essentially all of the explainable variance in participants' judgments (Figure~\ref{fig:post-play-fun-model}). Reward for thinking and game balance generally matter less for a fun game in people's post-play judgments compare to game length (Figure~\ref{fig:post-play-fun-model}); however, we caveat that the post-play judgments are generally more variable (as noted above in the split-halfs).

\begin{table}[!h]
\centering
\begin{tabular}[t]{lccc}
\toprule
Added Feature & F & p & $\Delta$AIC\\
\midrule
Board Size & 7.651 & 9.00e-03 & 6.2\\
Approx Novelty & 1.838 & 1.84e-01 & 0.1 \\
Binary Traits & 2.043 & 8.32e-02 & 2.5 \\
\bottomrule
\end{tabular}
\caption{\textbf{Assessing impact of non-simulation based features on funness model.} Comparing the inclusion of non-simulation based linguistic features into the funness model fit to peoples' post-play funness judgments. $\Delta$AIC is AIC$_\text{sim-only}$ - AIC$_\text{expanded}$ (higher indicates better from including the additional feature). There is a slight effect from incorporating either board size or all binary game traits (in contrast to adding an aggregate ``count'' of the number of traits that are ``on'' for any game, i.e., ``approximate novelty''). It is possible the potential benefit of incorporating board size is related to the seemingly higher role game length plays in people's funness judgments post-play.}
\label{tab:addtl-features-post}
\end{table}

\subsubsection{Evaluating games after one round of watching}

\begin{figure}
    \centering
    \includegraphics[width=1.0\linewidth]{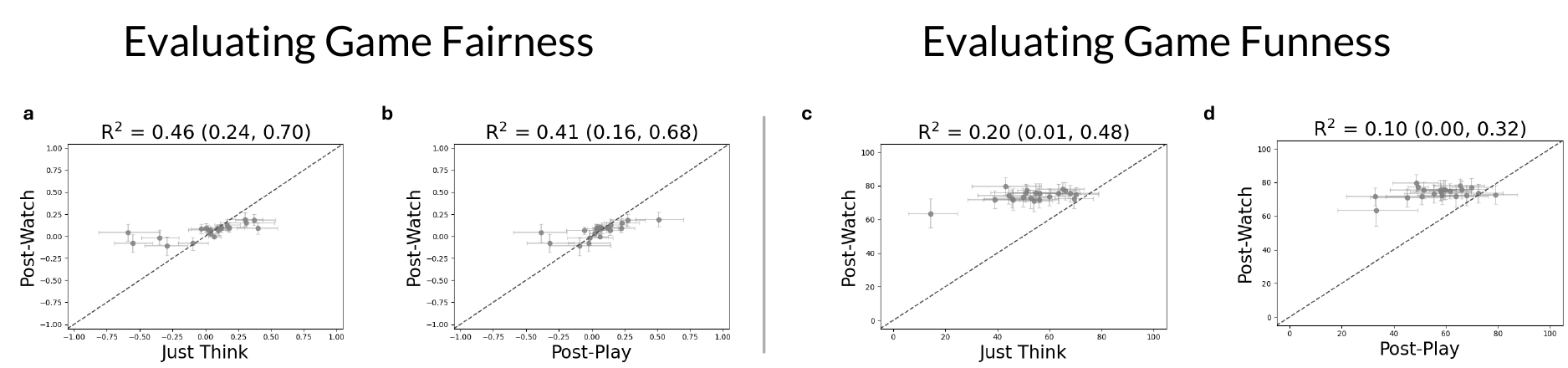}
    \caption{\textbf{Judgments after watching a match.} Comparing post-watch judgments of predicted payoff against those made in the \textbf{a,} ``just think'' and \textbf{b,} ``play'' experiments, on only the subset of $21$ games considered in the watch experiment, as well as the predicted funness for \textbf{c,} ``just think'' and \textbf{d,} after play. Participants in the post-watch experiment are generally more variable with each other; the noisier responses may be due to not having been held for $30$ to $60$ seconds before submitting their judgment like the other studies.}
    \label{fig:payoff-fun-w-watch}
\end{figure}

Participants' judgments after watching a single match are highly noisy for both payoff and funness queries (split-half $R^2 = 0.54$ [95\% CI: $0.30, 0.76$] and $R^2 = 0.07$ [95\% CI: $0.0001, 0.27$] for post-watch payoff and funness, respectively). In contrast to both the ``just think'' and post-play funness ratings which saw Tic-Tac-Toe rated around $50$ (average fun rating of $51.4 \pm 29.5$ SD for ``just think,'' and average fun rating of $49.4 \pm 21.8$ SD for post-play), the watchers rated Tic-Tac-Toe at an average of $77.6 \pm 16.6$ SD on funness. We observe that most funness scores are inflated and payoffs collapse towards zero (Figure~\ref{fig:payoff-fun-w-watch}).  We posit that the increased variability in the post-watch participants' judgments may have arisen from participants not having external pressure to ``force'' some degree of thinking. Notably, participants in the watch study were not held from submitting their answers until one minute passed as in the ``just think'' experiment (nor given an optional scratchpad) before making their judgments, which may have impacted effort. Participants in the post-play experiment also were required to spend time before responding (at least $30$ seconds); however, post-play participants were not given an optional scratchpad after they played their game. Future work should explore the role of forced response time on effort in encouraging thinking about game evaluations.

\section{Exploratory analyses with a intermediate depth model}

While our primary model focuses on highly computation-bounded, single-step look-ahead, our model class can naturally be extended to capture behavior of some game reasoners who may think more. We next consider a variant of the Intuitive Gamer model that incorporates the potential heuristic value of likely subsequent states in addition to the current state. This intermediate depth (Intuitive Gamer depth-3) model---so-called because it considers the current state, the opponent's likely response, and the possible responses to those actions---represents a moderate increase in computational load, though still requires far less computation than the Expert Gamer model.

\subsection{Model definition}

The primary difference of the Intuitive Gamer depth-3 model and our primary model is the value function. Concretely, the value assigned to a position under the Intuitive Gamer depth-3 model is as follows:
\begin{equation}
\label{eq:depth-3}
\tilde{\mathcal{V}}^3(s_t, a_t) = \tilde{\mathcal{V}}(s_t, a_t) + \beta \times \mathcal{V}^{\text{next}}(s_t, a_t).
\end{equation}

To compute $\mathcal{V}^{\text{next}}(s_t, a_t)$ we first apply the action $a_t$ to obtain the subsequent state $s_{t+1}$ and pass the turn to the opponent. We then assume that the opponent takes the most likely action under the Intuitive Gamer model, or the action that maximizes $\tilde{\mathcal{V}}(s_{t+1}, a_{t+1})$. This results in the state $s_{t+2}$ and returns play to the original player. We then say that
$$\tilde{\mathcal{V}}^{\text{next}}(s_t, a_t) = \text{max}_{a \in \mathcal{A}} \text{ } \tilde{\mathcal{V}}(s_{t+2}, a).$$

In special cases where either the original player or opponent move twice in a row, we either decrease or increase the depth of the simulation by one move. In each case, $s_{t+2}$ is the state in which the original player takes another turn. Actions are selected via softmax sampling over $\mathcal{V}^3$, as in the standard Intuitive Gamer model. For our experiments, we set the discount factor $\beta$ to 0.5.

\subsection{Results}

\begin{figure}
    \centering
    \includegraphics[width=0.5\linewidth]{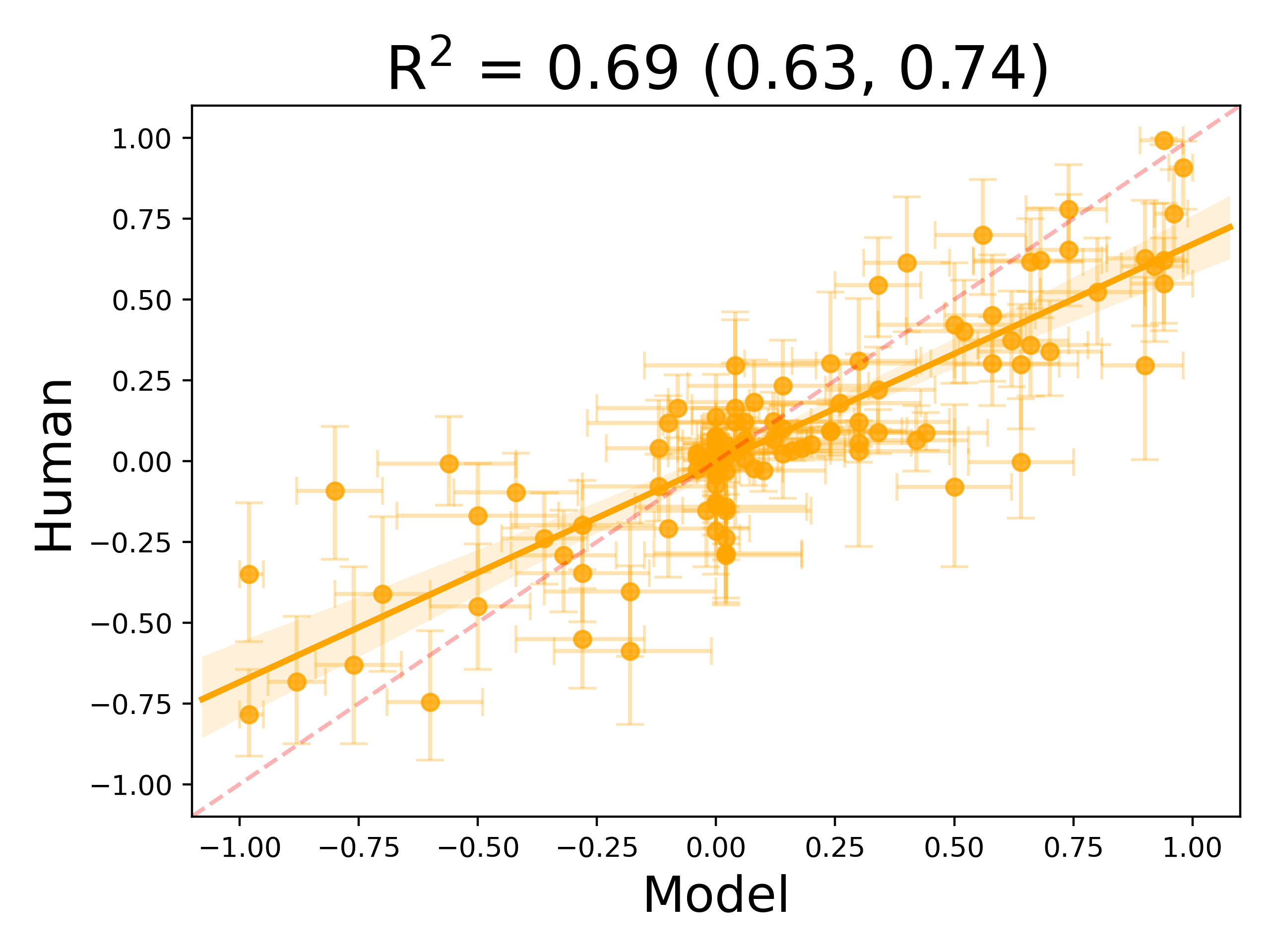}
    \caption{\textbf{Intermediate depth payoff predictions.} Non-flat (depth-$3$) variant of the Intuitive Gamer model's predicted payoffs for the $121$ games in the ``just think'' experiment, compared against people's predicted payoffs. Error bars depict 95\% CIs over bootstrapped human means and bootstrapped model-predicted payoffs under $k=6$ simulated games for $20$ simulated participants.}
    \label{fig:depth-3-predicted-payoff}
\end{figure}

\begin{figure}[h!]
    \centering
    \includegraphics[width=1.0\linewidth]{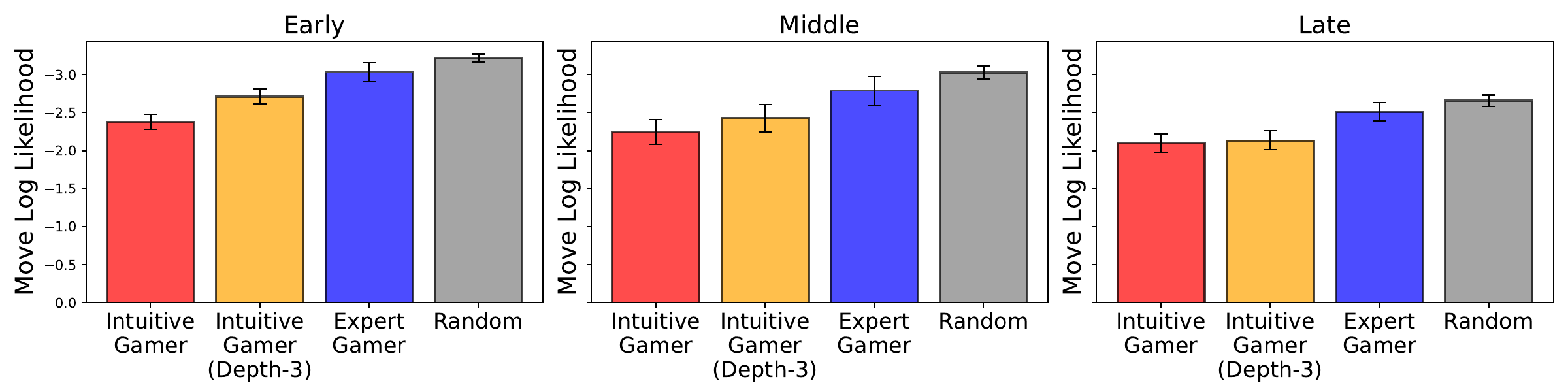}
    \caption{\textbf{``Human-human play'' analyses broken down by game stage.} The bar plots show the log likelihood of an observed subjects' move under our different models. Error bars depict 95\% bootstrapped confidence intervals around the mean log-likelihood over all games. }
\label{fig:game-stages-depth}
\end{figure}

\begin{figure}[h!]
    \centering
    \includegraphics[width=1.0\linewidth]{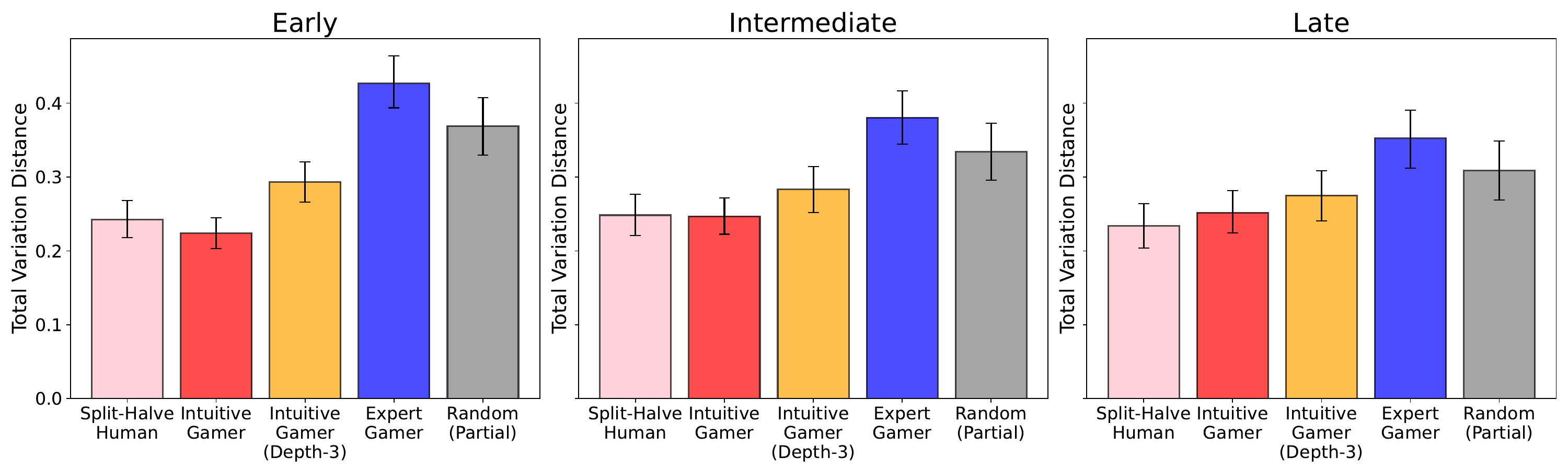}
    \caption{\textbf{``Watch-and-predict'' analyses broken down by game stage.} The bar plots show the Total Variation Distance (lower is better) of model distributions to the human watchers' distribution, as well as compared to the split-half human watcher TVD.}
\label{fig:game-stages-depth-watch}
\end{figure}

We report predicted payoffs under the Intuitive Gamer depth-$3$ model in \autoref{fig:depth-3-predicted-payoff}, as well as aggregate match log likelihoods and measure of distributional alignment (under TVD) to the play and ``watch-and-predict'' data, respectively (\autoref{fig:game-stages-depth} and \autoref{fig:game-stages-depth-watch}). As expected, this depth-$3$ model is a slightly worse fit to human behavior than the standard flat (shallow; depth-$1$) Intuitive Gamer, though it still outperforms the expert model in many cases. Notably, the differences between the depth-$3$ and depth-$1$ models largely disappear in board states toward the end of a game (see \autoref{fig:game-stages-depth}, right). This might be due to the fact that there are fewer possible moves in such states (making a deeper search easier to complete) or because the final outcome is more salient (encouraging the expenditure of more computational resources on search). We leave further analysis of these possibilities to future work.

\section{Game list}

We also include the full list of the $121$ games we explored in this study below. For each game, we include the average human funness rating and average human payoff prediction. We also include the averaged decomposed scores (i.e., $P(\text{draw})$ and $P(\text{P1 win})$, where $P(\text{win})$ is computed from each participants' $P(\text{draw})$ and $P(\text{P1 wins} | \text{not draw})$ as described in Section~\ref{sec:decomp-scores}.  

\begin{xltabular}{\textwidth}{
>{\centering\arraybackslash}p{1.2cm}X
>{\centering\arraybackslash}p{0.8cm}
>{\centering\arraybackslash}p{0.8cm}
>{\centering\arraybackslash}p{1cm}
>{\centering\arraybackslash}p{1.0cm}}
\toprule
\textbf{Board} & \textbf{Rules} & \textbf{Fun} & \textbf{Payoff} & \textbf{P(Draw)} & \textbf{P(P1 Wins)} \\
\midrule
\endhead
5x5 & 3 pieces in a row loses. & 72.4 & 0.2 & 0.4 & 0.4 \\
10x10 & 4 pieces in a row wins. & 70.2 & 0.1 & 0.5 & 0.3 \\
5x10 & 4 pieces in a row wins. & 70.1 & 0.1 & 0.6 & 0.3 \\
4x4 & 3 pieces in a row loses. & 69.4 & 0.0 & 0.5 & 0.2 \\
10x10 & 5 pieces in a row wins. The first player can place 2 pieces as their first move, while the second player can only place 1 piece as their first move. & 69.2 & 0.3 & 0.4 & 0.5 \\
5x5 & 3 pieces in a row wins. & 68.1 & 0.3 & 0.3 & 0.5 \\
10x10 & 4 pieces in a row wins. However, a player can only win by making a diagonal row. Horizontal and vertical rows do not count. & 66.7 & 0.2 & 0.6 & 0.3 \\
10x10 & 5 pieces in a row wins. & 66.2 & 0.2 & 0.5 & 0.4 \\
7x7 & 4 pieces in a row wins. & 65.1 & 0.2 & 0.5 & 0.4 \\
7x7 & 4 pieces in a row wins. The first player can place 2 pieces as their first move, while the second player can only place 1 piece as their first move. & 63.8 & 0.4 & 0.4 & 0.5 \\
10x10 & 5 pieces in a row wins. The second player can place 2 pieces as their first move, while the first player can only place 1 piece as their first move. & 63.6 & -0.1 & 0.5 & 0.2 \\
7x7 & 4 pieces in a row wins. However, a player can only win by making a diagonal row. Horizontal and vertical rows do not count. & 63.5 & 0.1 & 0.6 & 0.2 \\
5x5 & 4 pieces in a row loses. & 62.6 & 0.1 & 0.7 & 0.2 \\
10x10 & 5 pieces in a row wins. However, a player can only win by making a diagonal row. Horizontal and vertical rows do not count. & 61.8 & 0.1 & 0.6 & 0.2 \\
4x12 & 4 pieces in a row wins. & 61.6 & 0.1 & 0.6 & 0.3 \\
InfxInf & 5 pieces in a row wins. & 61.3 & 0.1 & 0.3 & 0.4 \\
7x7 & Each player needs 4 pieces in a row to win. The first player cannot win by making a diagonal row (only horizontal and vertical rows count), but the second player does not have this restriction. & 61.1 & -0.1 & 0.5 & 0.2 \\
10x10 & 4 pieces in a row wins. The first player can place 2 pieces as their first move, while the second player can only place 1 piece as their first move. & 60.2 & 0.6 & 0.2 & 0.7 \\
5x10 & 5 pieces in a row wins. & 60.0 & 0.0 & 0.8 & 0.1 \\
4x4 & 3 pieces in a row wins. & 59.5 & 0.5 & 0.3 & 0.6 \\
10x10 & 3 pieces in a row loses. & 59.4 & 0.0 & 0.4 & 0.3 \\
7x7 & 4 pieces in a row loses. & 58.9 & -0.0 & 0.6 & 0.2 \\
10x10 & 4 pieces in a row wins. The second player can place 2 pieces as their first move, while the first player can only place 1 piece as their first move. & 58.4 & -0.2 & 0.5 & 0.2 \\
10x10 & 5 pieces in a row loses. & 57.6 & 0.1 & 0.5 & 0.3 \\
10x10 & 4 pieces in a row wins. However, a player cannot win by making a diagonal row. Only horizontal and vertical rows count. & 57.0 & 0.2 & 0.5 & 0.4 \\
10x10 & Each player needs 5 pieces in a row to win. The first player cannot win by making a diagonal row (only horizontal and vertical rows count), but the second player does not have this restriction. & 56.4 & -0.3 & 0.4 & 0.1 \\
5x5 & The first player needs 4 pieces in a row to win, but the second player only needs 3 pieces in a row to win. & 56.2 & -0.6 & 0.2 & 0.1 \\
5x5 & 4 pieces in a row wins. & 54.8 & 0.1 & 0.7 & 0.2 \\
5x5 & 3 pieces in a row wins. However, a player cannot win by making a diagonal row. Only horizontal and vertical rows count. & 54.6 & 0.3 & 0.4 & 0.5 \\
10x10 & 5 pieces in a row wins. However, a player cannot win by making a diagonal row. Only horizontal and vertical rows count. & 54.5 & 0.1 & 0.7 & 0.2 \\
5x5 & 4 pieces in a row wins. The first player can place 2 pieces as their first move, while the second player can only place 1 piece as their first move. & 54.1 & 0.3 & 0.4 & 0.4 \\
5x5 & 3 pieces in a row wins. However, a player can only win by making a diagonal row. Horizontal and vertical rows do not count. & 53.9 & 0.1 & 0.5 & 0.3 \\
10x10 & Each player needs 4 pieces in a row to win. The first player cannot win by making a diagonal row (only horizontal and vertical rows count), but the second player does not have this restriction. & 52.9 & -0.2 & 0.4 & 0.2 \\
4x9 & 4 pieces in a row wins. & 52.8 & 0.0 & 0.5 & 0.2 \\
5x5 & Each player needs 3 pieces in a row to win. The first player can only win by making a diagonal row, but the second player does not have this restriction. & 52.6 & 0.1 & 0.2 & 0.5 \\
10x10 & 6 pieces in a row wins. & 51.8 & 0.0 & 0.7 & 0.2 \\
10x10 & 3 pieces in a row wins. However, a player cannot win by making a diagonal row. Only horizontal and vertical rows count. & 51.4 & 0.5 & 0.2 & 0.7 \\
3x3 & 3 pieces in a row wins. & 51.4 & 0.1 & 0.7 & 0.2 \\
3x5 & 3 pieces in a row wins. & 51.1 & 0.2 & 0.6 & 0.3 \\
7x7 & 4 pieces in a row wins. However, a player cannot win by making a diagonal row. Only horizontal and vertical rows count. & 50.9 & 0.1 & 0.5 & 0.3 \\
4x6 & 4 pieces in a row wins. & 50.6 & 0.1 & 0.7 & 0.2 \\
7x7 & 4 pieces in a row wins. The second player can place 2 pieces as their first move, while the first player can only place 1 piece as their first move. & 50.5 & -0.0 & 0.4 & 0.3 \\
10x10 & Each player needs 5 pieces in a row to win. The first player can only win by making a diagonal row, but the second player does not have this restriction. & 50.3 & -0.2 & 0.4 & 0.2 \\
3x3 & 3 pieces in a row wins. The second player can place 2 pieces as their first move, while the first player can only place 1 piece as their first move. & 50.1 & -0.2 & 0.5 & 0.2 \\
4x4 & 3 pieces in a row wins. However, a player cannot win by making a diagonal row. Only horizontal and vertical rows count. & 50.0 & 0.4 & 0.3 & 0.5 \\
10x10 & 8 pieces in a row wins. & 49.2 & 0.0 & 0.7 & 0.2 \\
4x4 & Each player needs 3 pieces in a row to win. The first player cannot win by making a diagonal row (only horizontal and vertical rows count), but the second player does not have this restriction. & 49.0 & -0.1 & 0.4 & 0.3 \\
5x5 & Each player needs 4 pieces in a row to win. The first player cannot win by making a diagonal row (only horizontal and vertical rows count), but the second player does not have this restriction. & 48.6 & -0.0 & 0.7 & 0.2 \\
10x10 & Each player needs 3 pieces in a row to win. The first player can only win by making a diagonal row, but the second player does not have this restriction. & 48.2 & 0.0 & 0.2 & 0.4 \\
5x5 & 4 pieces in a row wins. The second player can place 2 pieces as their first move, while the first player can only place 1 piece as their first move. & 46.0 & -0.1 & 0.7 & 0.1 \\
5x5 & 5 pieces in a row wins. The second player can place 2 pieces as their first move, while the first player can only place 1 piece as their first move. & 45.9 & -0.1 & 0.7 & 0.1 \\
7x7 & Each player needs 4 pieces in a row to win. The first player can only win by making a diagonal row, but the second player does not have this restriction. & 45.6 & -0.3 & 0.5 & 0.1 \\
10x10 & 3 pieces in a row wins. However, a player can only win by making a diagonal row. Horizontal and vertical rows do not count. & 44.9 & 0.3 & 0.4 & 0.5 \\
3x10 & 3 pieces in a row wins. & 44.5 & 0.4 & 0.3 & 0.5 \\
10x10 & 4 pieces in a row loses. & 44.4 & 0.0 & 0.5 & 0.3 \\
5x5 & The first player needs 5 pieces in a row to win, but the second player only needs 4 pieces in a row to win. & 44.4 & -0.2 & 0.5 & 0.1 \\
3x3 & 3 pieces in a row loses. & 43.2 & -0.0 & 0.8 & 0.1 \\
5x5 & 5 pieces in a row wins. However, a player cannot win by making a diagonal row. Only horizontal and vertical rows count. & 41.7 & 0.0 & 0.9 & 0.0 \\
8x8 & 3 pieces in a row wins. & 41.3 & 0.5 & 0.2 & 0.7 \\
10x10 & The first player needs 10 pieces in a row to win, but the second player only needs 9 pieces in a row to win. & 41.2 & -0.0 & 0.7 & 0.1 \\
4x4 & 3 pieces in a row wins. However, a player can only win by making a diagonal row. Horizontal and vertical rows do not count. & 41.2 & 0.1 & 0.8 & 0.1 \\
10x10 & 7 pieces in a row wins. & 40.9 & 0.1 & 0.7 & 0.2 \\
10x10 & Each player needs 4 pieces in a row to win. The first player can only win by making a diagonal row, but the second player does not have this restriction. & 40.9 & -0.2 & 0.4 & 0.2 \\
5x5 & Each player needs 3 pieces in a row to win. The first player cannot win by making a diagonal row (only horizontal and vertical rows count), but the second player does not have this restriction. & 40.6 & -0.0 & 0.4 & 0.3 \\
InfxInf & 3 pieces in a row wins. & 40.1 & 0.6 & 0.2 & 0.7 \\
10x10 & 10 pieces in a row wins. However, a player cannot win by making a diagonal row. Only horizontal and vertical rows count. & 40.1 & 0.0 & 0.8 & 0.1 \\
4x6 & 3 pieces in a row wins. & 39.7 & 0.6 & 0.2 & 0.7 \\
10x10 & 3 pieces in a row wins. & 39.2 & 0.6 & 0.1 & 0.7 \\
7x7 & The first player needs 4 pieces in a row to win, but the second player only needs 3 pieces in a row to win. & 39.1 & -0.6 & 0.2 & 0.1 \\
4x12 & 5 pieces in a row wins. & 38.3 & 0.0 & 0.8 & 0.1 \\
10x10 & The first player needs 5 pieces in a row to win, but the second player only needs 4 pieces in a row to win. & 38.1 & -0.1 & 0.3 & 0.3 \\
9x9 & 3 pieces in a row wins. & 37.9 & 0.6 & 0.1 & 0.7 \\
5x5 & 5 pieces in a row loses. & 37.2 & 0.0 & 0.7 & 0.1 \\
10x10 & The first player needs 4 pieces in a row to win, but the second player only needs 3 pieces in a row to win. & 37.2 & -0.4 & 0.2 & 0.2 \\
10x10 & 10 pieces in a row wins. The first player can place 2 pieces as their first move, while the second player can only place 1 piece as their first move. & 36.8 & 0.1 & 0.7 & 0.2 \\
7x7 & 3 pieces in a row wins. & 36.4 & 0.3 & 0.3 & 0.5 \\
10x10 & 10 pieces in a row wins. The second player can place 2 pieces as their first move, while the first player can only place 1 piece as their first move. & 36.2 & 0.0 & 0.6 & 0.2 \\
5x5 & Each player needs 5 pieces in a row to win. The first player cannot win by making a diagonal row (only horizontal and vertical rows count), but the second player does not have this restriction. & 35.7 & -0.1 & 0.6 & 0.1 \\
10x10 & 9 pieces in a row wins. & 35.6 & 0.0 & 0.8 & 0.1 \\
5x5 & Each player needs 4 pieces in a row to win. The first player can only win by making a diagonal row, but the second player does not have this restriction. & 35.4 & -0.3 & 0.6 & 0.1 \\
5x5 & 5 pieces in a row wins. & 35.4 & 0.1 & 0.7 & 0.2 \\
3x3 & 3 pieces in a row wins. However, a player cannot win by making a diagonal row. Only horizontal and vertical rows count. & 35.0 & 0.0 & 0.9 & 0.1 \\
5x5 & 4 pieces in a row wins. However, a player can only win by making a diagonal row. Horizontal and vertical rows do not count. & 34.8 & 0.0 & 0.8 & 0.1 \\
4x4 & Each player needs 3 pieces in a row to win. The first player can only win by making a diagonal row, but the second player does not have this restriction. & 34.0 & -0.3 & 0.6 & 0.1 \\
10x10 & Each player needs 3 pieces in a row to win. The first player cannot win by making a diagonal row (only horizontal and vertical rows count), but the second player does not have this restriction. & 33.9 & 0.3 & 0.2 & 0.5 \\
5x10 & 6 pieces in a row wins. & 33.9 & 0.0 & 0.7 & 0.2 \\
InfxInf & 10 pieces in a row wins. & 33.9 & 0.1 & 0.5 & 0.3 \\
5x5 & 5 pieces in a row wins. The first player can place 2 pieces as their first move, while the second player can only place 1 piece as their first move. & 33.8 & 0.0 & 0.8 & 0.1 \\
10x10 & Each player needs 10 pieces in a row to win. The first player cannot win by making a diagonal row (only horizontal and vertical rows count), but the second player does not have this restriction. & 32.9 & -0.0 & 0.8 & 0.1 \\
6x6 & 3 pieces in a row wins. & 32.4 & 0.4 & 0.3 & 0.5 \\
10x10 & 10 pieces in a row loses. & 32.3 & -0.0 & 0.9 & 0.0 \\
5x5 & 3 pieces in a row wins. The second player can place 2 pieces as their first move, while the first player can only place 1 piece as their first move. & 32.2 & -0.4 & 0.3 & 0.1 \\
2x10 & 3 pieces in a row wins. & 32.0 & 0.1 & 0.7 & 0.2 \\
4x4 & 3 pieces in a row wins. The first player can place 2 pieces as their first move, while the second player can only place 1 piece as their first move. & 32.0 & 0.7 & 0.1 & 0.8 \\
10x10 & 3 pieces in a row wins. The second player can place 2 pieces as their first move, while the first player can only place 1 piece as their first move. & 31.8 & -0.3 & 0.3 & 0.2 \\
3x3 & Each player needs 3 pieces in a row to win. The first player cannot win by making a diagonal row (only horizontal and vertical rows count), but the second player does not have this restriction. & 30.3 & -0.0 & 0.7 & 0.1 \\
4x9 & 5 pieces in a row wins. & 29.9 & 0.0 & 0.8 & 0.1 \\
5x5 & 4 pieces in a row wins. However, a player cannot win by making a diagonal row. Only horizontal and vertical rows count. & 29.2 & 0.1 & 0.6 & 0.2 \\
4x6 & 5 pieces in a row wins. & 29.0 & -0.0 & 0.9 & 0.0 \\
10x10 & 3 pieces in a row wins. The first player can place 2 pieces as their first move, while the second player can only place 1 piece as their first move. & 28.8 & 0.8 & 0.1 & 0.8 \\
4x4 & 3 pieces in a row wins. The second player can place 2 pieces as their first move, while the first player can only place 1 piece as their first move. & 28.1 & -0.5 & 0.3 & 0.1 \\
10x10 & 10 pieces in a row wins. & 26.1 & 0.0 & 0.9 & 0.1 \\
5x5 & Each player needs 5 pieces in a row to win. The first player can only win by making a diagonal row, but the second player does not have this restriction. & 25.8 & -0.1 & 0.7 & 0.1 \\
3x3 & 3 pieces in a row wins. However, a player can only win by making a diagonal row. Horizontal and vertical rows do not count. & 25.1 & 0.0 & 0.9 & 0.1 \\
5x5 & 5 pieces in a row wins. However, a player can only win by making a diagonal row. Horizontal and vertical rows do not count. & 25.1 & 0.0 & 0.8 & 0.1 \\
10x10 & The first player needs 3 pieces in a row to win, but the second player only needs 2 pieces in a row to win. & 24.7 & -0.8 & 0.1 & 0.1 \\
3x3 & 3 pieces in a row wins. The first player can place 2 pieces as their first move, while the second player can only place 1 piece as their first move. & 24.6 & 0.5 & 0.3 & 0.6 \\
4x4 & The first player needs 3 pieces in a row to win, but the second player only needs 2 pieces in a row to win. & 23.2 & -0.6 & 0.1 & 0.2 \\
3x3 & 2 pieces in a row wins. & 22.5 & 0.8 & 0.1 & 0.8 \\
3x3 & Each player needs 3 pieces in a row to win. The first player can only win by making a diagonal row, but the second player does not have this restriction. & 21.1 & -0.2 & 0.7 & 0.0 \\
5x5 & 3 pieces in a row wins. The first player can place 2 pieces as their first move, while the second player can only place 1 piece as their first move. & 20.8 & 0.7 & 0.2 & 0.7 \\
2x5 & 3 pieces in a row wins. & 20.7 & 0.0 & 0.8 & 0.1 \\
10x10 & 10 pieces in a row wins. However, a player can only win by making a diagonal row. Horizontal and vertical rows do not count. & 20.4 & 0.0 & 0.7 & 0.1 \\
10x10 & Each player needs 10 pieces in a row to win. The first player can only win by making a diagonal row, but the second player does not have this restriction. & 20.2 & -0.2 & 0.6 & 0.1 \\
3x3 & The first player needs 3 pieces in a row to win, but the second player only needs 2 pieces in a row to win. & 16.1 & -0.7 & 0.1 & 0.1 \\
1x10 & 3 pieces in a row wins. & 14.4 & 0.0 & 0.8 & 0.1 \\
1x5 & 3 pieces in a row wins. & 11.8 & 0.0 & 0.9 & 0.1 \\
10x10 & 2 pieces in a row wins. & 11.8 & 0.9 & 0.0 & 1.0 \\
5x5 & The first player needs 3 pieces in a row to win, but the second player only needs 2 pieces in a row to win. & 8.8 & -0.7 & 0.1 & 0.1 \\
1x5 & 2 pieces in a row wins. & 8.5 & 0.6 & 0.3 & 0.6 \\
5x5 & 2 pieces in a row wins. & 6.5 & 1.0 & 0.0 & 1.0 \\
\bottomrule
\end{xltabular}

\putbib[main]  
\end{bibunit}


\begin{thebibliography}{10}
\expandafter\ifx\csname url\endcsname\relax
  \def\url#1{\texttt{#1}}\fi
\expandafter\ifx\csname urlprefix\endcsname\relax\def\urlprefix{URL }\fi
\providecommand{\bibinfo}[2]{#2}
\providecommand{\eprint}[2][]{\url{#2}}

\bibitem{chase1973mind}
\bibinfo{author}{Chase, W.~G.} \& \bibinfo{author}{Simon, H.~A.}
\newblock \bibinfo{title}{The mind's eye in chess}.
\newblock In \emph{\bibinfo{booktitle}{Visual information processing}}, \bibinfo{pages}{215--281} (\bibinfo{publisher}{Elsevier}, \bibinfo{year}{1973}).

\bibitem{campbell2002deep}
\bibinfo{author}{Campbell, M.}, \bibinfo{author}{Hoane~Jr, A.~J.} \& \bibinfo{author}{Hsu, F.-h.}
\newblock \bibinfo{title}{{Deep Blue}}.
\newblock \emph{\bibinfo{journal}{{Artificial Intelligence}}} \textbf{\bibinfo{volume}{134}}, \bibinfo{pages}{57--83} (\bibinfo{year}{2002}).

\bibitem{gobet2004moves}
\bibinfo{author}{Gobet, F.}, \bibinfo{author}{Retschitzki, J.} \& \bibinfo{author}{de~Voogt, A.}
\newblock \emph{\bibinfo{title}{Moves in mind: The psychology of board games}} (\bibinfo{publisher}{Psychology Press}, \bibinfo{year}{2004}).

\bibitem{silver2016mastering}
\bibinfo{author}{Silver, D.} \emph{et~al.}
\newblock \bibinfo{title}{Mastering the game of {Go} with deep neural networks and tree search}.
\newblock \emph{\bibinfo{journal}{Nature}} \textbf{\bibinfo{volume}{529}}, \bibinfo{pages}{484--489} (\bibinfo{year}{2016}).

\bibitem{silver2017mastering_go}
\bibinfo{author}{Silver, D.} \emph{et~al.}
\newblock \bibinfo{title}{Mastering the game of go without human knowledge}.
\newblock \emph{\bibinfo{journal}{Nature}} \textbf{\bibinfo{volume}{550}}, \bibinfo{pages}{354--359} (\bibinfo{year}{2017}).

\bibitem{van2023expertise}
\bibinfo{author}{van Opheusden, B.} \emph{et~al.}
\newblock \bibinfo{title}{Expertise increases planning depth in human gameplay}.
\newblock \emph{\bibinfo{journal}{Nature}} \bibinfo{pages}{1--6} (\bibinfo{year}{2023}).

\bibitem{cleveland1907psychology}
\bibinfo{author}{Cleveland, A.~A.}
\newblock \bibinfo{title}{The psychology of chess and of learning to play it}.
\newblock \emph{\bibinfo{journal}{The American Journal of Psychology}} \textbf{\bibinfo{volume}{18}}, \bibinfo{pages}{269--308} (\bibinfo{year}{1907}).

\bibitem{shannon1950}
\bibinfo{author}{Shannon, C.~E.}
\newblock \bibinfo{title}{Xxii. programming a computer for playing chess}.
\newblock \emph{\bibinfo{journal}{The London, Edinburgh, and Dublin Philosophical Magazine and Journal of Science}} \textbf{\bibinfo{volume}{41}}, \bibinfo{pages}{256--275} (\bibinfo{year}{1950}).

\bibitem{newell1958chess}
\bibinfo{author}{Newell, A.}, \bibinfo{author}{Shaw, J.~C.} \& \bibinfo{author}{Simon, H.~A.}
\newblock \bibinfo{title}{Chess-playing programs and the problem of complexity}.
\newblock \emph{\bibinfo{journal}{IBM Journal of Research and Development}} \textbf{\bibinfo{volume}{2}}, \bibinfo{pages}{320--335} (\bibinfo{year}{1958}).

\bibitem{mnih2015human}
\bibinfo{author}{Mnih, V.} \emph{et~al.}
\newblock \bibinfo{title}{Human-level control through deep reinforcement learning}.
\newblock \emph{\bibinfo{journal}{nature}} \textbf{\bibinfo{volume}{518}}, \bibinfo{pages}{529--533} (\bibinfo{year}{2015}).

\bibitem{tsividis2021human}
\bibinfo{author}{Tsividis, P.~A.} \emph{et~al.}
\newblock \bibinfo{title}{Human-level reinforcement learning through theory-based modeling, exploration, and planning}.
\newblock \emph{\bibinfo{journal}{arXiv preprint arXiv:2107.12544}}  (\bibinfo{year}{2021}).

\bibitem{yannakakis2018artificial}
\bibinfo{author}{Yannakakis, G.~N.} \& \bibinfo{author}{Togelius, J.}
\newblock \emph{\bibinfo{title}{{Artificial Intelligence and Games}}} (\bibinfo{publisher}{Springer}, \bibinfo{year}{2018}).

\bibitem{simon1988skill}
\bibinfo{author}{Simon, H.} \& \bibinfo{author}{Chase, W.}
\newblock \bibinfo{title}{Skill in chess}.
\newblock In \emph{\bibinfo{booktitle}{Computer chess compendium}}, \bibinfo{pages}{175--188} (\bibinfo{publisher}{Springer}, \bibinfo{year}{1988}).

\bibitem{silver2017mastering}
\bibinfo{author}{Silver, D.} \emph{et~al.}
\newblock \bibinfo{title}{Mastering chess and shogi by self-play with a general reinforcement learning algorithm}.
\newblock \emph{\bibinfo{journal}{arXiv preprint arXiv:1712.01815}}  (\bibinfo{year}{2017}).

\bibitem{meta2022human}
\bibinfo{author}{FAIR} \emph{et~al.}
\newblock \bibinfo{title}{Human-level play in the game of diplomacy by combining language models with strategic reasoning}.
\newblock \emph{\bibinfo{journal}{Science}} \textbf{\bibinfo{volume}{378}}, \bibinfo{pages}{1067--1074} (\bibinfo{year}{2022}).

\bibitem{newell_simon1972human}
\bibinfo{author}{Newell, A.} \& \bibinfo{author}{Simon, H.~A.}
\newblock \emph{\bibinfo{title}{Human problem solving}} (\bibinfo{publisher}{Prentice-Hall}, \bibinfo{address}{Englewood Cliffs, NJ}, \bibinfo{year}{1972}).

\bibitem{charness1991expertise}
\bibinfo{author}{Charness, N.}
\newblock \bibinfo{title}{Expertise in chess: The balance between knowledge and search}.
\newblock In \bibinfo{editor}{Ericsson, K.~A.} \& \bibinfo{editor}{Smith, J.} (eds.) \emph{\bibinfo{booktitle}{Toward a general theory of expertise: Prospects and limits}}, \bibinfo{pages}{39--63} (\bibinfo{publisher}{{Cambridge University Press}}, \bibinfo{address}{Cambridge}, \bibinfo{year}{1991}).

\bibitem{kocsis2006bandit}
\bibinfo{author}{Kocsis, L.} \& \bibinfo{author}{Szepesvari, C.}
\newblock \bibinfo{title}{Bandit based {Monte-Carlo} planning}.
\newblock In \emph{\bibinfo{booktitle}{{European Conference on Machine Learning}}}, \bibinfo{pages}{282--293} (\bibinfo{organization}{Springer}, \bibinfo{year}{2006}).

\bibitem{gigerenzer2004fast}
\bibinfo{author}{Gigerenzer, G.}
\newblock \bibinfo{title}{Fast and frugal heuristics: The tools of bounded rationality}.
\newblock \emph{\bibinfo{journal}{Blackwell handbook of judgment and decision making}} \textbf{\bibinfo{volume}{62}}, \bibinfo{pages}{88} (\bibinfo{year}{2004}).

\bibitem{lieder2020resource}
\bibinfo{author}{Lieder, F.} \& \bibinfo{author}{Griffiths, T.~L.}
\newblock \bibinfo{title}{Resource-rational analysis: Understanding human cognition as the optimal use of limited computational resources}.
\newblock \emph{\bibinfo{journal}{Behavioral and Brain Sciences}} \textbf{\bibinfo{volume}{43}}, \bibinfo{pages}{e1} (\bibinfo{year}{2020}).

\bibitem{griffiths2008categorization}
\bibinfo{author}{Griffiths, T.~L.}, \bibinfo{author}{Sanborn, A.~N.}, \bibinfo{author}{Canini, K.~R.} \& \bibinfo{author}{Navarro, D.~J.}
\newblock \bibinfo{title}{Categorization as nonparametric {Bayesian} density estimation}.
\newblock \emph{\bibinfo{journal}{The probabilistic mind: Prospects for {Bayesian} cognitive science}} \bibinfo{pages}{303--328} (\bibinfo{year}{2008}).

\bibitem{sanborn2010rational}
\bibinfo{author}{Sanborn, A.~N.}, \bibinfo{author}{Griffiths, T.~L.} \& \bibinfo{author}{Navarro, D.~J.}
\newblock \bibinfo{title}{Rational approximations to rational models: alternative algorithms for category learning.}
\newblock \emph{\bibinfo{journal}{Psychological review}} \textbf{\bibinfo{volume}{117}}, \bibinfo{pages}{1144} (\bibinfo{year}{2010}).

\bibitem{vul2014one}
\bibinfo{author}{Vul, E.}, \bibinfo{author}{Goodman, N.}, \bibinfo{author}{Griffiths, T.~L.} \& \bibinfo{author}{Tenenbaum, J.~B.}
\newblock \bibinfo{title}{One and done? {Optimal} decisions from very few samples}.
\newblock \emph{\bibinfo{journal}{Cognitive science}} \textbf{\bibinfo{volume}{38}}, \bibinfo{pages}{599--637} (\bibinfo{year}{2014}).

\bibitem{icard2016subjective}
\bibinfo{author}{Icard, T.}
\newblock \bibinfo{title}{Subjective probability as sampling propensity}.
\newblock \emph{\bibinfo{journal}{Review of Philosophy and Psychology}} \textbf{\bibinfo{volume}{7}}, \bibinfo{pages}{863--903} (\bibinfo{year}{2016}).

\bibitem{zhu2020bayesian}
\bibinfo{author}{Zhu, J.-Q.}, \bibinfo{author}{Sanborn, A.~N.} \& \bibinfo{author}{Chater, N.}
\newblock \bibinfo{title}{The bayesian sampler: Generic bayesian inference causes incoherence in human probability judgments.}
\newblock \emph{\bibinfo{journal}{Psychological review}} \textbf{\bibinfo{volume}{127}}, \bibinfo{pages}{719} (\bibinfo{year}{2020}).

\bibitem{klein1993recognition}
\bibinfo{author}{Klein, G.~A.} \emph{et~al.}
\newblock \bibinfo{title}{A recognition-primed decision (rpd) model of rapid decision making}  (\bibinfo{year}{1993}).

\bibitem{coulom2006efficient}
\bibinfo{author}{Coulom, R.}
\newblock \bibinfo{title}{Efficient selectivity and backup operators in monte-carlo tree search}.
\newblock In \emph{\bibinfo{booktitle}{International conference on computers and games}}, \bibinfo{pages}{72--83} (\bibinfo{organization}{Springer}, \bibinfo{year}{2006}).

\bibitem{genesereth2014general}
\bibinfo{author}{Genesereth, M.} \& \bibinfo{author}{Thielscher, M.}
\newblock \emph{\bibinfo{title}{General Game Playing}}.
\newblock Synthesis Lectures on Artificial Intelligence and Machine Learning (\bibinfo{publisher}{Morgan \& Claypool Publishers}, \bibinfo{year}{2014}).

\bibitem{crowley1993flexible}
\bibinfo{author}{Crowley, K.} \& \bibinfo{author}{Siegler, R.~S.}
\newblock \bibinfo{title}{Flexible strategy use in young children's tic-tac-toe}.
\newblock \emph{\bibinfo{journal}{Cognitive Science}} \textbf{\bibinfo{volume}{17}}, \bibinfo{pages}{531--561} (\bibinfo{year}{1993}).

\bibitem{amir2022adaptive}
\bibinfo{author}{Amir, O.}, \bibinfo{author}{Tyomkin, L.} \& \bibinfo{author}{Hart, Y.}
\newblock \bibinfo{title}{Adaptive search space pruning in complex strategic problems}.
\newblock \emph{\bibinfo{journal}{PLoS Computational Biology}} \textbf{\bibinfo{volume}{18}}, \bibinfo{pages}{e1010358} (\bibinfo{year}{2022}).

\bibitem{myerson1983mechanism}
\bibinfo{author}{Myerson, R.~B.}
\newblock \bibinfo{title}{Mechanism design by an informed principal}.
\newblock \emph{\bibinfo{journal}{Econometrica: Journal of the Econometric Society}} \bibinfo{pages}{1767--1797} (\bibinfo{year}{1983}).

\bibitem{koster2022human}
\bibinfo{author}{Koster, R.} \emph{et~al.}
\newblock \bibinfo{title}{Human-centred mechanism design with democratic ai}.
\newblock \emph{\bibinfo{journal}{Nature Human Behaviour}} \textbf{\bibinfo{volume}{6}}, \bibinfo{pages}{1398--1407} (\bibinfo{year}{2022}).

\bibitem{manning2026general}
\bibinfo{author}{Manning, B.~S.} \& \bibinfo{author}{Horton, J.~J.}
\newblock \bibinfo{title}{General social agents}.
\newblock \bibinfo{type}{Tech. Rep.}, \bibinfo{institution}{National Bureau of Economic Research} (\bibinfo{year}{2026}).

\bibitem{camerer2003behavioral}
\bibinfo{author}{Camerer, C.~F.}
\newblock \emph{\bibinfo{title}{Behavioral Game Theory: Experiments in Strategic Interaction}} (\bibinfo{publisher}{Princeton University Press}, \bibinfo{address}{Princeton, NJ}, \bibinfo{year}{2003}).

\bibitem{carroll2019utility}
\bibinfo{author}{Carroll, M.} \emph{et~al.}
\newblock \bibinfo{title}{On the utility of learning about humans for human-{AI} coordination}.
\newblock \emph{\bibinfo{journal}{Advances in neural information processing systems}} \textbf{\bibinfo{volume}{32}} (\bibinfo{year}{2019}).

\bibitem{collins2024building}
\bibinfo{author}{Collins, K.~M.} \emph{et~al.}
\newblock \bibinfo{title}{Building machines that learn and think with people}.
\newblock \emph{\bibinfo{journal}{Nature Human Behavior}}  (\bibinfo{year}{2024}).

\bibitem{lantz2017depth}
\bibinfo{author}{Lantz, F.}, \bibinfo{author}{Isaksen, A.}, \bibinfo{author}{Jaffe, A.}, \bibinfo{author}{Nealen, A.} \& \bibinfo{author}{Togelius, J.}
\newblock \bibinfo{title}{Depth in strategic games.}
\newblock In \emph{\bibinfo{booktitle}{AAAI Workshops}} (\bibinfo{year}{2017}).

\bibitem{ryan2006motivational}
\bibinfo{author}{Ryan, R.~M.}, \bibinfo{author}{Rigby, C.~S.} \& \bibinfo{author}{Przybylski, A.}
\newblock \bibinfo{title}{The motivational pull of video games: A self-determination theory approach}.
\newblock \emph{\bibinfo{journal}{Motivation and emotion}} \textbf{\bibinfo{volume}{30}}, \bibinfo{pages}{344--360} (\bibinfo{year}{2006}).

\bibitem{brandle2025leveling}
\bibinfo{author}{Br{\"a}ndle, F.}, \bibinfo{author}{Wu, C.~M.} \& \bibinfo{author}{Schulz, E.}
\newblock \bibinfo{title}{Leveling up fun: Learning progress, expectations, and success influence enjoyment in video games}.
\newblock \emph{\bibinfo{journal}{Scientific Reports}} \textbf{\bibinfo{volume}{15}}, \bibinfo{pages}{34153} (\bibinfo{year}{2025}).

\bibitem{chu2023praise}
\bibinfo{author}{Chu, J.}, \bibinfo{author}{Tenenbaum, J.~B.} \& \bibinfo{author}{Schulz, L.~E.}
\newblock \bibinfo{title}{In praise of folly: flexible goals and human cognition}.
\newblock \emph{\bibinfo{journal}{Trends in Cognitive Sciences}}  (\bibinfo{year}{2023}).

\bibitem{lehrach2025code}
\bibinfo{author}{Lehrach, W.} \emph{et~al.}
\newblock \bibinfo{title}{Code world models for general game playing}.
\newblock \emph{\bibinfo{journal}{arXiv preprint arXiv:2510.04542}}  (\bibinfo{year}{2025}).

\bibitem{baker2017rational}
\bibinfo{author}{Baker, C.~L.}, \bibinfo{author}{Jara-Ettinger, J.}, \bibinfo{author}{Saxe, R.} \& \bibinfo{author}{Tenenbaum, J.~B.}
\newblock \bibinfo{title}{Rational quantitative attribution of beliefs, desires and percepts in human mentalizing}.
\newblock \emph{\bibinfo{journal}{Nature Human Behaviour}} \textbf{\bibinfo{volume}{1}}, \bibinfo{pages}{0064} (\bibinfo{year}{2017}).

\bibitem{ullman2017mind}
\bibinfo{author}{Ullman, T.~D.}, \bibinfo{author}{Spelke, E.}, \bibinfo{author}{Battaglia, P.} \& \bibinfo{author}{Tenenbaum, J.~B.}
\newblock \bibinfo{title}{Mind games: Game engines as an architecture for intuitive physics}.
\newblock \emph{\bibinfo{journal}{Trends in cognitive sciences}} \textbf{\bibinfo{volume}{21}}, \bibinfo{pages}{649--665} (\bibinfo{year}{2017}).

\bibitem{dasgupta2018remembrance}
\bibinfo{author}{Dasgupta, I.}, \bibinfo{author}{Schulz, E.}, \bibinfo{author}{Goodman, N.~D.} \& \bibinfo{author}{Gershman, S.~J.}
\newblock \bibinfo{title}{Remembrance of inferences past: Amortization in human hypothesis generation}.
\newblock \emph{\bibinfo{journal}{Cognition}} \textbf{\bibinfo{volume}{178}}, \bibinfo{pages}{67--81} (\bibinfo{year}{2018}).

\bibitem{kolb2014experiential}
\bibinfo{author}{Kolb, D.~A.}
\newblock \emph{\bibinfo{title}{Experiential learning: Experience as the source of learning and development}} (\bibinfo{publisher}{FT press}, \bibinfo{year}{2014}).

\bibitem{rutledge2014computational}
\bibinfo{author}{Rutledge, R.~B.}, \bibinfo{author}{Skandali, N.}, \bibinfo{author}{Dayan, P.} \& \bibinfo{author}{Dolan, R.~J.}
\newblock \bibinfo{title}{A computational and neural model of momentary subjective well-being}.
\newblock \emph{\bibinfo{journal}{{Proceedings of the National Academy of Sciences}}} \textbf{\bibinfo{volume}{111}}, \bibinfo{pages}{12252--12257} (\bibinfo{year}{2014}).

\bibitem{nguyen2019games}
\bibinfo{author}{Nguyen, C.~T.}
\newblock \bibinfo{title}{Games and the art of agency}.
\newblock \emph{\bibinfo{journal}{Philosophical Review}} \textbf{\bibinfo{volume}{128}}, \bibinfo{pages}{423--462} (\bibinfo{year}{2019}).

\bibitem{hintikka1981logic}
\bibinfo{author}{Hintikka, J.}
\newblock \bibinfo{title}{On the logic of an interrogative model of scientific inquiry}.
\newblock \emph{\bibinfo{journal}{Synthese}} \bibinfo{pages}{69--83} (\bibinfo{year}{1981}).

\bibitem{laudan1978progress}
\bibinfo{author}{Laudan, L.}
\newblock \emph{\bibinfo{title}{Progress and its problems: Towards a theory of scientific growth}}, vol. \bibinfo{volume}{282} (\bibinfo{publisher}{Univ of California Press}, \bibinfo{year}{1978}).

\bibitem{thurston1995proof}
\bibinfo{author}{Thurston, W.~P.}
\newblock \bibinfo{title}{On proof and progress in mathematics}.
\newblock \emph{\bibinfo{journal}{For the learning of mathematics}} \textbf{\bibinfo{volume}{15}}, \bibinfo{pages}{29--37} (\bibinfo{year}{1995}).

\end{thebibliography}


\begin{thebibliography}{10}
\expandafter\ifx\csname url\endcsname\relax
  \def\url#1{\texttt{#1}}\fi
\expandafter\ifx\csname urlprefix\endcsname\relax\def\urlprefix{URL }\fi
\providecommand{\bibinfo}[2]{#2}
\providecommand{\eprint}[2][]{\url{#2}}

\bibitem{van2023expertise}
\bibinfo{author}{van Opheusden, B.} \emph{et~al.}
\newblock \bibinfo{title}{Expertise increases planning depth in human gameplay}.
\newblock \emph{\bibinfo{journal}{Nature}} \bibinfo{pages}{1--6} (\bibinfo{year}{2023}).

\bibitem{palan2018prolific}
\bibinfo{author}{Palan, S.} \& \bibinfo{author}{Schitter, C.}
\newblock \bibinfo{title}{Prolific.ac---{A} subject pool for online experiments}.
\newblock \emph{\bibinfo{journal}{Journal of Behavioral and Experimental Finance}} \textbf{\bibinfo{volume}{17}}, \bibinfo{pages}{22--27} (\bibinfo{year}{2018}).

\bibitem{almaatouq2021empirica}
\bibinfo{author}{Almaatouq, A.} \emph{et~al.}
\newblock \bibinfo{title}{Empirica: a virtual lab for high-throughput macro-level experiments}.
\newblock \emph{\bibinfo{journal}{Behavior Research Methods}} \textbf{\bibinfo{volume}{53}}, \bibinfo{pages}{2158--2171} (\bibinfo{year}{2021}).

\bibitem{amir2022adaptive}
\bibinfo{author}{Amir, O.}, \bibinfo{author}{Tyomkin, L.} \& \bibinfo{author}{Hart, Y.}
\newblock \bibinfo{title}{Adaptive search space pruning in complex strategic problems}.
\newblock \emph{\bibinfo{journal}{PLoS Computational Biology}} \textbf{\bibinfo{volume}{18}}, \bibinfo{pages}{e1010358} (\bibinfo{year}{2022}).

\bibitem{crowley1993flexible}
\bibinfo{author}{Crowley, K.} \& \bibinfo{author}{Siegler, R.~S.}
\newblock \bibinfo{title}{Flexible strategy use in young children's tic-tac-toe}.
\newblock \emph{\bibinfo{journal}{Cognitive Science}} \textbf{\bibinfo{volume}{17}}, \bibinfo{pages}{531--561} (\bibinfo{year}{1993}).

\bibitem{silver2016mastering}
\bibinfo{author}{Silver, D.} \emph{et~al.}
\newblock \bibinfo{title}{Mastering the game of {Go} with deep neural networks and tree search}.
\newblock \emph{\bibinfo{journal}{Nature}} \textbf{\bibinfo{volume}{529}}, \bibinfo{pages}{484--489} (\bibinfo{year}{2016}).

\bibitem{cao2019uct}
\bibinfo{author}{Cao, X.} \& \bibinfo{author}{Lin, Y.}
\newblock \bibinfo{title}{Uct-adp progressive bias algorithm for solving gomoku}.
\newblock In \emph{\bibinfo{booktitle}{2019 IEEE Symposium Series on Computational Intelligence (SSCI)}}, \bibinfo{pages}{50--56} (\bibinfo{organization}{IEEE}, \bibinfo{year}{2019}).

\bibitem{sheoran2022solving}
\bibinfo{author}{Sheoran, K.}, \bibinfo{author}{Dhand, G.}, \bibinfo{author}{Dabas, M.}, \bibinfo{author}{Dahiya, N.} \& \bibinfo{author}{Pushparaj, P.}
\newblock \bibinfo{title}{Solving connect 4 using optimized minimax and monte carlo tree search}.
\newblock \emph{\bibinfo{journal}{Advances and Applications in Mathematical Sciences}} \textbf{\bibinfo{volume}{21}}, \bibinfo{pages}{3303--3313} (\bibinfo{year}{2022}).

\bibitem{lihongxun2025gobang}
\bibinfo{author}{Li, H.}
\newblock \bibinfo{title}{{gobang: JavaScript Gobang AI based on Alpha-Beta Pruning}}.
\newblock \bibinfo{howpublished}{\url{https://github.com/lihongxun945/gobang}} (\bibinfo{year}{2025}).

\bibitem{luce1959individual}
\bibinfo{author}{Luce, R.~D.}
\newblock \emph{\bibinfo{title}{Individual choice behavior}}, vol.~\bibinfo{volume}{4} (\bibinfo{publisher}{Wiley New York}, \bibinfo{year}{1959}).

\bibitem{train2009discrete}
\bibinfo{author}{Train, K.~E.}
\newblock \emph{\bibinfo{title}{Discrete choice methods with simulation}} (\bibinfo{publisher}{Cambridge University Press}, \bibinfo{year}{2009}).

\bibitem{franke2023softmax}
\bibinfo{author}{Franke, M.} \& \bibinfo{author}{Degen, J.}
\newblock \bibinfo{title}{The softmax function: Properties, motivation, and interpretation}.
\newblock \bibinfo{howpublished}{OSF Preprints} (\bibinfo{year}{2023}).

\bibitem{rubner1998metric}
\bibinfo{author}{Rubner, Y.}, \bibinfo{author}{Tomasi, C.} \& \bibinfo{author}{Guibas, L.~J.}
\newblock \bibinfo{title}{A metric for distributions with applications to image databases}.
\newblock In \emph{\bibinfo{booktitle}{Sixth international conference on computer vision ({IEEE} Cat. No. 98CH36271)}}, \bibinfo{pages}{59--66} (\bibinfo{organization}{{IEEE}}, \bibinfo{year}{1998}).

\bibitem{althofer2003computer}
\bibinfo{author}{Althofer, I.}
\newblock \bibinfo{title}{Computer-aided game inventing}.
\newblock \emph{\bibinfo{journal}{Friedrich Schiller University, Jena, Germany, Tech. Rep}}  (\bibinfo{year}{2003}).

\bibitem{browne2008automatic}
\bibinfo{author}{Browne, C.~B.}
\newblock \emph{\bibinfo{title}{Automatic generation and evaluation of recombination games}}.
\newblock Ph.D. thesis, \bibinfo{school}{Queensland University of Technology} (\bibinfo{year}{2008}).

\bibitem{todd2024gavel}
\bibinfo{author}{Todd, G.} \emph{et~al.}
\newblock \bibinfo{title}{Gavel: Generating games via evolution and language models}.
\newblock \emph{\bibinfo{journal}{Advances in Neural Information Processing Systems}} \textbf{\bibinfo{volume}{37}}, \bibinfo{pages}{110723--110745} (\bibinfo{year}{2024}).

\bibitem{chu2023praise}
\bibinfo{author}{Chu, J.}, \bibinfo{author}{Tenenbaum, J.~B.} \& \bibinfo{author}{Schulz, L.~E.}
\newblock \bibinfo{title}{In praise of folly: flexible goals and human cognition}.
\newblock \emph{\bibinfo{journal}{Trends in Cognitive Sciences}}  (\bibinfo{year}{2023}).

\bibitem{lantz2017depth}
\bibinfo{author}{Lantz, F.}, \bibinfo{author}{Isaksen, A.}, \bibinfo{author}{Jaffe, A.}, \bibinfo{author}{Nealen, A.} \& \bibinfo{author}{Togelius, J.}
\newblock \bibinfo{title}{Depth in strategic games.}
\newblock In \emph{\bibinfo{booktitle}{AAAI Workshops}} (\bibinfo{year}{2017}).

\bibitem{coulom2006efficient}
\bibinfo{author}{Coulom, R.}
\newblock \bibinfo{title}{Efficient selectivity and backup operators in monte-carlo tree search}.
\newblock In \emph{\bibinfo{booktitle}{International conference on computers and games}}, \bibinfo{pages}{72--83} (\bibinfo{organization}{Springer}, \bibinfo{year}{2006}).

\bibitem{genesereth2014general}
\bibinfo{author}{Genesereth, M.} \& \bibinfo{author}{Thielscher, M.}
\newblock \emph{\bibinfo{title}{General Game Playing}}.
\newblock Synthesis Lectures on Artificial Intelligence and Machine Learning (\bibinfo{publisher}{Morgan \& Claypool Publishers}, \bibinfo{year}{2014}).

\bibitem{silver2017mastering}
\bibinfo{author}{Silver, D.} \emph{et~al.}
\newblock \bibinfo{title}{Mastering chess and shogi by self-play with a general reinforcement learning algorithm}.
\newblock \emph{\bibinfo{journal}{arXiv preprint arXiv:1712.01815}}  (\bibinfo{year}{2017}).

\bibitem{uiterwijk2019solving}
\bibinfo{author}{Uiterwijk, J.~W.}
\newblock \bibinfo{title}{Solving strong and weak 4-in-a-row}.
\newblock In \emph{\bibinfo{booktitle}{2019 {IEEE} conference on games ({CIG})}}, \bibinfo{pages}{1--8} (\bibinfo{organization}{{IEEE}}, \bibinfo{year}{2019}).

\bibitem{nassar2016taming}
\bibinfo{author}{Nassar, M.~R.} \& \bibinfo{author}{Frank, M.~J.}
\newblock \bibinfo{title}{Taming the beast: extracting generalizable knowledge from computational models of cognition}.
\newblock \emph{\bibinfo{journal}{Current opinion in behavioral sciences}} \textbf{\bibinfo{volume}{11}}, \bibinfo{pages}{49--54} (\bibinfo{year}{2016}).

\end{thebibliography}


\begin{thebibliography}{1}
\expandafter\ifx\csname url\endcsname\relax
  \def\url#1{\texttt{#1}}\fi
\expandafter\ifx\csname urlprefix\endcsname\relax\def\urlprefix{URL }\fi
\providecommand{\bibinfo}[2]{#2}
\providecommand{\eprint}[2][]{\url{#2}}

\bibitem{van2023expertise}
\bibinfo{author}{van Opheusden, B.} \emph{et~al.}
\newblock \bibinfo{title}{Expertise increases planning depth in human gameplay}.
\newblock \emph{\bibinfo{journal}{Nature}} \bibinfo{pages}{1--6} (\bibinfo{year}{2023}).

\bibitem{browne2012survey}
\bibinfo{author}{Browne, C.~B.} \emph{et~al.}
\newblock \bibinfo{title}{A survey of monte carlo tree search methods}.
\newblock \emph{\bibinfo{journal}{{IEEE} Transactions on Computational Intelligence and {AI} in Games}} \textbf{\bibinfo{volume}{4}}, \bibinfo{pages}{1--43} (\bibinfo{year}{2012}).

\bibitem{openai2024openaio1card}
\bibinfo{author}{{OpenAI Team}} \emph{et~al.}
\newblock \bibinfo{title}{{OpenAI} o1 {System Card}} (\bibinfo{year}{2024}).
\newblock \urlprefix\url{https://arxiv.org/abs/2412.16720}.
\newblock \eprint{2412.16720}.

\bibitem{menendez1997jensen}
\bibinfo{author}{Men{\'e}ndez, M.~L.}, \bibinfo{author}{Pardo, J.~A.}, \bibinfo{author}{Pardo, L.} \& \bibinfo{author}{Pardo, M. d.~C.}
\newblock \bibinfo{title}{The {J}ensen-{S}hannon divergence}.
\newblock \emph{\bibinfo{journal}{Journal of the Franklin Institute}} \textbf{\bibinfo{volume}{334}}, \bibinfo{pages}{307--318} (\bibinfo{year}{1997}).

\end{thebibliography}
\end{document}